\documentclass{article}
\usepackage[margin=0.8in]{geometry}
\usepackage{amsmath,amssymb,amsfonts,algorithmic}
\usepackage{graphicx,setspace,multirow,url,hyperref,bm}
\usepackage{authblk}

\newcommand{\R}{\ensuremath{\mathbb{R}}}
\newcommand{\journal}[1]{}
\usepackage[sorting = none, backend = biber, style=numeric-comp]{biblatex}

\begin{document}

\title{Initial State Encoding via Reverse Quantum Annealing and h-gain Features}

\author[1]{Elijah Pelofske\thanks{Email: epelofske@lanl.gov}}
\author[2]{Georg Hahn}
\author[1,3]{Hristo Djidjev}

\affil[1]{Los Alamos National Laboratory, CCS-3 Information Sciences}
\affil[2]{Harvard University, T.H.\ Chan School of Public Health}
\affil[3]{Institute of Information and Communication Technologies, Bulgarian Academy of Sciences, Sofia, Bulgaria}

\date{\vspace{-6ex}}

\maketitle

\begin{abstract}
Quantum annealing is a specialized type of quantum computation that aims to use quantum fluctuations in order to obtain global minimum solutions of combinatorial optimization problems. Programmable D-Wave quantum annealers are available as cloud computing resources which allow users low level access to quantum annealing control features. In this paper, we are interested in improving the quality of the solutions returned by a quantum annealer by encoding an initial state into the annealing process. We explore two D-Wave features allowing one to encode such an initial state: the reverse annealing and the h-gain features. Reverse annealing (RA) aims to refine a known solution following an anneal path starting with a classical state representing a good solution, going backwards to a point where a transverse field is present, and then finishing the annealing process with a forward anneal. The h-gain (HG) feature allows one to put a time-dependent weighting scheme on linear ($h$) biases of the Hamiltonian, and we demonstrate that this feature likewise can be used to bias the annealing to start from an initial state. We also consider a hybrid method consisting of a backward phase resembling RA, and a forward phase using the HG initial state encoding. Importantly, we investigate the idea of iteratively applying RA and HG to a problem, with the goal of monotonically improving on an initial state that is not optimal. The HG encoding technique is evaluated on a variety of input problems including the edge-weighted Maximum Cut problem and the vertex-weighted Maximum Clique problem, demonstrating that the HG technique is a viable alternative to RA for some problems. We also investigate how the iterative procedures perform for both RA and HG initial state encoding on random whole-chip spin glasses with the native hardware connectivity of the D-Wave Chimera and Pegasus chips.
\end{abstract}
\section{Introduction}
\label{sec:introduction}
Quantum annealing is a form of specialized quantum computation that uses quantum fluctuations in order to search for the global minimum of a combinatorial optimization problem \cite{Kadowaki1998, Das2008, Morita2008, Hauke2020, Farhi2000, Chakrabarti2023}. programmable quantum annealers, available as cloud computing resources, are manufactured by D-Wave Systems, Inc., using superconducting flux qubits \cite{Johnson2011, Lanting2014, Boixo2013, Boixo2014, Mandra2016, King2022}. Quantum annealers are designed to minimize quadratic forms (called Hamiltonian functions), defined by
\begin{align}
    Q(x_1,\ldots,x_n) = \sum_{i=1}^n h_i x_i + \sum_{i<j} J_{ij} x_i x_j.
    \label{eq:hamiltonian}
\end{align}

The variables $x_i$ in eq.~\eqref{eq:hamiltonian} are unknown and take binary values only. The coefficients $h_i \in \R$ (linear weights) and $J_{ij} \in \R$ (quadratic couplers) are chosen by the user to define the problem under investigation, where $i,j \in \{1,\ldots,n\}$. If the variables $x_i \in \{0,1\}$ then eq.~\eqref{eq:hamiltonian} is called a \textit{QUBO} (\textit{quadratic unconstrained binary optimization}) problem, and if $x_i \in \{-1,+1\}$ it is called an \textit{Ising problem}, and importantly QUBO and Ising formulations are equivalent. Many important NP-hard problems map to simple minimization problems of the type of eq.~\eqref{eq:hamiltonian}, see \cite{Lucas2014}.

With newer generations of the D-Wave quantum annealer, an increasing number of features have been added to the machines that allow the user to obtain greater control over the annealing process. Such features include the spin reversal transform, customized anneal schedules, or anneal offsets for individual qubits. In this contribution, we focus on two of the latest features, called reverse annealing and \textit{time-dependent gain in linear biases}, also known as \textit{h-gain}. Both reverse annealing and h-gain protocols are specified as time dependent schedules over the course of the annealing process. 

This article investigates how both the reverse annealing (RA) and h-gain (HG) features can be used to improve the quality of an initial (suboptimal) solution, which is used to seed the annealing process. In fact, the purpose of RA is to guide the annealing process from a known classical (suboptimal) state to a point where more of the tranvserve field Hamltonian is present, before reversing direction again and continuing with a standard forward anneal (optionally with more complicated schedules, or simply to pause at an anneal fraction for some duration). We show that the HG feature allows one to achieve a similar goal as RA, though through a different mechanism. For both methods, one hopes that seeding the anneal with a solution that is close to optimal will help the annealer to transition to a better minimum, thereby improving the best found solution.

The HG feature allows one to put an additional weight on the linear terms in eq.~\eqref{eq:hamiltonian} in a time-dependent way. The feature was introduced to better study freeze-out points \cite{Johnson2011} and phase transitions in spin glasses \cite{Harris2018}. We show in this work that HG can also be used to bias the annealing process towards an initially computed solution. However, in contrast to RA, we only use a forward anneal. The mechanism we employ works as follows. We assume an Ising formulation in eq.~\eqref{eq:hamiltonian} with no linear terms. We then add to the given Ising formulation a new linear term that serves as a bias towards the known initial solution. Using the HG feature, we can put maximal weight on the linear terms at the start of the annealing process that biases the anneal towards the initial state, and decrease the HG strength during the annealing process to zero in order to make the annealer explore nearby solutions. An extension of this idea to Ising formulations with linear terms is presented as well. Finally, we explore schedules that combine both RA and HG in that they use a backward phase resembling an RA step and a forward phase that uses our HG idea.

Both the RA and HG techniques investigated in this work have a variety of tunable parameters. In particular, the RA feature requires the specification of annealing time duration and anneal schedule. The HG feature likewise requires annealing time,  HG schedule governing the time-dependent linear biases, and up to two additional parameters used to scale the appended linear terms in relation to the existing terms in eq.~\eqref{eq:hamiltonian}. To tune those parameters, we employ a Bayesian optimization framework \cite{Mockus1974}. The details of all optimizations being performed are given in the article, together with the best anneal schedules we found for RA and HG to encode initial solutions. These insights may prove valuable for users programming reverse annealing and h-gain schedules on D-Wave quantum annealers.

This article is a journal version and substantial extension of the conference paper of \cite{Pelofske_2020}, published in the \textit{QCE20 IEEE International Conference on Quantum Computing and Engineering}. In addition to the conference version, the present journal article also considers an application of RA and HG to Quantum Evolution Monte Carlo (QEMC), also known as iterative reverse annealing. We demonstrate that both RA and HG can be employed in quantum annealing to encode initial states in each iteration. We investigate the performance of such QEMC algorithms on random spin glasses with native D-Wave connectivity across three different D-Wave QPU's (Quantum Processing Units). The evaluated D-Wave QPU's have native hardware graphs known as Chimera (specifically with size $C_{16}$) \cite{zbinden2020embedding} and Pegasus (specifically with size $P_{16}$) \cite{dattani2019pegasus, boothby2020nextgeneration, zbinden2020embedding}. 

The article is structured as follows. We start with a brief literature review (Section~\ref{sec:litreview}) before introducing details on how to encode an initial state prior to an annealing process using the RA and/or HG features (Section~\ref{sec:methods}). That section also describes how we employ RA and HG in connection with QEMC. Section~\ref{sec:experiments} presents experimental results for the edge-weighted Maximum Cut problem (Section~\ref{sec:experiments_maxcut}), the vertex-weighted Maximum Clique problem (Section~\ref{sec:experiments_maxclique}), as well as QEMC (Section~\ref{sec:QEMC_experiments}). The article concludes with a discussion in Section~\ref{sec:discussion}. Certain tuning parameters we employed in our experiments are given in the appendix.

\section{Previous work}
\label{sec:litreview}
The new HG feature has not received too much attention in the literature to date. However, RA has been studied by several authors. Its idea was first introduced in \cite{Perdomo2011} under the name of \textit{sombrero adiabatic quantum computation}. Using tests on 3-SAT instances, the authors noted that the performance of RA was dependent on the Hamming distance between the planted initial solution and the optimal solution. However, the methodology of \cite{Perdomo2011} differs from the dissipative protocol actually used on the D-Wave devices in their \textit{reverse annealing} feature. The latter is closer to the protocol of \cite{Chancellor_2017}, who demonstrates how sequential calls to quantum annealers can be used to construct analogues of population annealing and parallel tempering which use quantum searches as subroutines.

The term "reverse annealing" is not uniquely defined in the literature and might refer to distinct protocols. An overview of those protocols can be found in \cite{Callison2022}.

Since its introduction, RA has been used in a wide range of practical applications, ranging from matrix factorization \cite{Golden2021} to first attempts on neural networks \cite{Rocutto2021} and portfolio optimization \cite{Grant2021}.

A theoretical contribution can be found in \cite{Ohkuwa2018}, where authors study conditions under which RA can lead to improvements over standard annealing for the fully connected $p$-spin model. They  present a theoretical framework to characterize such cases, but remark that their results do not necessarily apply to experimental setups where RA is performed adiabatically and in a thermal environment.

An application of adiabatic RA, which is a forward anneal similar to the HG version studied here, and iterative RA as used in the D-Wave annealer, is analyzed in \cite{Yamashiro2019}, using direct numerical integration of the time-dependent Schr{\"o}dinger equation. The authors find a theoretical speed-up of adiabatic RA over QA for mean-field-type $p$-spin models. However, they also find that iterative RA as used by D-Wave does not provide such advantage in theory, which is attributed to the fact that D-Wave is not a closed system, meaning that the theoretical results may not apply. However, a later publication \cite{Passarelli2022} concludes that indeed, standard quantum annealing can outperform adiabatic reverse annealing with decoherence.

A considerable speedup of RA over QA in a real-world application (portfolio optimization) is reported in \cite{Venturelli2019}, under the condition that RA starts from a planted heuristic solution.

Some differences of open system dynamics versus closed system RA is studied empirically in \cite{Passarelli2020} using $3$-spin models. The authors observe that RA with a pause converges to the ground state with a higher success probability than purely closed system RA, prompting the authors to conjecture that the open system dynamics makes RA work in devices such as the D-Wave annealer.

An interesting recent development in the field pertains to biased search protocols. Those work by initializing the anneal in an unequal (inhomogeneous) superposition of the possible states, thereby biasing the annealing dynamics into the desired solution using longitudinal fields \cite{QianHeng2013, Grass2019}, which is a very similar idea to the D-Wave hardware h-gain field (except that the h-gain field is uniformly applied to all qubits).

\section{Methods}
\label{sec:methods}
This section describes the techniques we use to encode an initial solution, both via the D-Wave's RA feature (Section~\ref{sec:reverse_annealing}), and with the help of suitably chosen additional linear term in connection with the HG feature (Sections~\ref{sec:hgain} and \ref{sec:hgain_linear}). Section~\ref{sec:ra+hg_method} describes the combined technique of RA+HG. Section~\ref{sec:parameters} states the final Hamiltonian we optimize and briefly describes the Bayesian optimization framework used to optimize the reverse quantum annealing and h-gain schedules. Section~\ref{sec:QEMC} defines the iterative state encoding procedure, also known as QEMC, when applied using reverse quantum annealing and the h-gain initial state encoding method. 

\subsection{Anneal paths based on reverse annealing}
\label{sec:reverse_annealing}

\begin{figure*}
    \centering
    \includegraphics[width=0.95\textwidth]{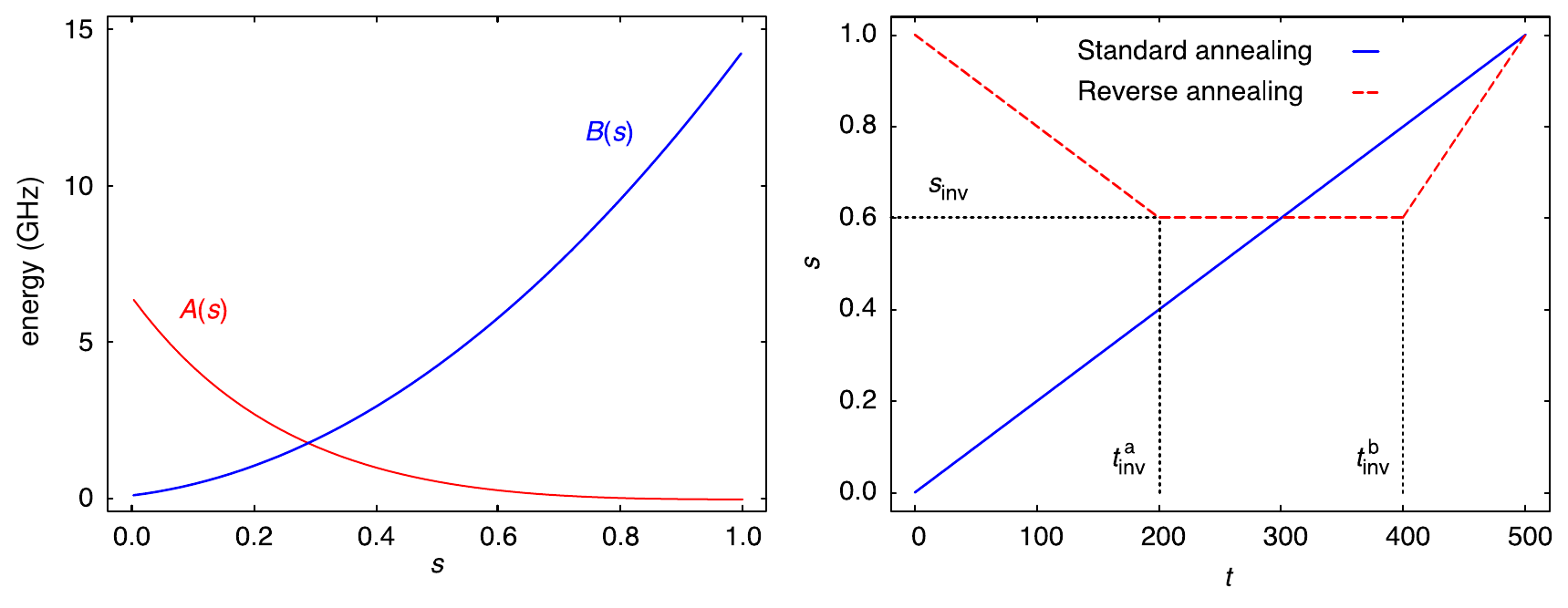}
    \caption{Left: Functions $A(s)$ and $B(s)$ controlling the annealing process, where $s \in [0,1]$ is the anneal fraction. Right: Progression of the anneal fraction $s$ for standard forward and reverse annealing with pause as a function of time $t \in [0,500]$ microseconds. Figure adapted from~\cite{Passarelli2020}. Note that the functions of $A(s)$ and $B(s)$ will change slightly depending on the quantum annealer. 
    \label{fig:RA}}
\end{figure*}

In a standard forward anneal (FA), all qubits are prepared in an equal superposition of all   states, as determined by the transverse field portion of the system's Hamiltonian. During annealing, the amplitude of the transverse field is being decreased towards $0$, while the Hamiltonian is slowly transformed into a Hamiltonian corresponding to the Ising problem being minimized. Specifically, the evolution of D-Wave's quantum system is described by the following time-dependent Hamiltonian
\begin{align}
    \nonumber
    H(s)=&-\frac{A(s)}{2}\Big(\sum_{i=1}^n \hat{\sigma}^{(i)}_x\Big) \\ &+\frac{B(s)}{2}\Big(\sum_{i=1}^nh_i\hat{\sigma}_z^{(i)} + \sum_{i\leq j} J_{ij} \hat{\sigma}_z^{(i)} \hat{\sigma}_z^{(j)}\Big), \label{eq:FA}
\end{align}
where the first term having the prefactor $-A(s)/2$ is the transverse field and the term following the prefactor $B(s)/2$ is the Hamiltonian corresponding to the Ising model of eq.~\eqref{eq:hamiltonian},  and $\hat{\sigma}^{(i)}_x$ and $\hat{\sigma}^{(i)}_z$ are the Pauli $x$ and $z$ operators operating on qubit $i$. The specific functions $A(s)$ and $B(s)$ used for the D-Wave 2000Q machine at Los Alamos are shown on Figure~\ref{fig:RA} (left). These functions are indexed by a parameter $s \in [0,1]$ called the \textit{anneal fraction}, which itself is a function $s(t)$ of the time $t$. In the case of the FA, it is given as $s(t)=t/T$, where $T$ is the total annealing time.

In contrast to FA, reverse annealing (RA) starts with a precomputed classical solution that is expected to be much closer in quality to an optimal one than a random starting point. Then, a two-stage process is initiated (see the red curve in Figure~\ref{fig:RA}, right), during which quantum fluctuations are first increased by reducing the anneal fraction from $s=1$ to a value $s_\text{inv}\in(0,1)$ at time $t_\text{inv}^a$. After the turning point is reached, and after an optional pause until time $t_\text{inv}^b$, the anneal follows again the path of a standard forward anneal from $s_\text{inv}$ up to $s=1$ at full annealing time $T$. Careful choices of the turning point and the initial state can lead to improvements in the solution compared to a forward anneal, see~\cite{Perdomo2011, King2018}.

\subsection{Anneal paths based on the h-gain schedule}
\label{sec:hgain}
The feature of a \textit{time-dependent gain in Hamiltonian linear biases} allows the user to have more control of the annealing process by biasing linear terms of an Ising model with the help of a time-dependent function $g(t)$ as follows:
\begin{align}
    \nonumber
    H_{\mathrm{HG}}(s) &= - \frac{A(s)}{2} \Big( \sum_{i=1}^n \hat{\sigma}_x^{(i)} \Big)\\
    &+ \frac{B(s)}{2} \Big( \sum_{i=1}^n g(t) h_i \hat{\sigma}_z^{(i)} + \sum_{i>j} J_{ij} \hat{\sigma}_z^{(i)} \hat{\sigma}_z^{(j)} \Big),
    \label{eq:hgain}
\end{align}
see~\cite{Harris2018}. Compared to eq.~\eqref{eq:FA}, we see that the linear terms of the Ising model in eq.~\eqref{eq:hgain} are weighted with a function $g(t)$, specified by the user, which controls the time-dependent gain for the linear terms. In our implementation, we initialize the function with $g(0) \in [0,5]$ (5 being the largest value allowed for D-Wave 2000Q) and decrease it to $g(T)=0$ using up to $20$ points on the schedule. The specification of the HG feature is actually more general than the way we use it in this work. For instance, the function $g(t)$ may actually return values in $[-5,5]$, it does not need to be monotonic, there is a (machine-dependent) upper bound of $500$ for the slope between changes in the schedule, and a (machine-dependent) upper bound of $20$ on the number of points determining the schedule~\cite{TechnicalDescriptionDwave}.

Our aim is to employ the HG feature to encode an initial solution at the start of the annealing process. Assume we are given an Ising problem of the type of eq.~\eqref{eq:hamiltonian} with no linear term, i.e., $h_i=0$ for all $i$. The idea lays in the observation that, for a fixed initial value $\bm{x^{(0)}} = (x_1^0,\ldots,x_n^0) \in \{-1,+1\}^n$, the minimum of the special Ising function containing only linear terms,
\begin{equation}
    \label{eq:H0}
    h(\bm{x}) = \sum_{i=1}^n (-x_i^0) x_i    
\end{equation}
for $\bm{x}=(x_1,\ldots,x_n)$, is equal to $-n$, and it occurs at $\bm{x}=\bm{x^{(0)}}$. Hence we can define $h_i=-x_i^0$ for $i=1,\dots,n$ and use a HG anneal schedule of the type of eq.~\eqref{eq:hgain}. By putting a large weight on the linear terms at the start of the anneal using the function $g(t)$, we bias the annealing solution towards our planted solution $\bm x^{(0)}$. Over the course of the anneal, the HG bias (the function $g(t)$ in eq.~\eqref{eq:hgain}) is decreased towards $0$, thus allowing the annealing process to move away from the planted solution and to explore alternative solutions in its neighborhood.

However, in order for this idea to work, the original Ising model may not have a linear term, so we can create our own linear term to encode the initial solution. For instance, Maximum Cut, Graph Partitioning, and Number Partitioning are such NP-hard problems without linear terms~\cite{Lucas2014}. Most Ising formulations of NP-hard problems, however, seem to have linear terms. Next, we will show that, even for such problems, the HG approach can be applicable.

\subsection{Using HG for Ising problems containing linear terms}
\label{sec:hgain_linear}
For problems whose Ising formulations do have linear terms, we apply the following transformation to eliminate them. First, we homogenize the polynomial in eq.~\eqref{eq:hamiltonian} by converting the linear term into a quadratic one. This is achieved by introducing a new variable $z \in \{-1,+1\}$, which we call a \textit{slack variable}. The slack variable $z$ is multiplied with each linear term, thus transforming eq.~\eqref{eq:hamiltonian} into
\begin{align}
    Q'(\bm{x},z) = \sum_{i=1}^n h_i x_i z + \sum_{i<j} J_{ij} x_i x_j.
    \label{eq:H'}
\end{align}
Note that $Q$ can be recovered from $Q'$ by setting $z=1$. Now we can apply the method as discussed in Section~\ref{sec:hgain}. After the end of the annealing process, we ignore all solutions with $z=-1$. We can guide the annealing process to favor solutions with $z=1$ by using an appropriate HG bias (initial solution).

\subsection{Reverse annealing and h-gain combined}
\label{sec:ra+hg_method}
The ideas of RA and HG can actually be combined into a single D-Wave schedule. To be precise, given an initial solution $\bm x^{(0)}$ to be encoded, we first apply the methodology of Sections~\ref{sec:hgain} and \ref{sec:hgain_linear} to arrive at a new Ising model encoding $\bm x^{(0)}$. We then solve the new Ising model using an RA schedule, which specifies the anneal fraction $s$ as a function of time, combined with an HG schedule, which specifies the gain $g(t)$ as a function of time.

If the HG Hamiltonian computed in Sections~\ref{sec:hgain} and \ref{sec:hgain_linear} requires a slack variable $z$, we also need to supply an initial state for $z$ when running RA. In order to reinforce $z=1$, we simply set $z=1$ in the RA initial state additionally to $\bm{x} = \bm x^{(0)}$. Note that the two schedules for RA+HG contain a total number of five parameters, three parameters for the RA schedule and two parameters for the HG schedule.

\subsection{Final Hamiltonian and tuning of parameters for minor embedded combinatorial optimization problem instances}
\label{sec:parameters}
To arrive at an effective implementation of the RA and HG methods, we need to determine appropriate values for a set of parameters, some optional, others required.

For HG, optional parameters are the coefficients $h_i$ from eq.~\eqref{eq:hgain}, for which we have so far suggested only their sign in eq.~\eqref{eq:H0}. While choosing individual weights for each $h_i$ will result in highest accuracy, it is also the most difficult to accomplish and beyond the scope of this paper. Instead, we use a single coefficient $\alpha_1$ for $i=1,\dots,n$ and, in the case when we need to homogenize the input Ising model, another coefficient $\alpha_2$ for the new variable $z$.

Combining the above, we encode an initial state using the Ising model
\begin{align}
    Q_\mathrm{final}(\bm{x},z)=\alpha_1\Big(\sum_{i=1}^n (-x_i^0) x_i\Big) - \alpha_2 z + Q'(\bm{x},z),
    \label{eq:H_final}
\end{align}
which is a function of $x_1,\ldots,x_n$ and $z$. The two scaling constants $\alpha_1$ and $\alpha_2$ allow us to control the strength enforcing the bias towards the initial solution and the condition that $z=1$. If the Ising model under consideration in eq.~\eqref{eq:hamiltonian} does not have a linear term, no new variable $z$ is needed and thus $\alpha_2=0$ in eq.~\eqref{eq:H_final}. When implementing eq.~\eqref{eq:H_final} on the D-Wave QPU, we set the internal option \texttt{}{autoscale} to on (which is the default option), thus scaling all quadratic terms to the range $[-1,+1]$ when programmed on \texttt{DW\_2000Q\_LANL}.

Apart from $\alpha_1$ and $\alpha_2$ in eq.~\eqref{eq:H_final}, the parameters that are required for both RA and HG are the schedule parameters. For RA we specifically target reverse annealing schedules with a single pause. Therefore, for RA we need the values $t_\text{inv}^a$, $t_\text{inv}^b$, and $s_\text{inv}$, see Figure~\ref{fig:RA}, plus the total annealing time $T$. For HG we need the function $g(t)$ given as a polygonal line subject to D-Wave's restrictions on magnitude, angles, and number of points. While there is some previous work that can be used as a guide for setting the schedule, in the case of HG there is no such previous work. Hence, we apply an optimization procedure for choosing the HG parameters and, in order to make a fair comparison between RA and HG, we use the same method for choosing the RA parameters.

We employ the following bayesian optimization procedure to tune the aforementioned parameters for the minor embedded combinatorial optimization problems, all on the D-Wave quantum annealing device with chip id \texttt{DW\_2000Q\_LANL}. The tuning is done separately for the two classes of minor embedded problems (weighted maximum clique and weighted maximum cut) that we study in more detail in the experiments of Section~\ref{sec:experiments}, the Maximum Cut and the Maximum Clique problems, as follows:

\begin{enumerate}
    \item We first fix the anneal time $T$, and then the anneal schedule for RA. After having determined $T$, we fix the starting point $(t=0,s=1)$ and the end point $(t=T,s=1)$, see Figure~\ref{fig:RA} (right). As in Figure~\ref{fig:RA} (right), we decrease the anneal fraction $s$ to a point $(t_\text{inv}^a,s_\text{inv})$. We then allow for a pause, meaning we also allow a point $(t_\text{inv}^b,s_\text{inv})$ at the same $s_\text{inv}$. All in all, we need to determine four parameters for RA: $T, t^a_\text{inv}, t^b_\text{inv},$ and $s_\text{inv}$. In practice, the way that the bayesian optimization procedure for RA (with a pause) is parameterized is using two real numbers both in range $[0.1, 0.9]$ - the first real number specifies at what point in the anneal (as a proportion of the total annealing), and the second number specifies the pause duration (as a proportion out of the available annealing time after the pause begins). These two parameters can completely specify symmetric reverse annealing schedules with pauses. Combined, the extreme ranges for these two parameters could allow the bayesian optimizer to propose schedules which would cause the D-Wave quantum annealer to return an error due to the anneal schedule changing too quickly, assuming that the full range of annealing times could be applied. Therefore, for RA schedule bayesian optimization experiments we set the minimum allowed annealing time to be $100$ microseconds, this ensures that all schedules that the bayesian optimizer would propose are valid. Specifically, this constraint means that the extreme choices of the RA schedule parameters could result in a change from $s=0$ to $s=1$ in $1$ microsecond, which is within the hardware anneal schedule slope constraints for \texttt{DW\_2000Q\_LANL}. 
    \item Similarly, for the h-gain schedule, we first fix $T$ and then the schedule's end points, starting at $(0,5)$ and ending at $(T,0)$. We allow for one point in-between, $(h,t)$, where $h \in [0,5]$ and $t \in (0,1)$. Together, three parameters are required for HG, that is, $T, h, t$. Note that such a shape for an HG schedule is by no means optimal, but we want to keep the number of parameters low to have a more manageable search space. However, before determining the schedule parameters, we first determine the best scaling factors $\alpha_1$ and $\alpha_2$ in eq.~\eqref{eq:H_final}. If the Ising model under consideration in eq.~\eqref{eq:hamiltonian} only has quadratic terms, homogenizing the polynomial is not necessary and we thus only need to find $\alpha_1$ in the Hamiltonian of eq.~\eqref{eq:H_final}. Otherwise, both $\alpha_1$ and $\alpha_2$ are determined. In practice, on the hardware, there are limitations on the slope of the h-gain schedule. These maximum h-gain schedule slopes are device dependent, but on the quantum annealer that was used for the minor embedded combinatorial optimization problems, which is \texttt{DW\_2000Q\_LANL}, we constrained the bayesian optimization search space for the h-gain point during the anneal to be between $[0.01, 0.99]$ as a proportion out of the total annealing time. Without these constraints, it would be possible for the bayesian optimizer to propose schedules which would cause the D-Wave backend to return an error. Under this constraint, the smallest annealing time available on \texttt{DW\_2000Q\_LANL}, $1$ microsecond, is feasible for the bayesian optimization to be applied since this would cause at most an h-gain field strength change of $0.01$ microseconds, which was within the h-gain field parameters of the device. 
    \item For the combined technique of RA+HG, after having determined the scaling constants $\alpha_1$ and $\alpha_2$ and the total annealing time $T$, we are left with five parameters determining the schedules: $t^a_\text{inv}, t^b_\text{inv},$ and $s_\text{inv}$ for RA, and $h, t$ for HG.
\end{enumerate}

\begin{figure*}[th!]
    \centering
    \includegraphics[width=0.49\textwidth]{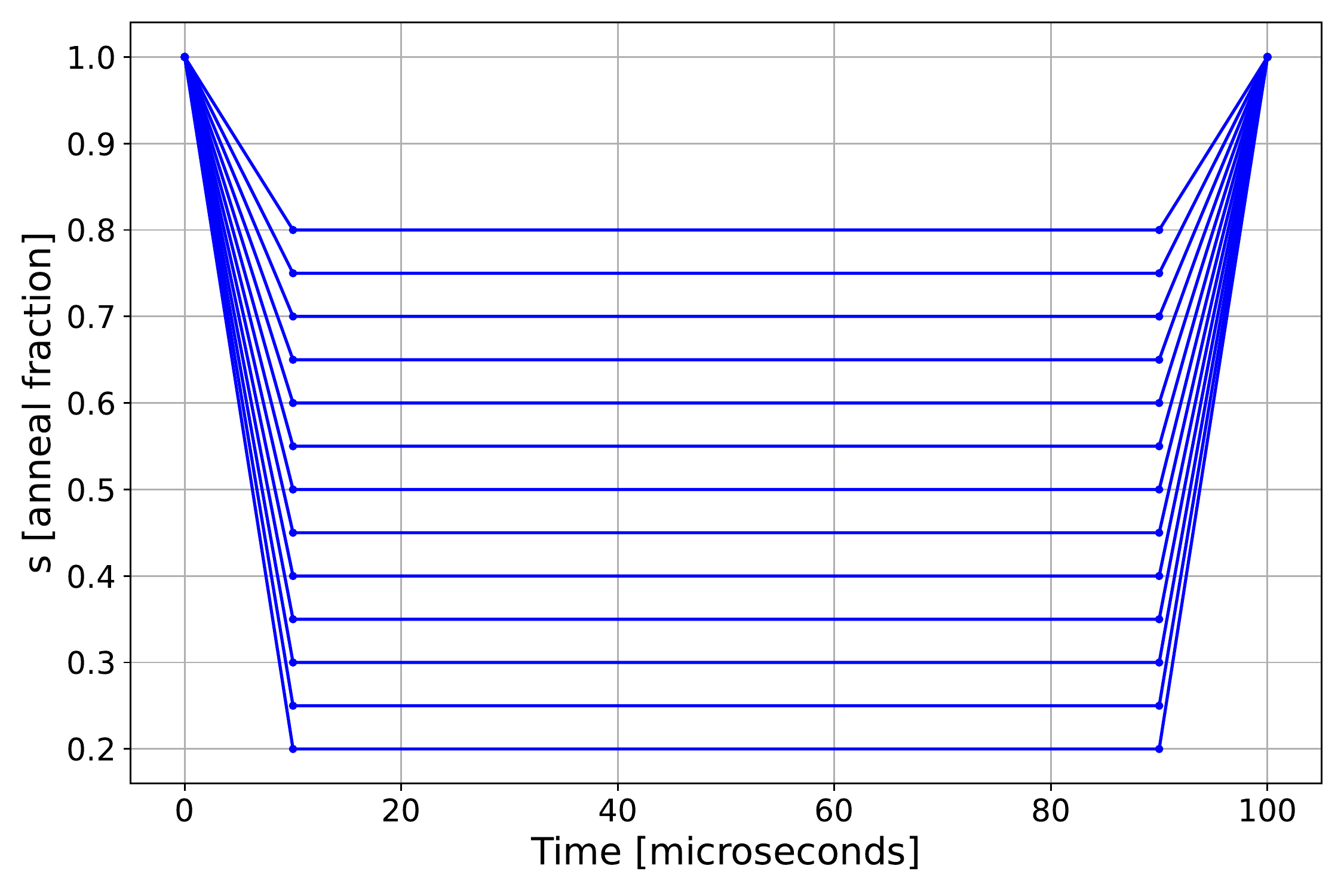}
    \includegraphics[width=0.49\textwidth]{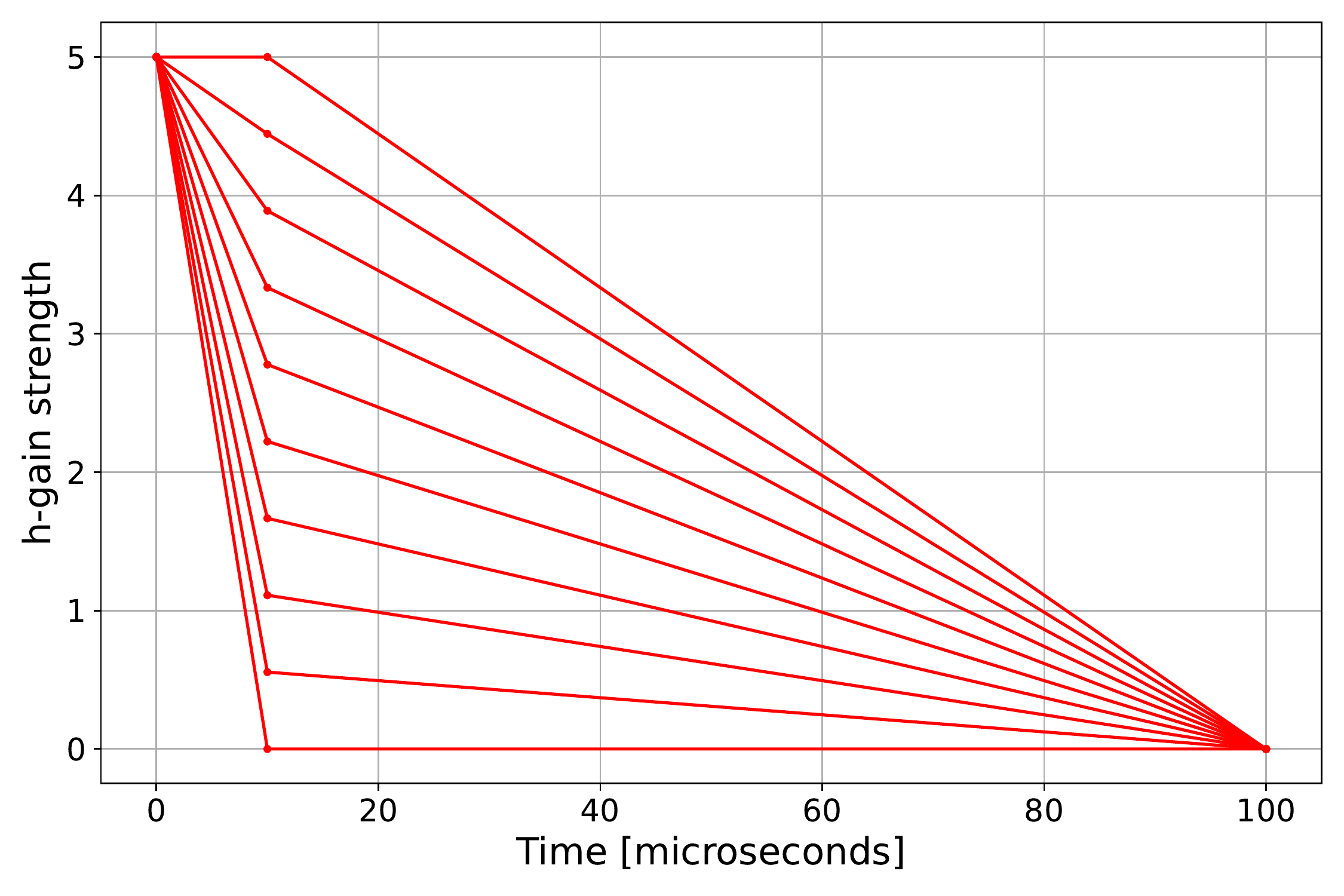}\\
    \includegraphics[width=0.49\textwidth]{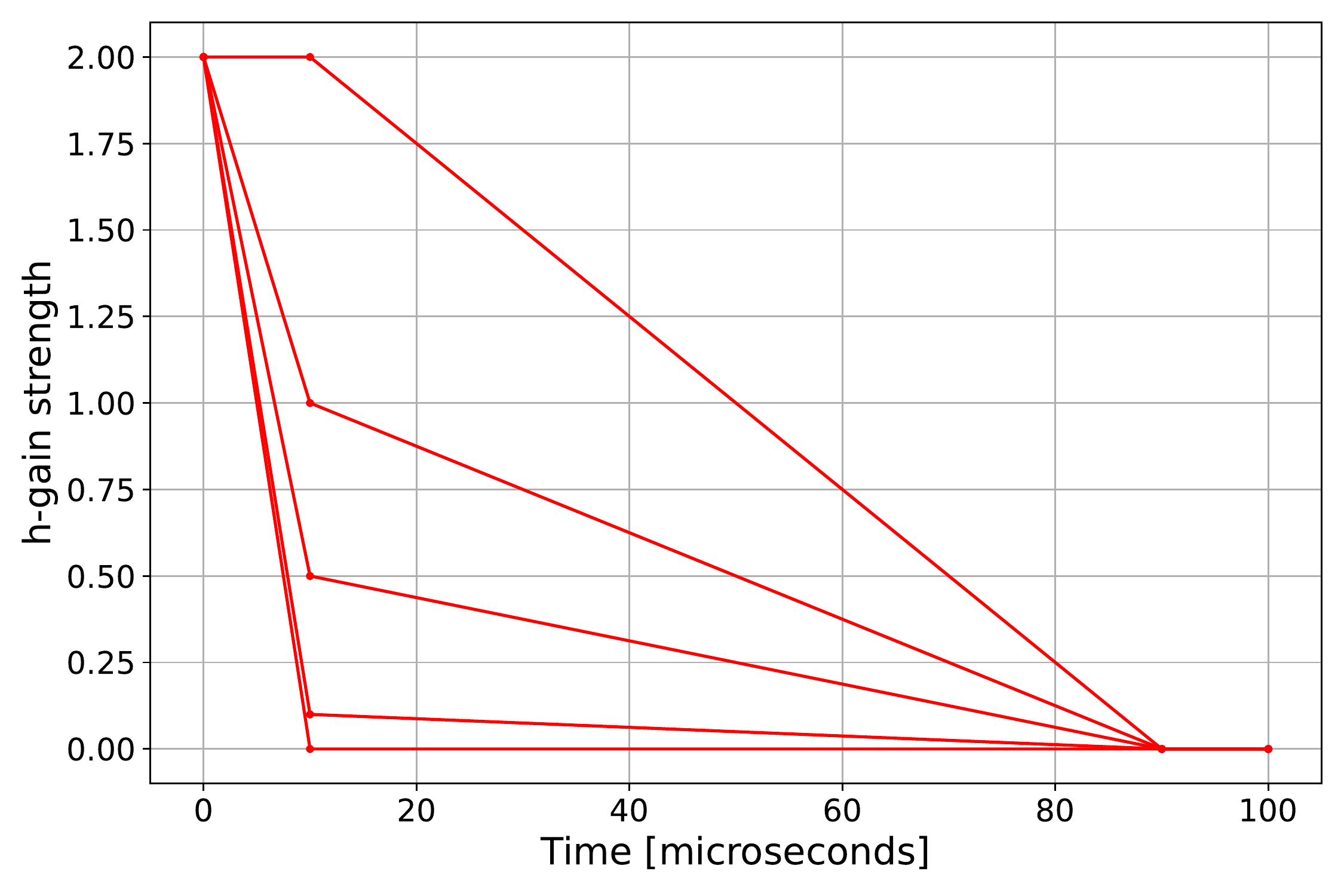}
    \includegraphics[width=0.49\textwidth]{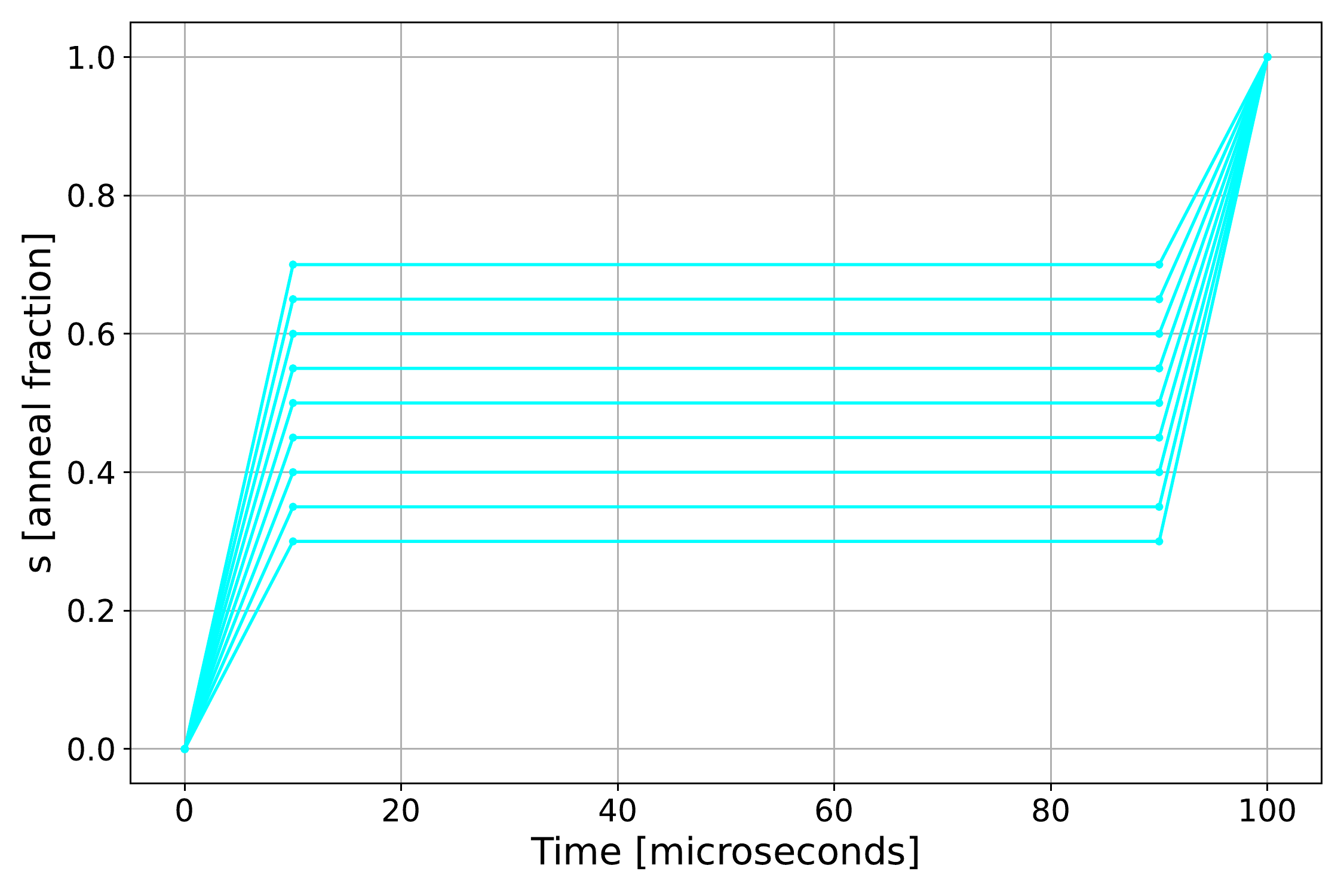}
    \caption{Range of RA schedules (top left), single point varied HG schedule with a fixed annealing time of $50$ microseconds (top right), single point varied HG schedule with an HG strength of $0$ at $90$ microseconds (bottom left), and forward anneal schedules with pauses (bottom right). The total annealing time for each schedule is $100$ microseconds. The annealing pauses for both RA and FA start at $10$ microseconds.}
    \label{fig:iterated_anneal_schedules}
\end{figure*}

To optimize all these parameters, we employ the Bayesian optimization tool of~\cite{bayesianopt}. Bayesian optimization \cite{Mockus1974, Mockus1977, Mockus1989} is a sequential optimization strategy to find the global optimum of a smooth function without the need for derivatives. An advantage of Bayesian optimization and the reason we chose it in this research is the fact that it also works with functions that are noisy, which is the case here since the function we optimize is based on the energy values returned by the D-Wave quantum annealer.

\subsection{Quantum evolution Monte Carlo}
\label{sec:QEMC}
A straightforward extension of the methodology presented in Sections~\ref{sec:reverse_annealing} and \ref{sec:hgain} is to iteratively apply the initial state encoding over some number of iterations, where the best solution from each previous iteration serves as the seed of the new anneal. This idea often referred to in the literature as \textit{quantum evolution Monte Carlo} (QEMC) \cite{King_2021, King2018, King_2021_scaling, PRXQuantum.1.020320, https://doi.org/10.48550/arxiv.2301.01853} also referred to as \textit{iterated reverse annealing} \cite{Yamashiro2019, PhysRevApplied.17.054033, PhysRevResearch.3.033006}. Initial state encodings are also known as warm starting in other contexts \cite{ash2020warm, Sack_2021, Egger_2021, https://doi.org/10.48550/arxiv.2207.05089}. Such approaches have been used for a variety of physics simulation computations on quantum annealers. By iteratively seeding each new anneal with the previously (best) obtained solution, one hopes to incrementally improve upon the solution quality. In the remainder of this article we will refer to this method as \texttt{QEMC}.

\begin{table*}[th!]
    \begin{center}
        \begin{tabular}{|l||l|l|l|}
        \hline
        D-Wave device chip ID & Topology name & Available qubits & Available couplers \\ 
        \hline
        \texttt{DW\_2000Q\_6} & Chimera $C_{16}$ & 2041 & 5974\\ 
        \texttt{Advantage\_system4.1} & Pegasus $P_{16}$ & 5627 & 40279\\ 
        \texttt{Advantage\_system6.1} & Pegasus $P_{16}$ & 5616 & 40135\\ 
        \hline
        \end{tabular}
    \end{center}
    \caption{Summary of D-Wave quantum annealing hardware used for the QEMC, or iterated reverse annealing, experiments. Note that each of these three devices have some hardware defects on the qubit topology which causes the available hardware (qubits and couplers) to be smaller than the ideal graph lattice structure.}
    \label{table:hardware_summary}
\end{table*}

We aim to employ both RA and HG to encode the initial states in QEMC before each new iteration. As in Section~\ref{sec:parameters}, these schedules have a variety of tuning parameters that we need to determine. Specifically, we test the following four different iterated state encoding methods:
\begin{enumerate}
    \item Reverse annealing with a single symmetric pause. Here we fix the anneal time and time for the ramps, but vary the anneal fraction $s$ at which the pause occurs. Example reverse annealing schedules are shown in Figure~\ref{fig:iterated_anneal_schedules} (top-left subplot).  
    \item HG initial state encoding only. We design the HG schedules to be monotonically decreasing h-gain value over time, where the first and last point of the schedule are fixed and the middle point can be varied both in terms of the time point and the HG strength. Figure~\ref{fig:iterated_anneal_schedules} (top-right subplot) shows the functional shape of these HG schedules, where we have fixed the time at which the change occurs to be $10$ microseconds into the $100$ microsecond anneal. The HG schedules are applied at the same time as the default linearly interpolated forward anneal schedule. The earlier in the schedule the HG field is set to $0$, the less influence the initial state has on the evolution of the annealing process. Conversely, the later in the annealing process the HG field is set to $0$, the greater the influence the initial state has on the evolution of the quantum state. Importantly, the maximum h-gain field used in Figure~\ref{fig:iterated_anneal_schedules} (top-right subplot) varies depending on the device, but the plot shows an example where the h-gain field change occurs at steps linearly spaced between $0$ and $5$. 
    \item Reverse quantum annealing schedule, with a symmetric pause, combined with the HG initial state encoding method. This method involves specifying a RA schedule as shown in Figure~\ref{fig:iterated_anneal_schedules} (top-left subplot), and an h-gain schedule as shown in Figure~\ref{fig:iterated_anneal_schedules} (bottom-left subplot). In this case, we modify the HG schedule slightly so that it reaches an HG strength of $0$ at the same time that the RA schedule begins to move back towards the readout state ($s=1$). However, the HG schedule is still monotonically decreasing.
    \item FA schedule with a symmetric pause combined with the HG initial state encoding method. In this case, we combine a FA schedule with a symmetric pause as shown in Figure~\ref{fig:iterated_anneal_schedules} (bottom-right subplot) with an HG schedule shown in Figure~\ref{fig:iterated_anneal_schedules} (bottom-left subplot). As in the previous case that combined FA and HG, the HG schedule here is constructed such that it reaches an HG strength of $0$ at the same time in the annealing process that the FA schedule stops pausing before it continues up to $s=1$. 
\end{enumerate}

Each of these four initial state encoding methods are tested on the three D-Wave quantum annealers shown in Table~\ref{table:hardware_summary}.

We apply QEMC to random Ising models (see Section~\ref{sec:introduction}), where the couplers in each problem are designed to match the hardware graph of each the three QPUs in Table~\ref{table:hardware_summary}, meaning that the test Ising models use the entire quantum annealer chip making the search space for the optimal variable assignments quite large. An example of those hardware graphs is given in Appendix~\ref{sec:hardware_graph_spin_glasses}.

We test three different weight precision levels: $10$ linearly spaced weights between $-1$ and $1$ (not including $0$), $100$ linearly spaced weights between $-1$ and $1$, and $200$ linearly spaced weights between $-1$ and $1$ (excluding $0$ so that all coefficients are nonzero). The purpose of this approach is to push the limits of the precision with which the Ising coefficients can be mapped onto the quantum annealing hardware of the D-Wave devices. All problem instances we test only contain quadratic terms, thus leaving the linear terms free to encode the initial state in our proposed HG state encoding method (Section~\ref{sec:hgain}).

\section{Experimental analysis}
\label{sec:experiments}
This section reports on a variety of experiments conducted to assess the performance of both RA and HG, the combined of RA+HG, as well as QEMC for improving a planted solution. After introducing the experimental setting (Section~\ref{sec:setting}), the main experiments are divided into three subsections. First, we investigate two important NP-hard problems, the weighted Maximum Cut problem (Section~\ref{sec:experiments_maxcut}) and the weighted Maximum Clique problem (Section~\ref{sec:experiments_maxclique}). Last, the assessment of QEMC in Section~\ref{sec:QEMC_experiments} is performed on random spin glass models.

The experiments on minor embedded combinatorial optimization problems in Sections \ref{sec:experiments_maxcut} and \ref{sec:experiments_maxclique} are performed using a D-Wave 2000Q quantum annealer located at Los Alamos National Laboratory with chip id \texttt{DW\_2000Q\_LANL}. The experiments in Section~\ref{sec:QEMC_experiments} are performed using three other D-Wave quantum annealers with chip ids \texttt{DW\_2000Q\_6}, \texttt{Advantage\_system4.1}, \texttt{Advantage\_system6.1}. 

\subsection{Experimental setting}
\label{sec:setting}
The structure of Sections~\ref{sec:experiments_maxcut} and \ref{sec:experiments_maxclique} is identical: We first fix the scaling constants in eq.~\eqref{eq:H_final} for HG before we determine a suitable annealing duration for applying each of the three methods. Afterwards, we employ Bayesian optimization to determine the best anneal schedule, parameterized as described in Section~\ref{sec:parameters}. Once both the annealing duration and the anneal schedule are found for each of the RA, HG, and RA+HG methods, we evaluate all three techniques with respect to either the cut value (for the edge-weighted Maximum Cut problem) or the clique weight (for the vertex-weighted Maximum Clique problem).

The experiments on weighted maximum cut and weighted maximum clique problems used Erd{\H o}s--R{\'e}nyi random graphs~\cite{ErdosRenyi1960} with probability/density parameter $p$, where $p \in \{0.1,0.2,\ldots,0.9\}$. Once the Ising model coefficients for the Maximum Cut or Maximum Clique problem are computed for each test graph, we embed it using the tool {\tt minorminer} \cite{Cai2014, minorminer} (generating random minor embeddings, produced with all default minorminer parameters) using a chain strength value of $2$ and the default settings on the D-Wave devices. Note that in general, without custom algorithms for structured minor embeddings \cite{boothby2016fast}, random minor embeddings are difficult to compute and NP-hard in general \cite{lobe2021minor, zbinden2020embedding, Lobe_2021}. Minor embeddings represent a logical problem graph on the hardware graph by linking together chains of qubits with strong ferromagnetic couplers to form a logical variable state, which ideally retains the same state during the annealing process and at readout. 

In order to have a baseline truth for comparing RA, HG, as well as RA+HG we proceed as follows. We generate random graphs with $65$ vertices, which is the largest random all-to-all minor embedding that can fit onto the \texttt{DW\_2000Q\_LANL} chip. This graph size ensures that even if the QUBO or Ising model is quite dense, there will still be a computable minor embedding. For each density,  $10$ of those graphs are fixed, along with their random minor embeddings; the random minor embeddings are computed individually for each graph, allowing the minorminer tool to reduce minor embedding chain lengths for sparser QUBOs or Ising models, and that embedding is then re-used when that graph is sampled again. We then perform $1000$ anneals of duration $1$ microsecond for each of the test instances. The best solution among those anneals is then taken as the baseline. When testing RA, HG and RA+HG, all values we report are averages over those $10$ graphs. Moreover, we generate another set of $10$ graphs for each density to use as a validation set. All samples (for the minor embedded maximum clique and maximum cut problems) are unembedded using the default method in the D-Wave SDK, which applies majority vote to chains which contain qubit state measurements that disagree on the logical state of the variable. 

Moreover, we employ the {\tt bayes\_opt} tool of~\cite{bayesianopt} using the following parameters: the number of points for random exploration is set to {\tt init\_points=100}, the number of iterations for optimization is set to {\tt n\_iter=200}, and the noise level is set to {\tt alpha=0.01}. The parameter {\tt alpha} indicates to the optimizer how noisy the optimization landscape is. Since D-Wave samples are quite noisy (in part simply due to the finite sampling effect), we observed that setting {\tt alpha} to a higher value, such as $0.01$, is favorable. However, we observe that large values of {\tt alpha} seem to cause an error in the optimizer, while smaller values lead to insufficient exploration of the optimization landscape.


In the experiments of Section~\ref{sec:QEMC_experiments}, we investigate random spin glass models which fit the D-Wave topology of the three annealers given in Table~\ref{table:hardware_summary}, sampled using iterated reverse annealing (also known as QEMC). The annealing time is always $100$ microseconds, and we read out $1000$ anneals. The readout thermalization and programming thermalization time were both set to $0$ microseconds. The boolean option to reduce intersample correlation was enabled. For experiments involving RA, the option \texttt{reinitialize\_state} was enabled. All schedules with a pause had a pause duration of $80$ microseconds, with $10$ microseconds for the ramp up and ramp down parts. The initial state for each random spin glass instance was determined by running a single job of $1000$ anneals at $100$ microsecond annealing time, using a standard forward anneal schedule. The sample with the best energy was used as the starting point for all iterative procedures. We repeat this across the different D-Wave devices (Table~\ref{table:hardware_summary}) and the different random spin glass coefficient precisions. In all QEMC applications we used exactly $20$ iterations, where the best sample found at the previous iteration seed the encoded states for the current iteration. 

The maximum HG strengths that can be applied are device dependent. For \texttt{DW\_2000Q\_LANL}, the maximum HG strength is $5$, for \texttt{DW\_2000Q\_6} the maximum HG strength is $5$, for \texttt{Advantage\_system4.1} it is $3$, for \texttt{Advantage\_system6.1} it is $4$. Note that although not used in these experiments, the maximum h-gain field strengths are sign-symmetric, so for example h-gain schedules with strengths of $-5$ are also valid for \texttt{DW\_2000Q\_LANL} and \texttt{DW\_2000Q\_6}. These maximum h-gain field strengths affect the range of possible h-gain schedules that can be applied for the h-gain state encoding methodology. Typically, for the QEMC experiments we will hold the h-gain schedules constant across the tested devices in order to perform a direct comparison. 

\begin{figure}[th]
    \centering
    \includegraphics[width=0.5\textwidth]{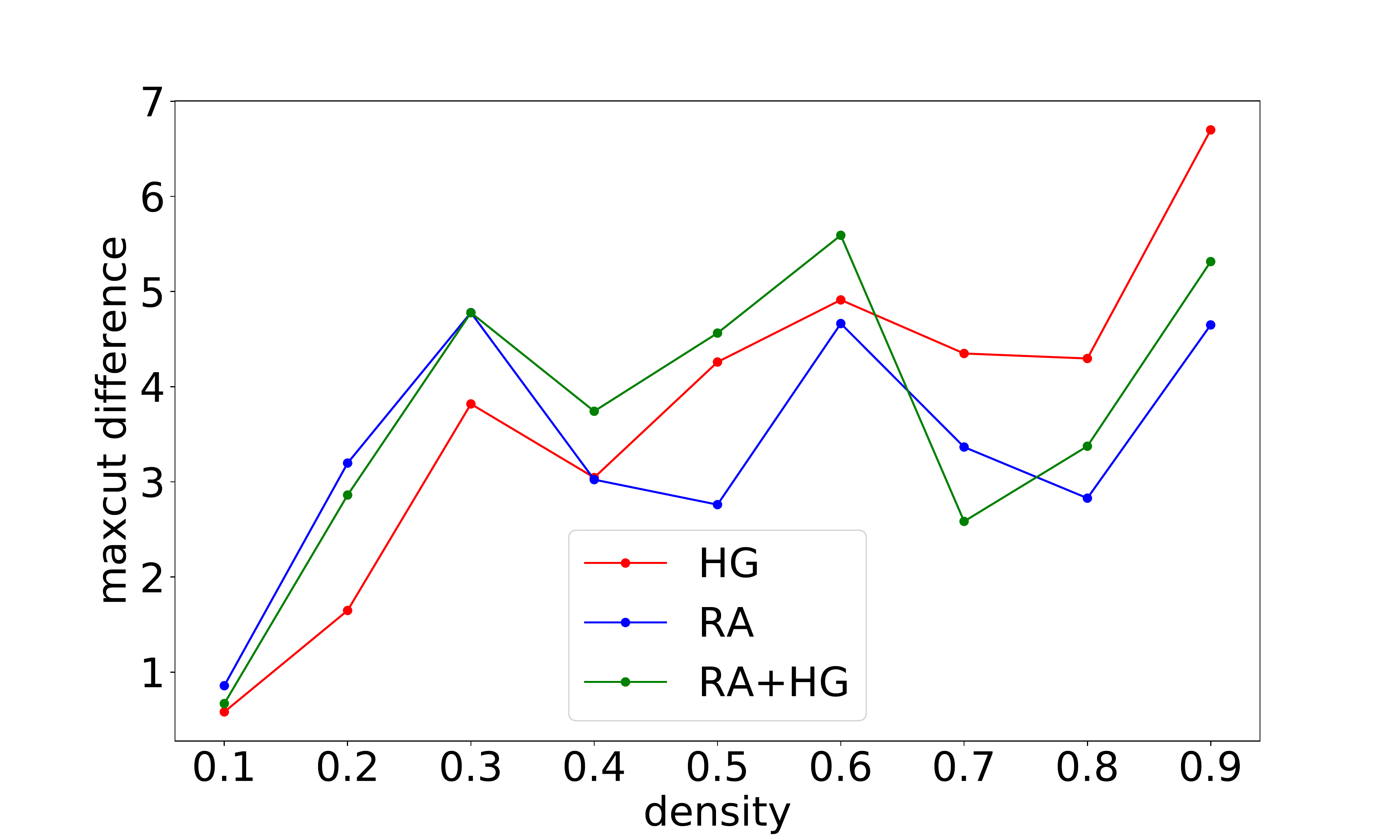}\hfill
    \caption{Comparison of RA, HG, and RA+HG with respect to the maximum cut improvement (the difference in maximum cut to the baseline value) per random graph density (x-axis). Best schedules obtained via Bayesian optimization. Plot uses a set of $10$ new (unseen) test graphs. Computed on \texttt{DW\_2000Q\_LANL}.}
    \label{fig:maxcut_densities_rerun}
\end{figure}

\begin{figure*}[th]
    \centering
    \includegraphics[width=0.5\textwidth]{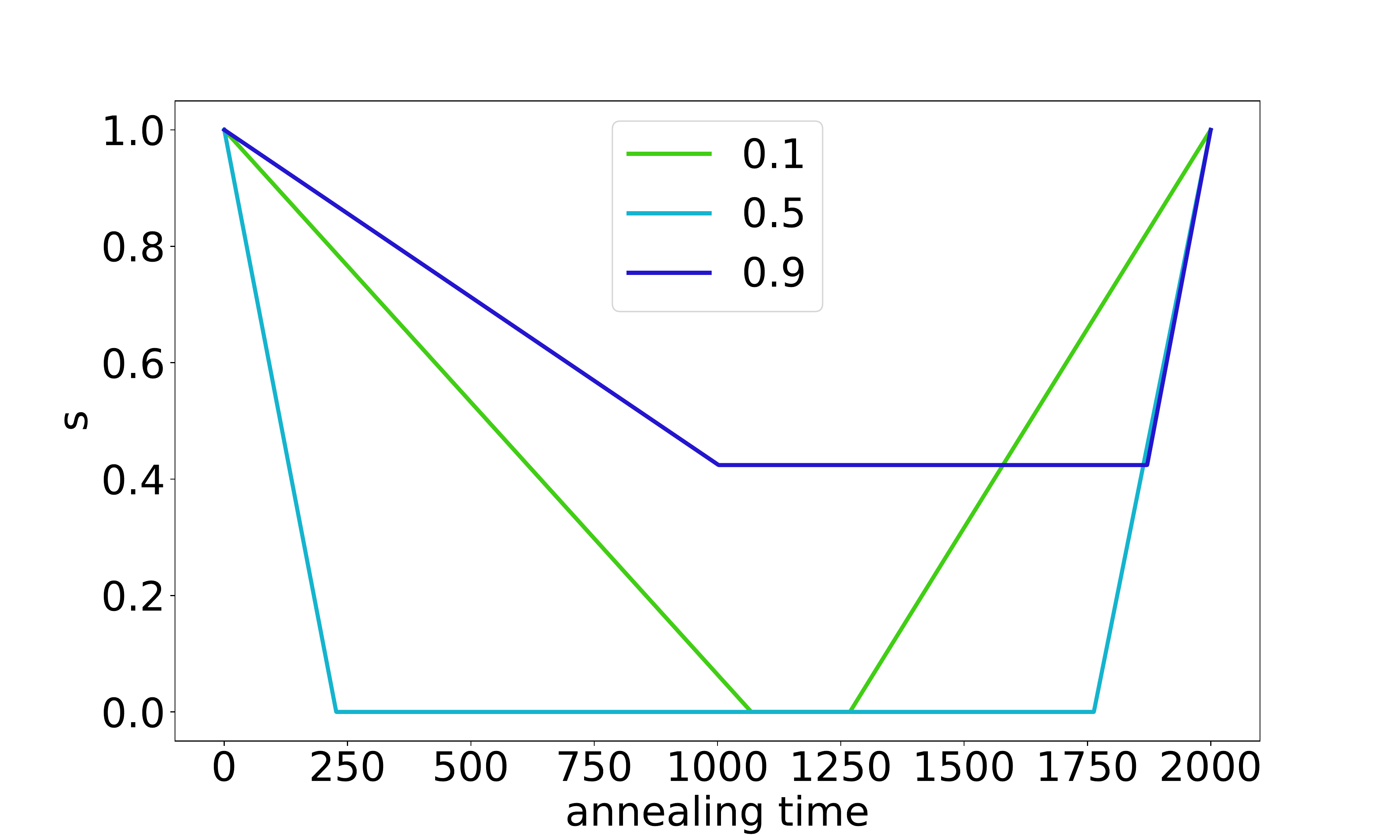}\hfill
    \includegraphics[width=0.5\textwidth]{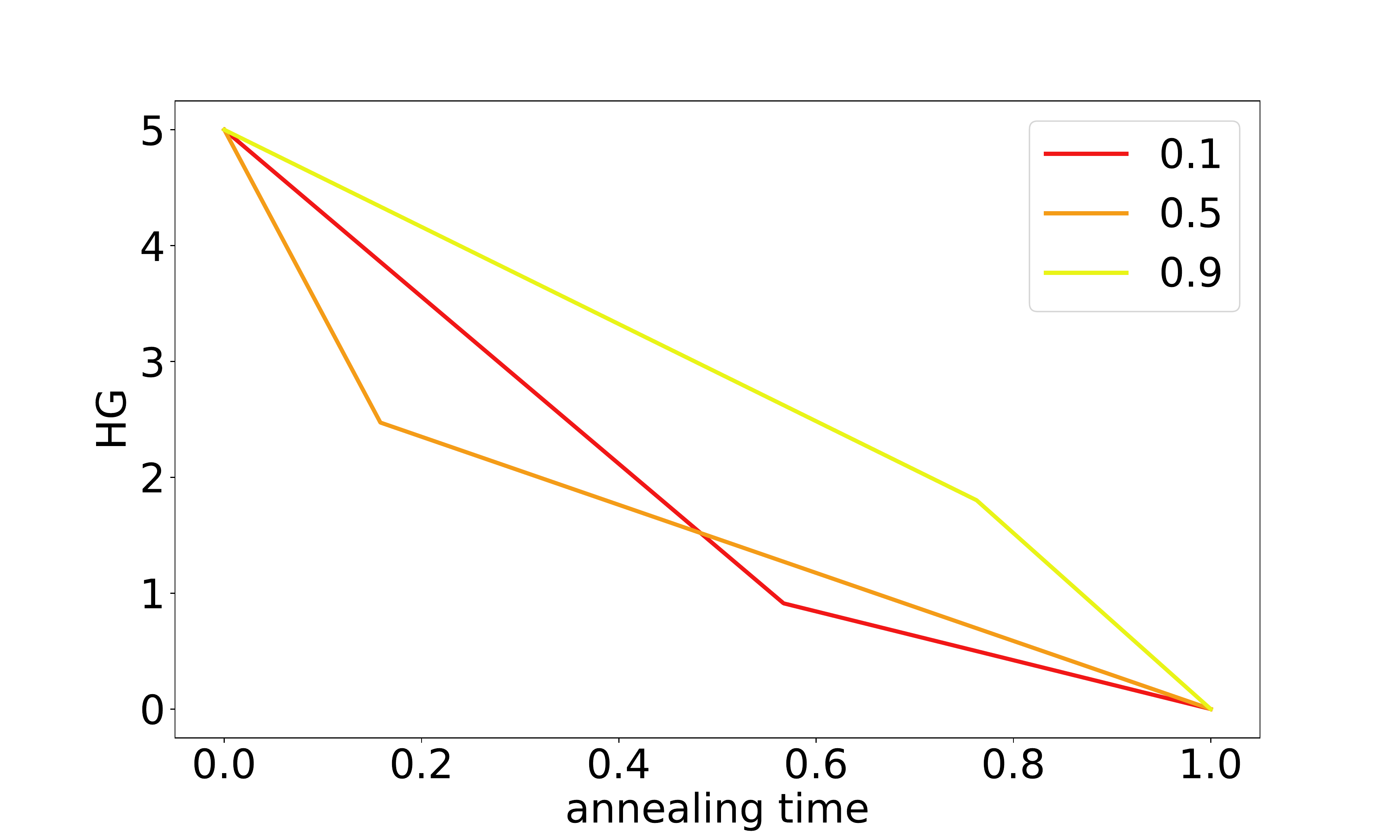}\hfill
    \caption{Maximum Cut problem. Best schedules for RA (left) and HG (right) for three different densities each, optimized for maximum cut difference. Each line is the best schedule for one density. These schedules were computed using the bayesian optimization approach, on \texttt{DW\_2000Q\_LANL}. }
    \label{fig:maxcut_schedules}
\end{figure*}

\subsection{Weighted Maximum Cut problem}
\label{sec:experiments_maxcut}
This section focuses on the edge weighted Maximum Cut problem, defined as follows. Given an undirected graph $G=(V,E)$ with edge weights $w(e)$ for each edge $e=(u,v)$ connecting two vertices $u,v \in V$, we define a \textit{cut} to be any partition of $V$ into the disjoint union $C_1 \cup C_2$, where $C_1 \subseteq V$ and $C_2=V \setminus C_1$. The set of cut edges, called \textit{cutset}, is defined as $\mathcal{E}=\{ e=(u,v) \in E: u \in C_1, v \in C_2\}$ and its weight is $\sum_{e \in \mathcal{E}} w(e)$. The \textit{weighted Maximum Cut problem} asks to find a cutset of maximum weight. The Ising formulation of the weighted Maximum Cut problem is obtained by modifying the (unweighted) formulation in~\cite{Hahn2017ReducingBQ}, resulting in
$$Q_{cut}(\bm{x})=\sum_{(i,j) \in E} w((i,j)) \cdot x_i x_j,$$
where $x_i, x_j \in \{-1,+1\}$. Since the Ising formulation of the Maximum Cut problem does not have linear terms, no slack variable $z$ is needed in the Ising formulation of eq.~\eqref{eq:H_final}. The scaling constant $\alpha_1$ for HG in eq.~\eqref{eq:H_final} is given in Appendix~\ref{sec:scaling_factors}.

For the weighted maximum cut problem graphs, random graphs are generating with edge probability $p \in \{0.1,0.2,\ldots,0.9\}$, and uniformly drawn edge weights in $(-1,1)$. 

Before comparing RA, HG, and RA+HG, a series of tuning parameters given in Section~\ref{sec:parameters} have to be determined. Details on these experiments can be found in Appendix~\ref{sec:tuning_maxcut}. The values of the scaling constants $\alpha_1$ and $\alpha_2$ we employ can be found in Table~\ref{tab:scaling_factors}. Moreover, we observe that an annealing duration of $2000$ microseconds works best for RA, while a $1$ microseconds anneal is best for HG (see Table~\ref{tab:maxcut}). When using RA+HG, we likewise employ an annealing time of $2000$ microseconds. The anneal schedules we use are given in Section~\ref{sec:experiments_maxcut_bayes}.

\subsubsection{Comparison of RA, HG, and RA+HG}
\label{sec:experiments_maxcut_comparison}
Having determined best schedule parameters for RA, HG, and RA+HG, we run the experiment again on 10 new graphs (per density) using these schedules. Figure~\ref{fig:maxcut_densities_rerun} shows results of this experiment. We observe that neither technique is uniformly better than the others. RA seems to be best for low densities, while HG and RA+HG perform best for high density graphs.

\begin{figure}[th]
    \centering
    \includegraphics[width=0.5\textwidth]{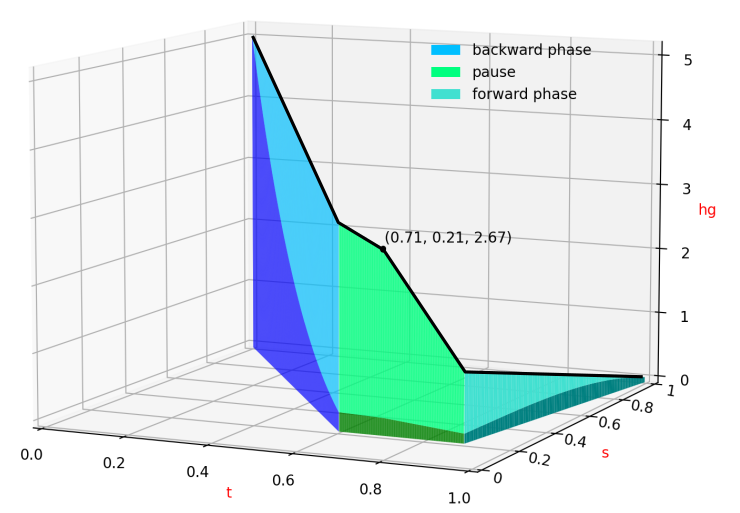}
    \caption{Illustration of an RA+HG schedule. Best schedule found for the weighted Maximum Cut problems for $p=0.3$ graph density using bayesian optimization, run on \texttt{DW\_2000Q\_LANL}. }
    \label{fig:ra+hg_3d}
\end{figure}

\subsubsection{Best schedules for RA, HG, and RA+HG}
\label{sec:experiments_maxcut_bestschedules}
It is interesting to look at the shape of some of the optimal schedules for RA and HG found by the Bayesian optimization. For this purpose, we visualize one example of a schedule for RA+HG.

Figure~\ref{fig:maxcut_schedules} shows the best schedules for RA and HG color-coded by density. For improved readability, we only display the schedules for $p \in \{0.1,0.5,0.9\}$.

We observe a pattern for the RA schedules in Figure~\ref{fig:maxcut_schedules} (left). In particular, when optimizing for maximum cut difference, RA schedules for low densities decrease down to an anneal fraction of zero, followed by a pause until roughly the midpoint of the anneal. In contrast, RA schedules for high densities only decrease to roughly an anneal fraction of $0.5$ at the midpoint of the anneal, followed by a pause until almost the full annealing time.

Similarly, a pattern can be observed for the HG schedules in Figure~\ref{fig:maxcut_schedules} (right). The HG schedules for low densities seem to have a steeper slope at the start of the anneal, and flatten off afterwards. In contrast, schedules for high densities seem to be closer to a straight line between the start point $(0,5)$ and the end point $(1,0)$.

An RA+HG anneal can be executed by sending to the D-Wave backend (in this case \texttt{DW\_2000Q\_LANL}) one RA schedule and one (independent) HG schedule. But while the RA aspect is easy to comprehend, the HG one is more difficult to grasp by just looking at the two component schedules because of the way the RA portion affects HG. Specifically, if $s=RA(t)$ and $g=HG(t)$ are the functions determined by the RA and HG schedules, respectively, then the real gain applied at time $t$ to the linear biases in eq.~\eqref{eq:hgain} is $(B(s)/2)HG(t)=B(RA(t))HG(t)/2$, where $B(s)$ is the function from eq.~\eqref{eq:FA}.

Figure~\ref{fig:ra+hg_3d} is given to help visualize the effect of an RA+HG schedule and the interplay between the parameters. The time $t$ is normalized in $[0,1]$. The RA component of the schedule, which has a pause for $t \in [0.6,0.89]$ at $s=0.21$ can be seen as a projection in the $t$-$s$ plane. The black line shows the HG values, specifically, the points $(t, s(t),hg(t))$ for $t\in(0,1)$.The HG schedule, which has middle point at $(t,hg)=(0.71,2.67)$, can be also seen as the lighter color projection in the $t$-$hg$ plane. The blue, green, and teal colors indicate the backward annealing, pause, and forward anneal phases, respectively. Finally, the cumulative gain applied to the linear biases at each time, which depends on both the values of HG and the annealing coefficient $B(s)$ from eq. (\ref{eq:hgain}), is represented by the darker colored portion of the plot. The annotated point shows the value of the h-gain (2.67) at the middle point of the HG schedule (at 0.71). To simplify the plot, function $B(s)$ has been normalized to $[0,1]$. We can see that the real gain applied during the pause and forward phases of the RA schedule stays mostly unchanged. These observations are, in a strict sense, valid for experimental setting of Section~\ref{sec:setting} only, although we anticipate them to hold true in greater generality for a broader class of problems.

\subsection{Weighted Maximum Clique problem}
\label{sec:experiments_maxclique}
We carry out a similar analysis for the vertex weighted Maximum Clique problem, defined as follows. For any graph $G=(V,E)$, a clique $C$ is a fully connected subset of vertices, i.e.\ $C \subseteq V$ such that $C \times C \subseteq E$. A maximum clique is a clique in $G$ of maximum size.

For the (vertex-)weighted version of the problem, we define a weight $w(v)$ for each vertex $v \in V$. The weight of a clique is accordingly defined as $w(C) = \sum_{v \in C} w(v)$. The weighted maximum clique problem asks for the clique $C \subseteq V$ having the largest weight $w(C)$. The QUBO formulation of the weighted Maximum Clique problem is obtained by modifying the (unweighted) formulation in~\cite{Chapuis2019}, resulting in
$$- \sum_{i=1}^n w(i) \cdot x_i + 2 \sum_{(i,j) \notin E} \max \{ w(i),w(j) \} \cdot x_i x_j,$$
where $x_i,x_j \in \{0,1\}$. We can convert the above QUBO formulation into an Ising problem using the equivalence given in \cite{Chapuis2019}. In contrast to the Maximum Cut problem investigated in Section~\ref{sec:experiments_maxcut}, the Maximum Clique formulation as an Ising model of the form of eq.~\eqref{eq:hamiltonian} does contain linear terms. We thus introduce a slack variable $z$ to homogenize the linear terms as in eq.~\eqref{eq:H'}, and add a new linear term encoding the initial solution as done in eq.~\eqref{eq:H_final}.

In the following experiments, we choose the vertex weights to be positive and randomly drawn from a uniform distribution in $(0.001,1)$.

As done for the weighted Maximum Cut problem, a set of parameters has to be determined for RA, HG, as well as RA+HG. Details on these experiments can be found in Appendix~\ref{sec:tuning_maxclique}. The values of the scaling constants $\alpha_1$ and $\alpha_2$ we employ can be found in Table~\ref{tab:scaling_factors}. Based on the parameter tuning conducted in Section~\ref{sec:tuning_maxclique}, we run RA with an annealing time of $2000$ microseconds in the remainder of this section, and HG with an annealing time of $1$ microseconds. For RA+HG we fix the annealing duration at $2000$ microseconds (see Table~\ref{tab:maxclique}). The schedules can be found in Section~\ref{sec:experiments_maxclique_bayes}.

\subsubsection{Comparison of RA, HG, and RA+HG}
\label{sec:experiments_maxclique_comparison}

\begin{figure}[th]
    \centering
    \includegraphics[width=0.5\textwidth]{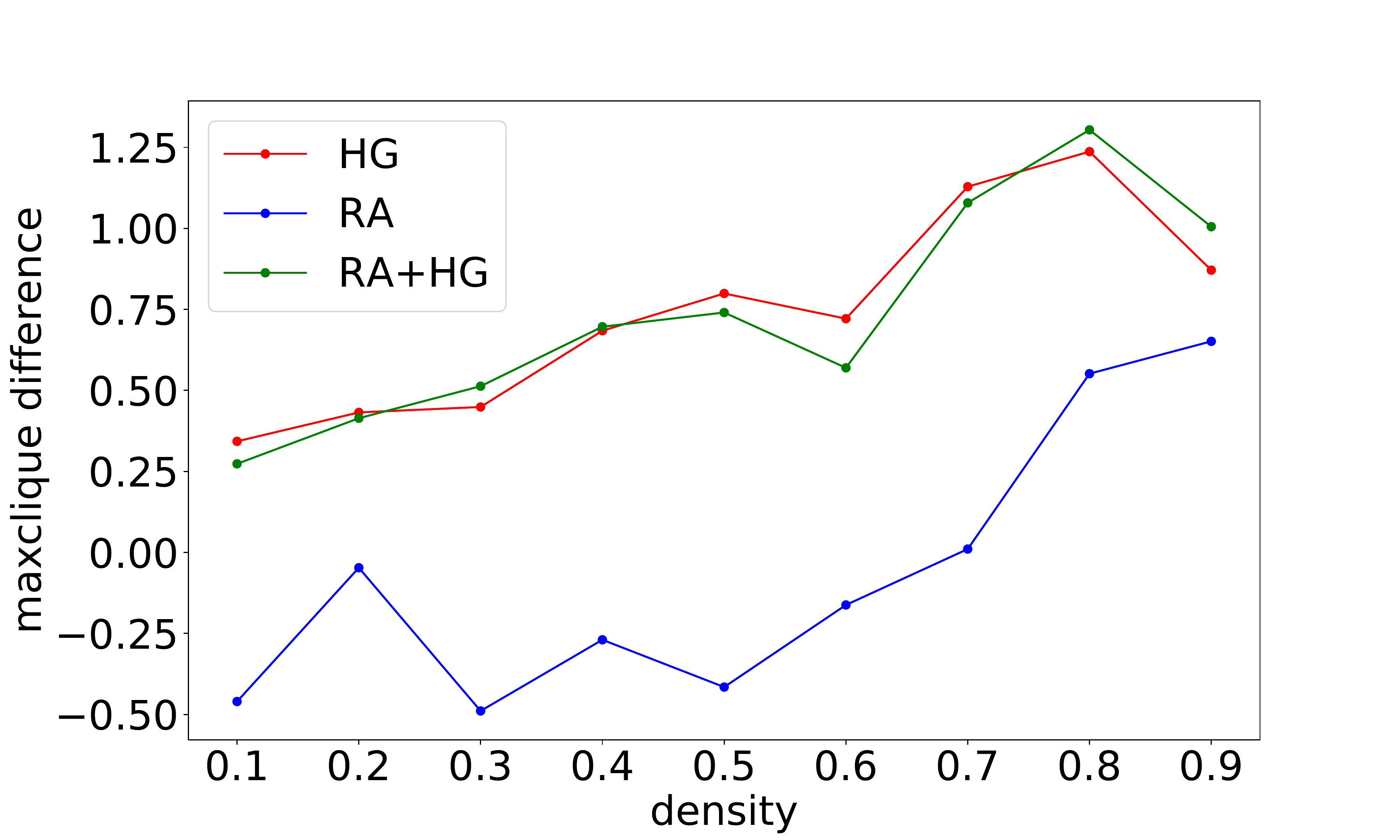}
    \caption{Comparison of RA, HG, and RA+HG with respect to the improvement in maximum clique weight (the difference in maximum clique weight to the baseline value) per random graph density (x-axis). Plot uses a set of $10$ new (unseen) test graphs. Computed on \texttt{DW\_2000Q\_LANL}.}
    \label{fig:maxclique_comparisons}
\end{figure}

As in Section~\ref{sec:experiments_maxcut_comparison}, we evaluate RA, HG as well as RA+HG after tuning the scaling factors, annealing durations, and schedules. Results are shown in Figure~\ref{fig:maxclique_comparisons} for $10$ new problems not used in the training set. We observe that the behavior of all three techniques is consistent: On the new problems, RA performs worst with the exception of graph density corresponding to $p=0.9$. Both HG and RA+HG perform very similarly and consistently better than RA, although they draw equal with RA for $p=0.9$.

This behavior is different from the equivalent experiment for Maximum Cut in Figure~\ref{fig:maxcut_densities_rerun}, where both HG and RA+HG were only marginally better than RA.

\begin{figure*}[ht!]
    \centering
    \includegraphics[width=0.32\textwidth]{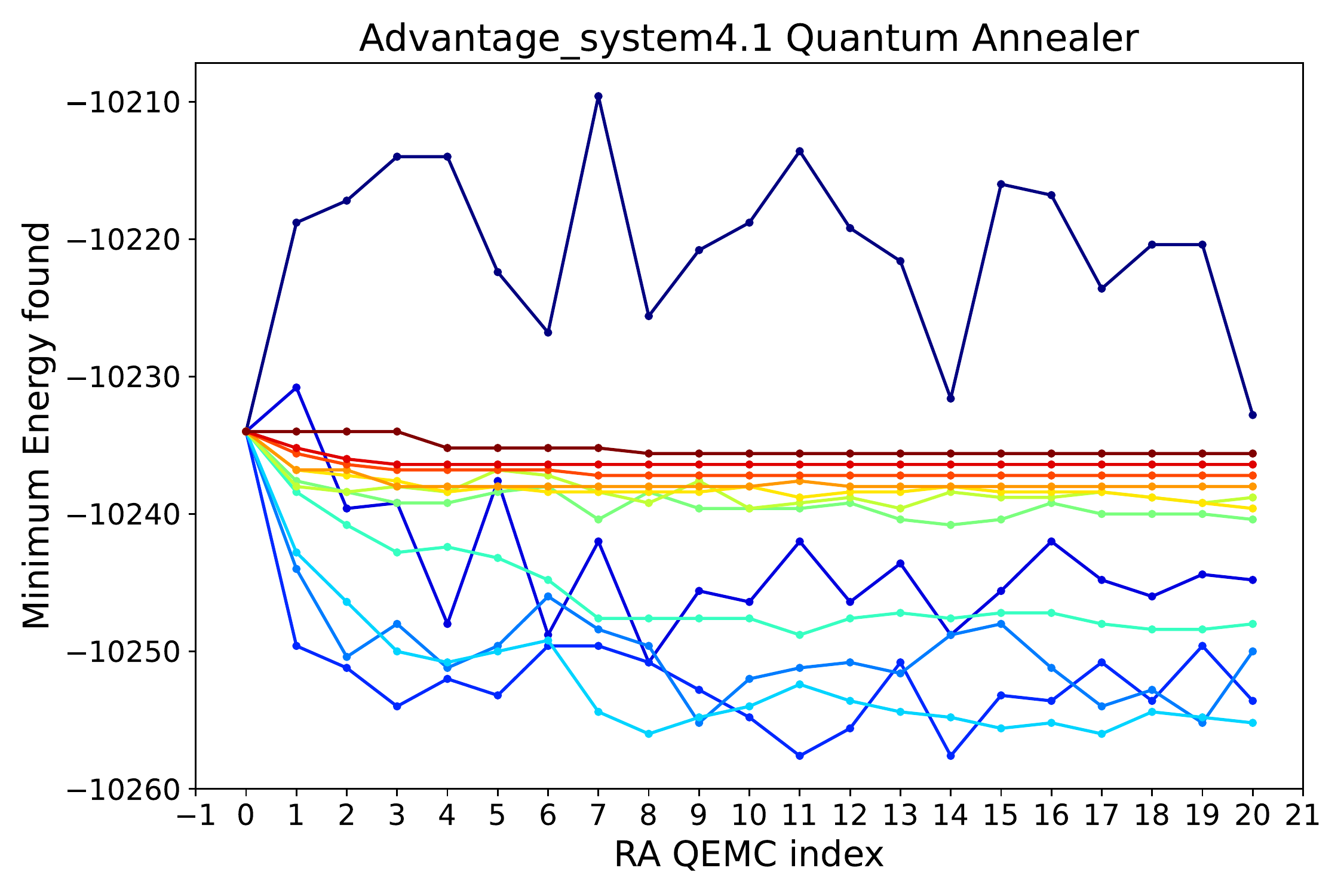}
    \includegraphics[width=0.32\textwidth]{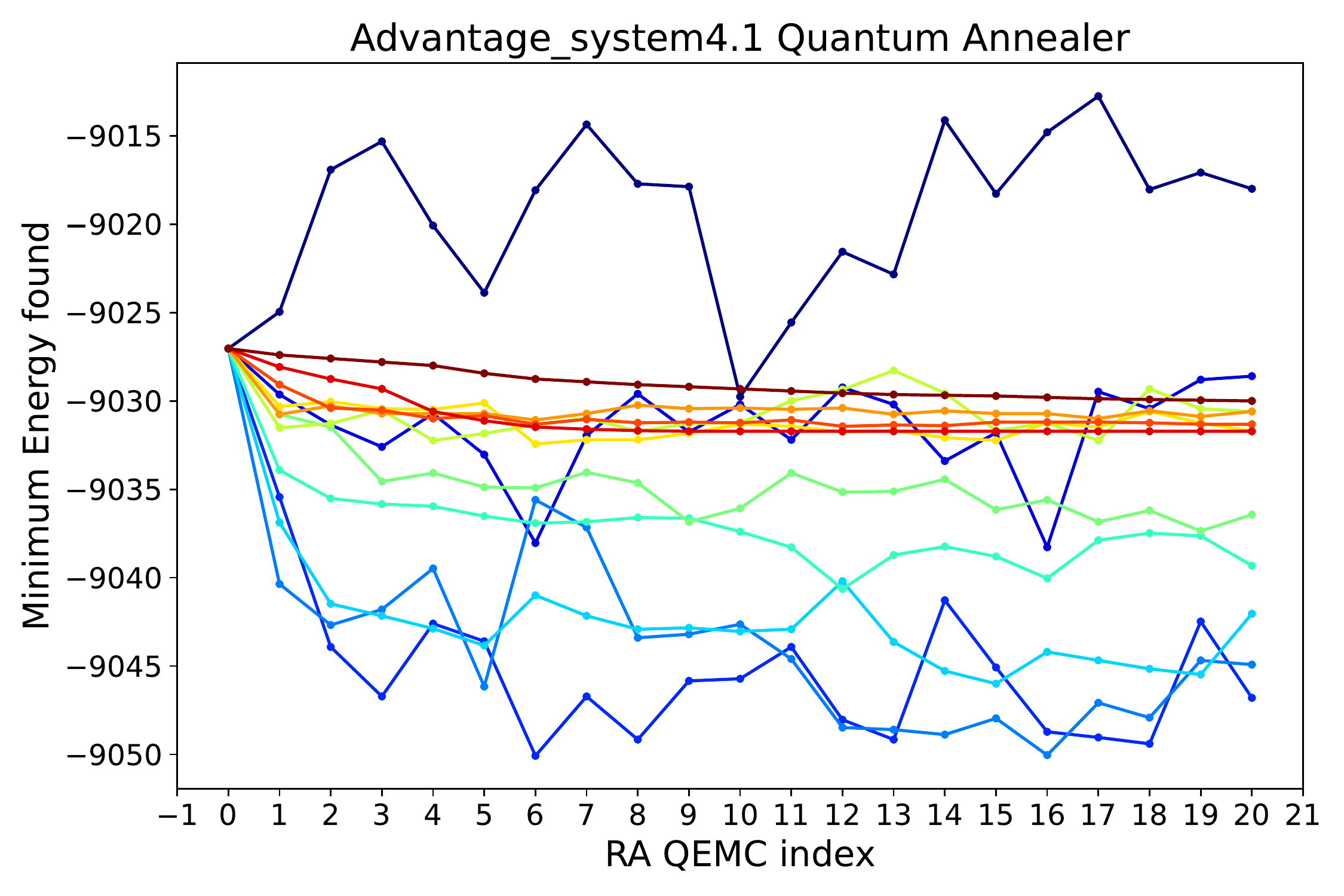}
    \includegraphics[width=0.32\textwidth]{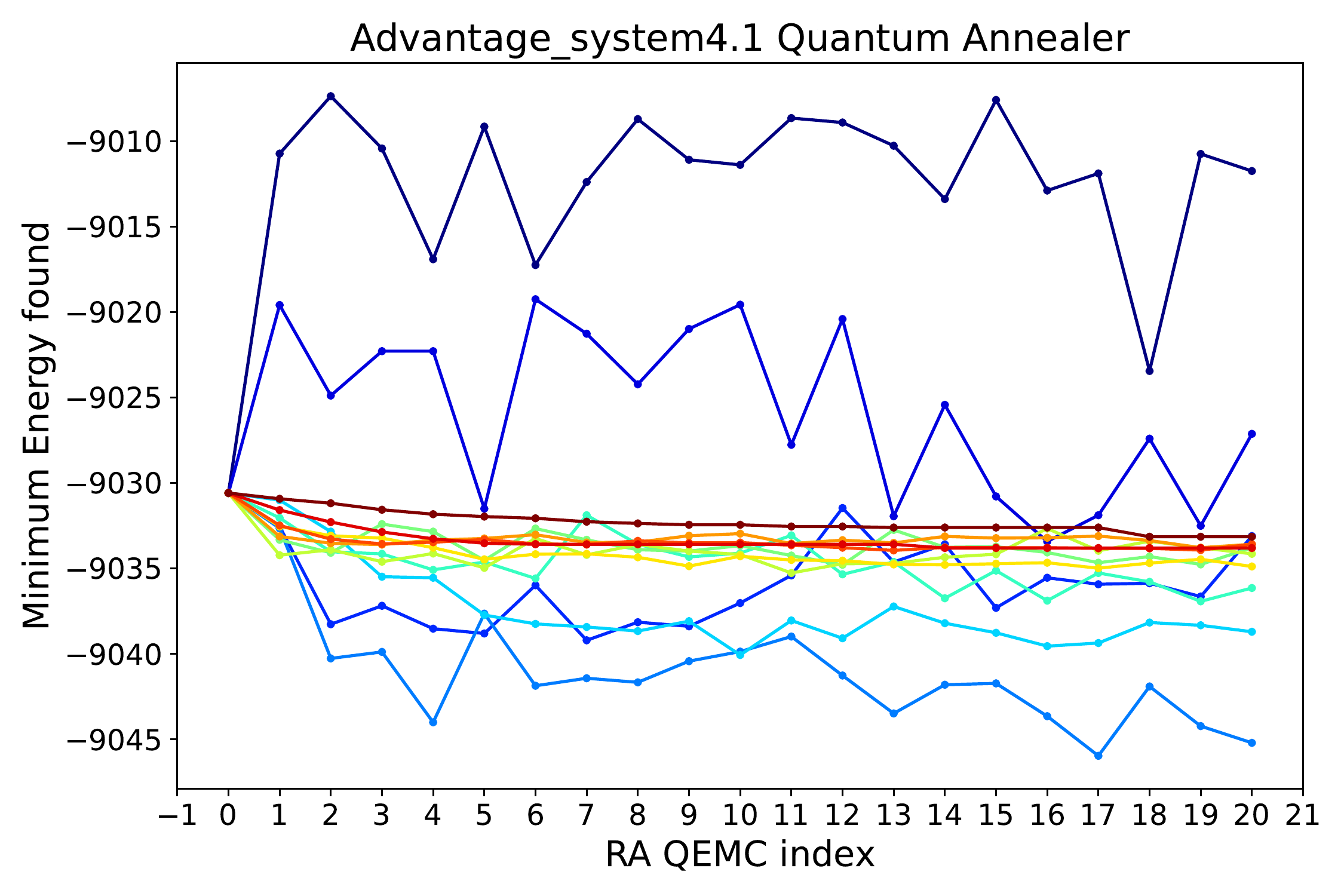}\\
    \includegraphics[width=0.32\textwidth]{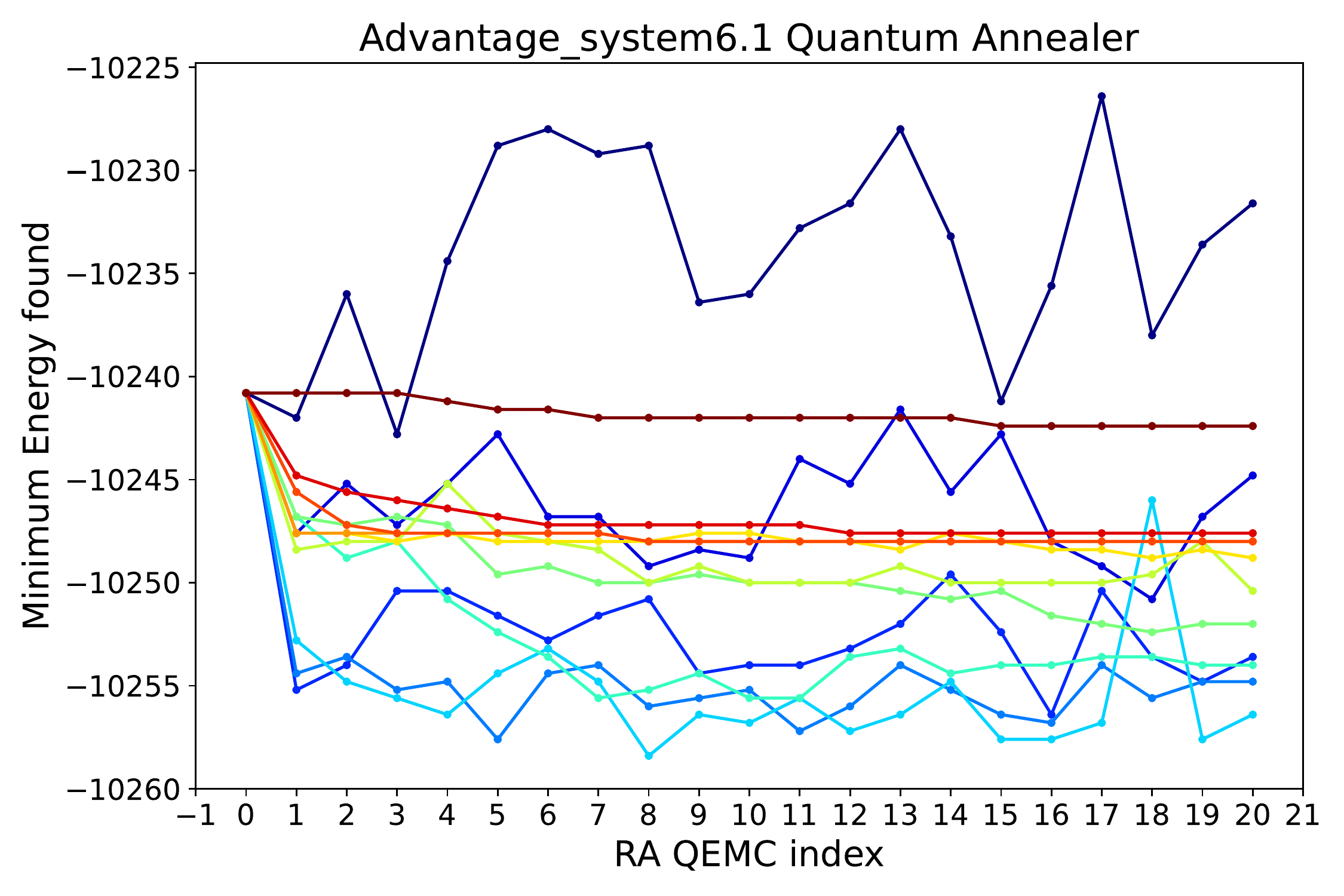}
    \includegraphics[width=0.32\textwidth]{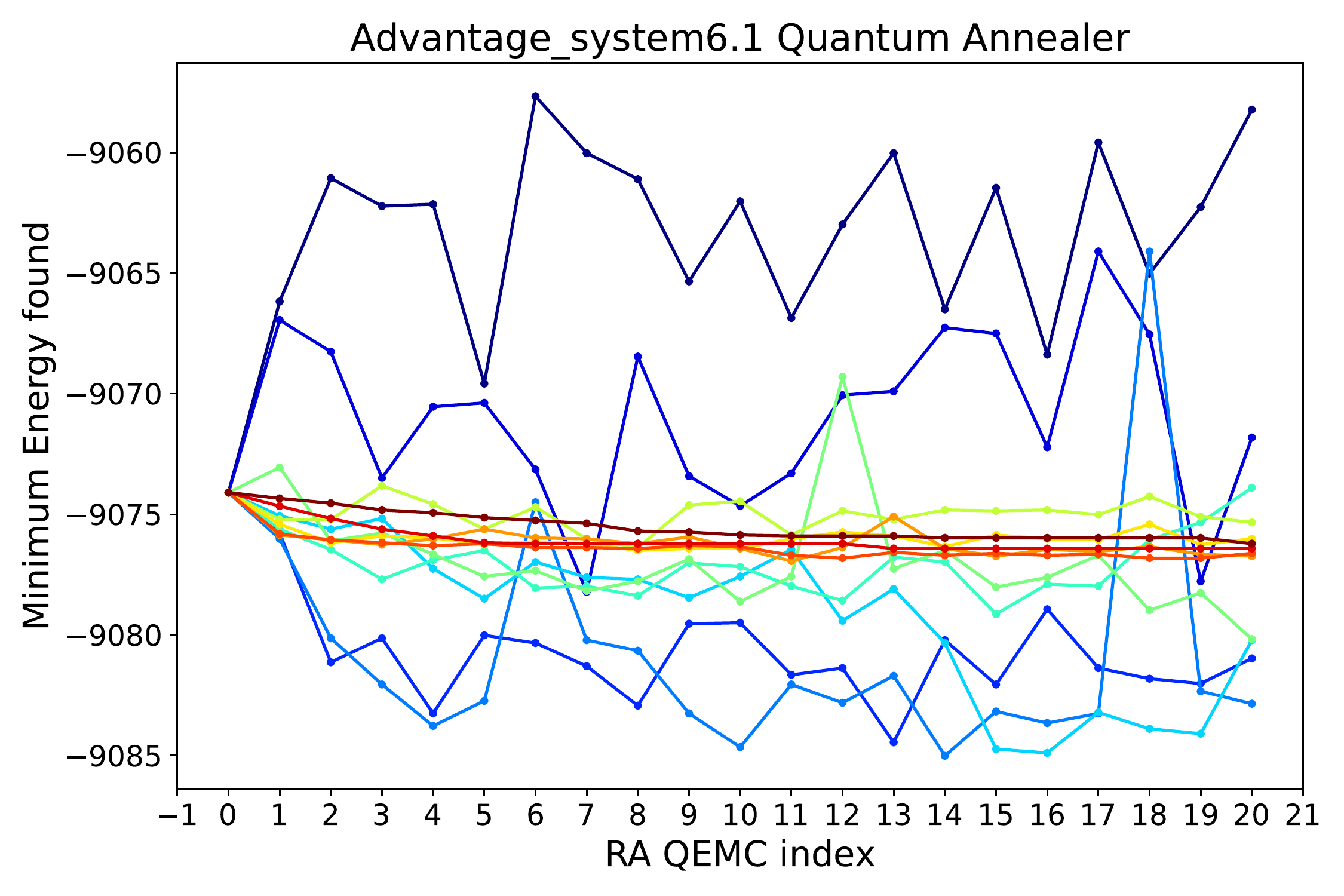}
    \includegraphics[width=0.32\textwidth]{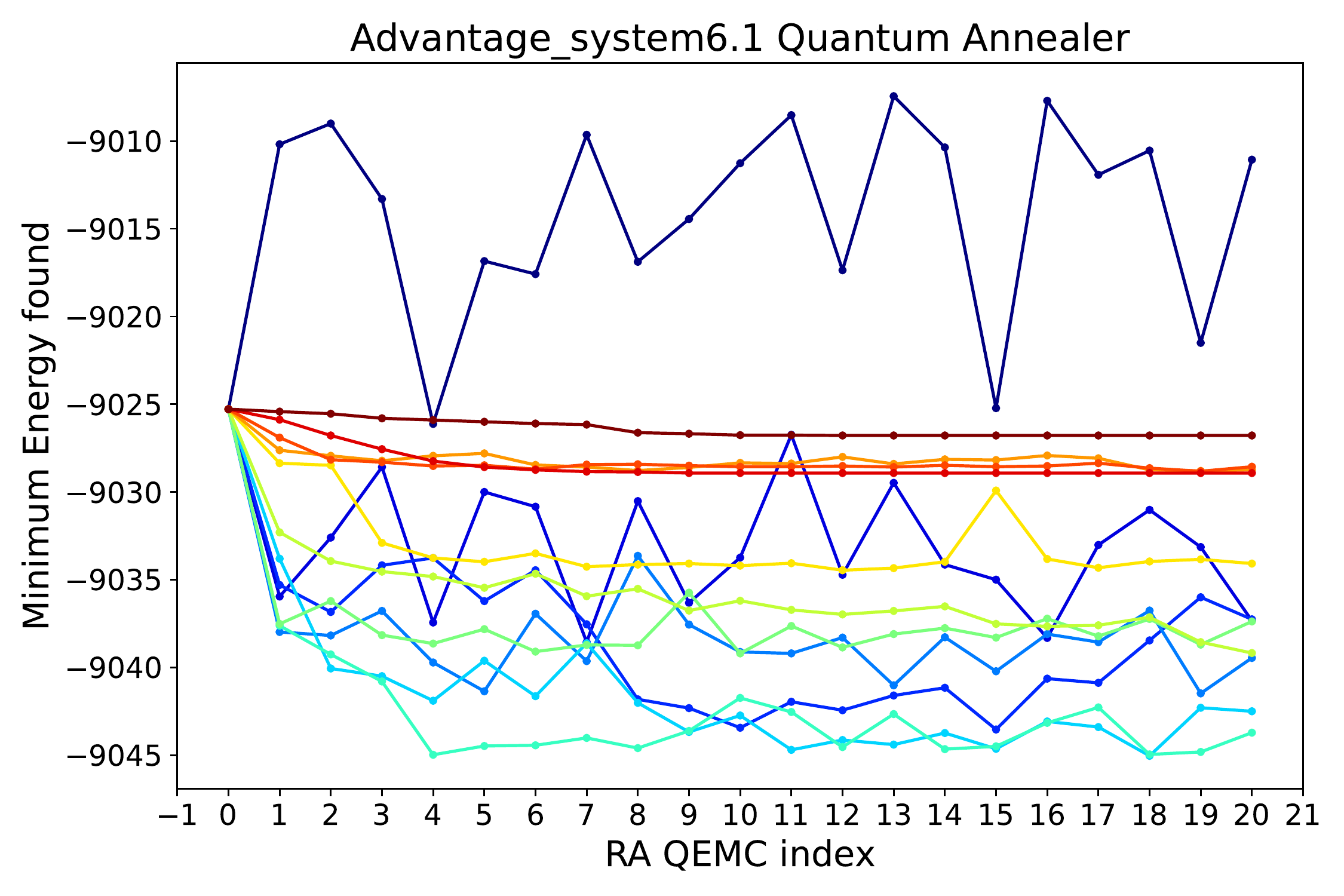}\\
    \includegraphics[width=0.32\textwidth]{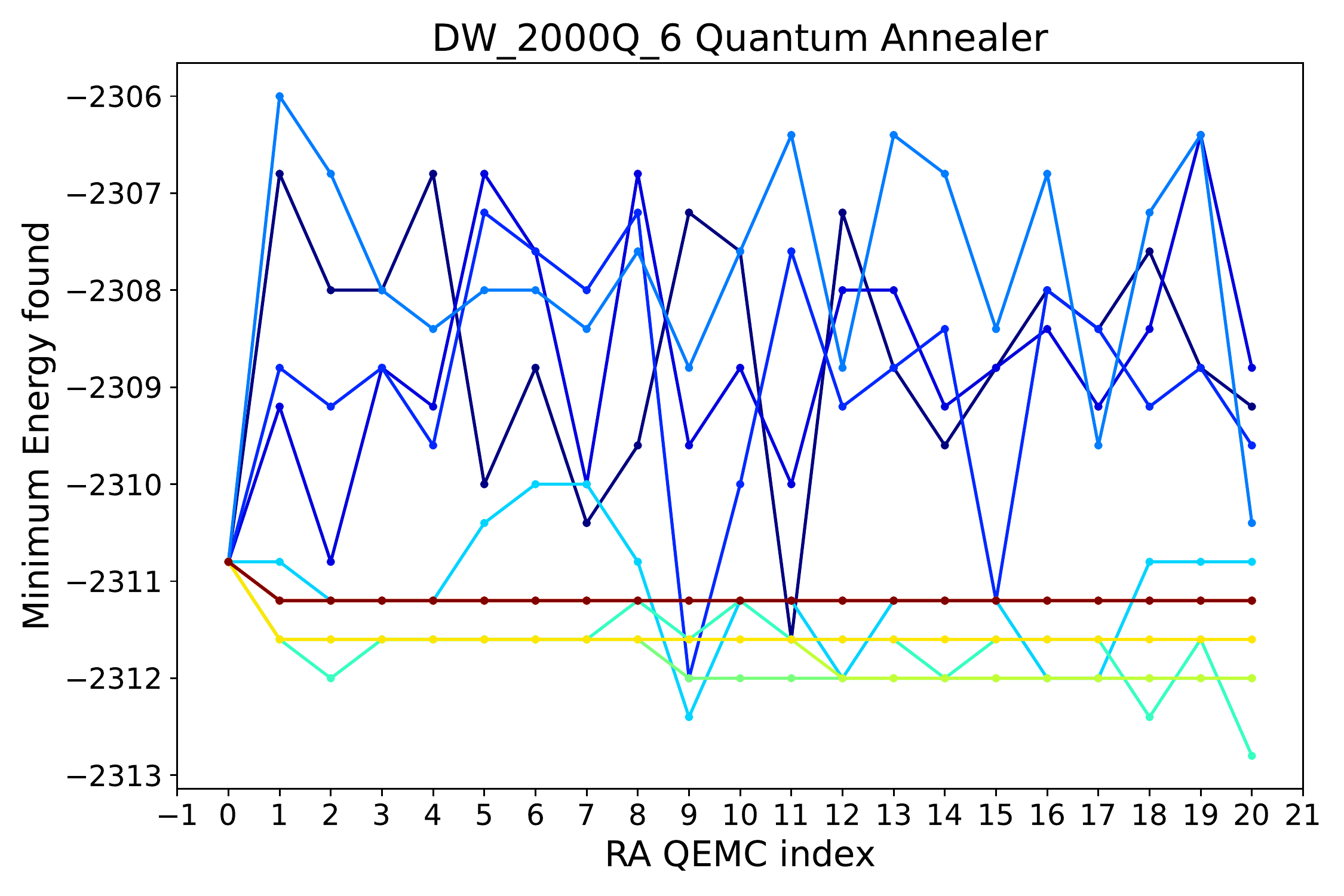}
    \includegraphics[width=0.32\textwidth]{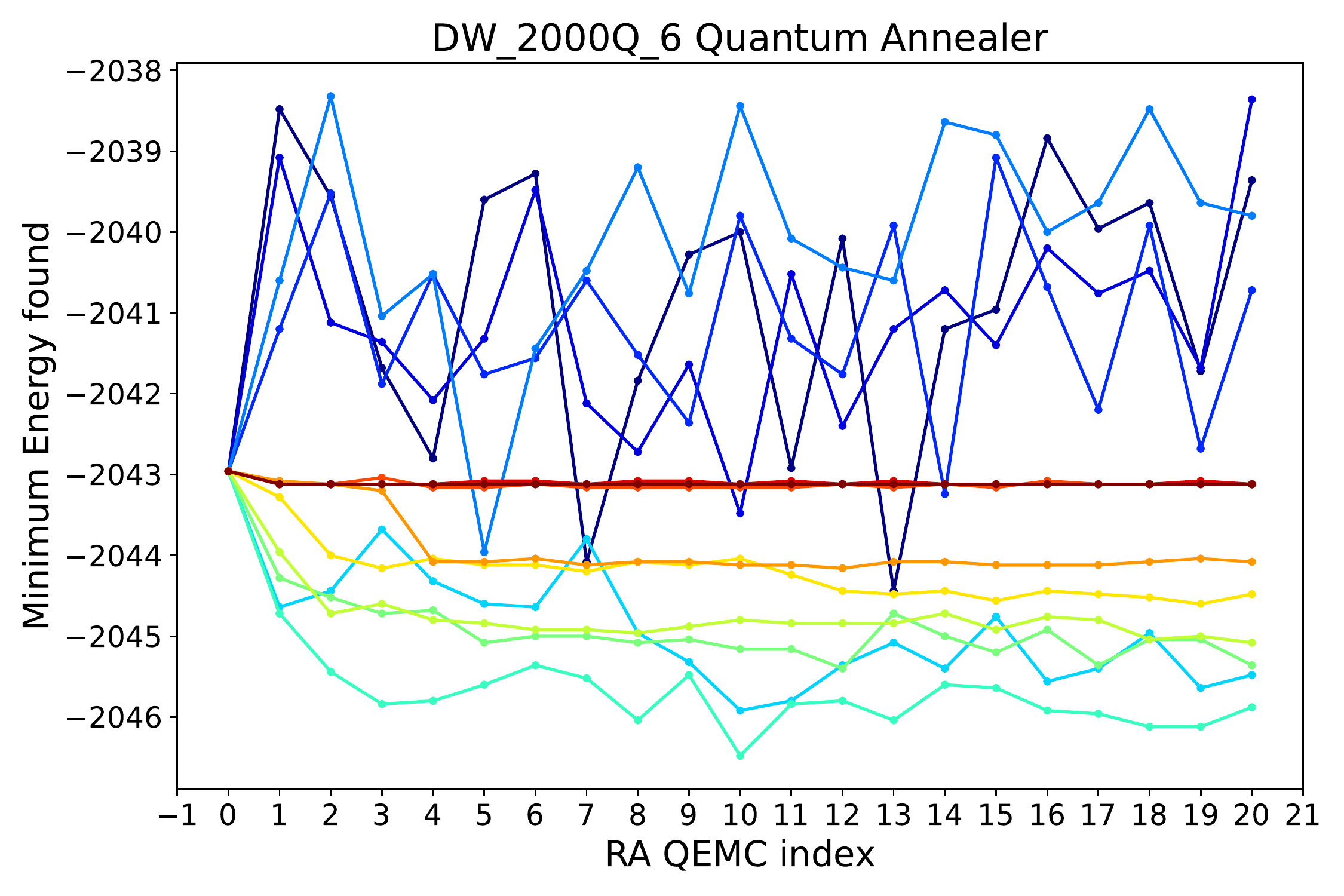}
    \includegraphics[width=0.32\textwidth]{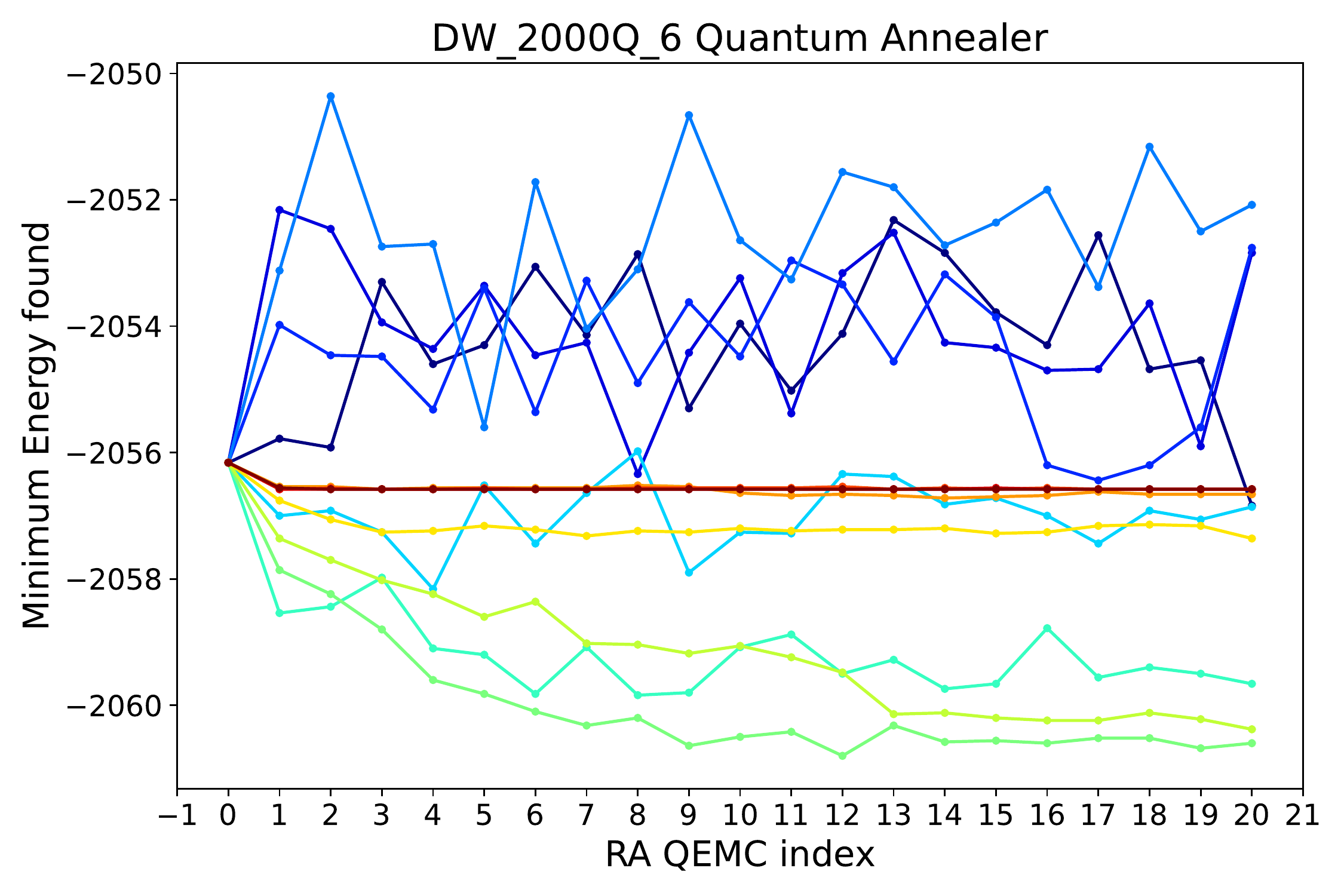}\\
    \includegraphics[width=0.5\textwidth]{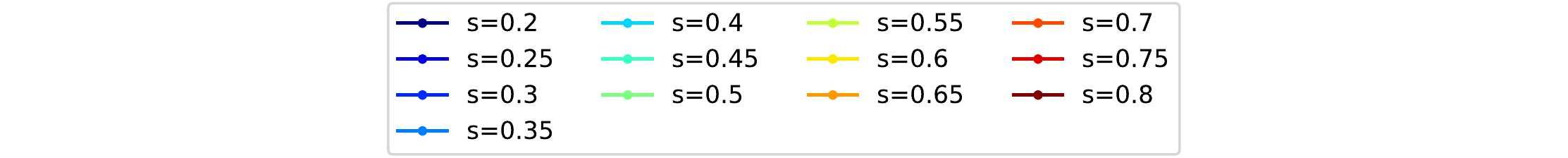}
    \caption{QEMC with reverse annealing only, where each subsequent step uses the lowest energy solution found from the previous step. The problem instances are random spin glasses on \texttt{Advantage\_system4.1} (top row), \texttt{Advantage\_system6.1} (middle row), and \texttt{DW\_2000Q\_6} (bottom row). Spin glasses generated with linearly spaced precision of $10$ (left column), $100$ (middle column), and $200$ (right column). The curves show different anneal fractions $s$ at which the symmetric reverse anneal is paused at, given in the legend. Annealing time of $100$ microseconds, and $1000$ anneals per step is used for these experiments. }
    \label{fig:RA_iterated_initial_state}
\end{figure*}

\subsection{Evaluation of QEMC}
\label{sec:QEMC_experiments}
We finally evaluate QEMC, that is, the technique of iteratively improving the sampled solutions. As outlined in Section~\ref{sec:QEMC}, four options are available for planting the solution in iteration. Those are the options evaluated in Sections~\ref{sec:experiments_maxcut} and \ref{sec:experiments_maxclique}, namely reverse annealing only, an h-gain field only (using the qubit coefficients to encode an initial state for the anneal), reverse annealing combined with h-gain (where the same initial states is encoded using both linear terms and the reverse annealing initial state), and forward annealing with a pause combined with h-gain initial state encoding. Each QEMC iteration uses $1000$ anneals and an annealing time of $100$ microseconds, and we plot the lowest energy state found at each iteration. The lowest energy sample found at each iteration then seeds the encoded state of the anneal for the next iteration (whether it is encoded using h-gain state encoding or reverse annealing). For good parameter choices, we would observe that there is a monotonic decrease in the energy, e.g. the solutions are getting progressively better for the minimization combinatorial optimization problem. Each QEMC experiment (e.g., with each unique Ising model on each device) is initialized with the best solution found from a normal forward anneal (e.g. with a linearly interpolated anneal schedule) from $1000$ anneals with an annelaing time of $100$ microseconds. Therefore, each set of QEMC experiments begins in exactly the same initial state.

These experiments are conducted on random spin glass models. As remarked in the literature \cite{Katzgraber2014, PhysRevX.5.019901, Mandra2017}, random spin glasses with the chimera structure have been shown to exhibit a zero temperature phase transition. Random spin glasses with Chimera connectivity are therefore not expected to be computationally difficult to sample from classically, which could be a reason why an advantage of quantum over classical optimization algorithms has not been demonstrated yet for this class of Ising models. However, there is also evidence that general spin glasses on these hardware graphs (Chimera and Pegasus) may provide advantageous sampling for very large system sizes \cite{jauma2023exploring}. 

\begin{figure*}[ht!]
    \centering
    \includegraphics[width=0.32\textwidth]{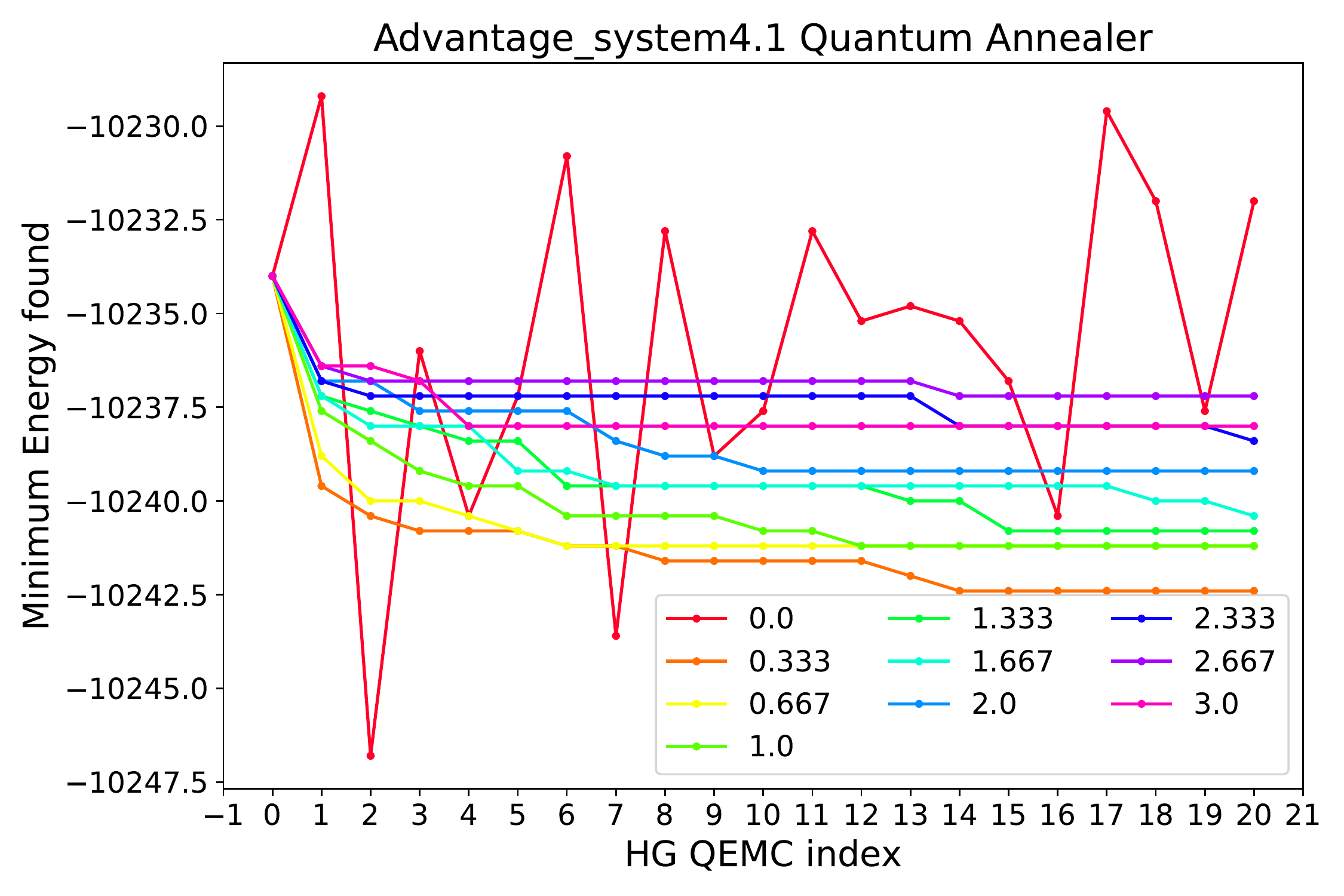}
    \includegraphics[width=0.32\textwidth]{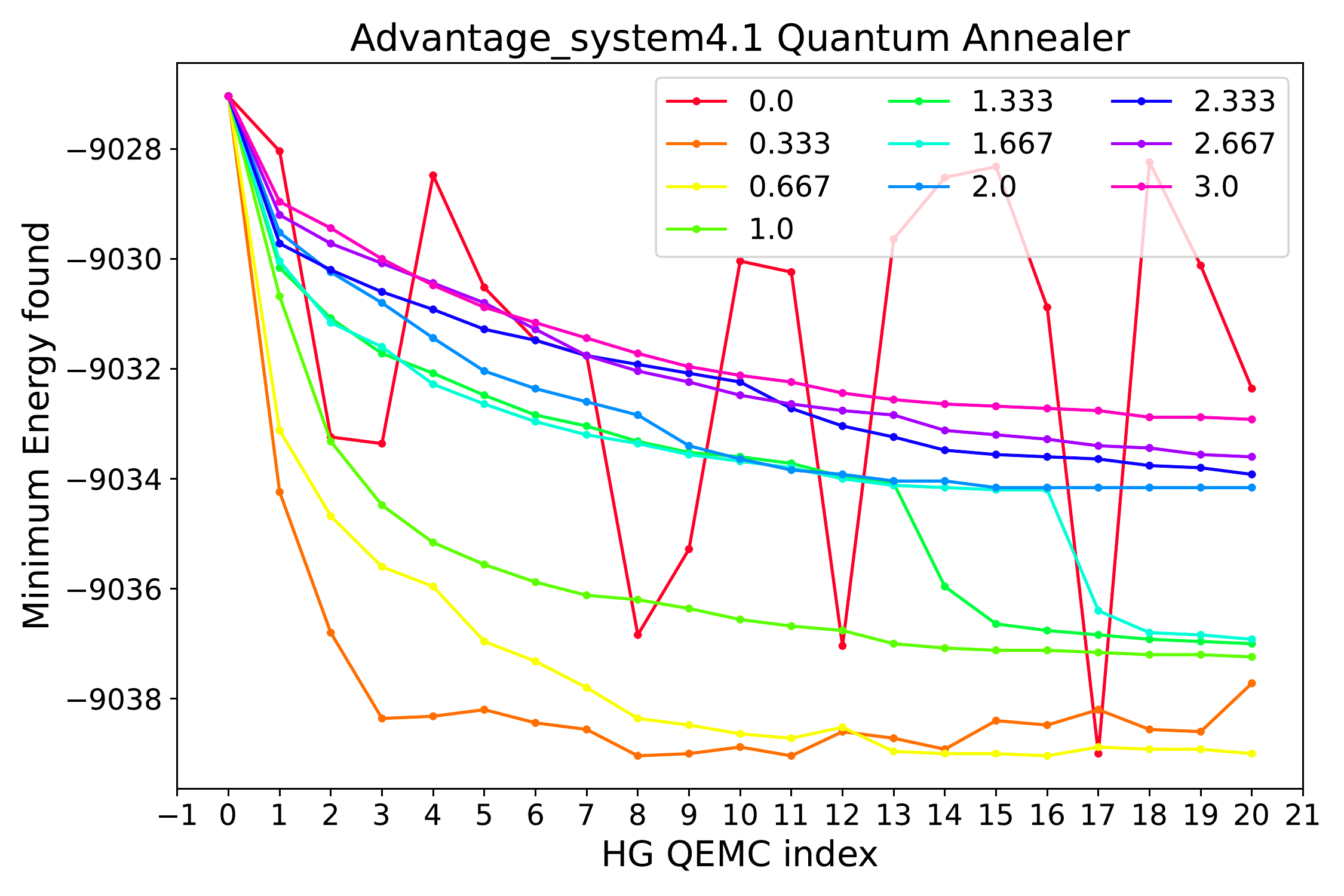}
    \includegraphics[width=0.32\textwidth]{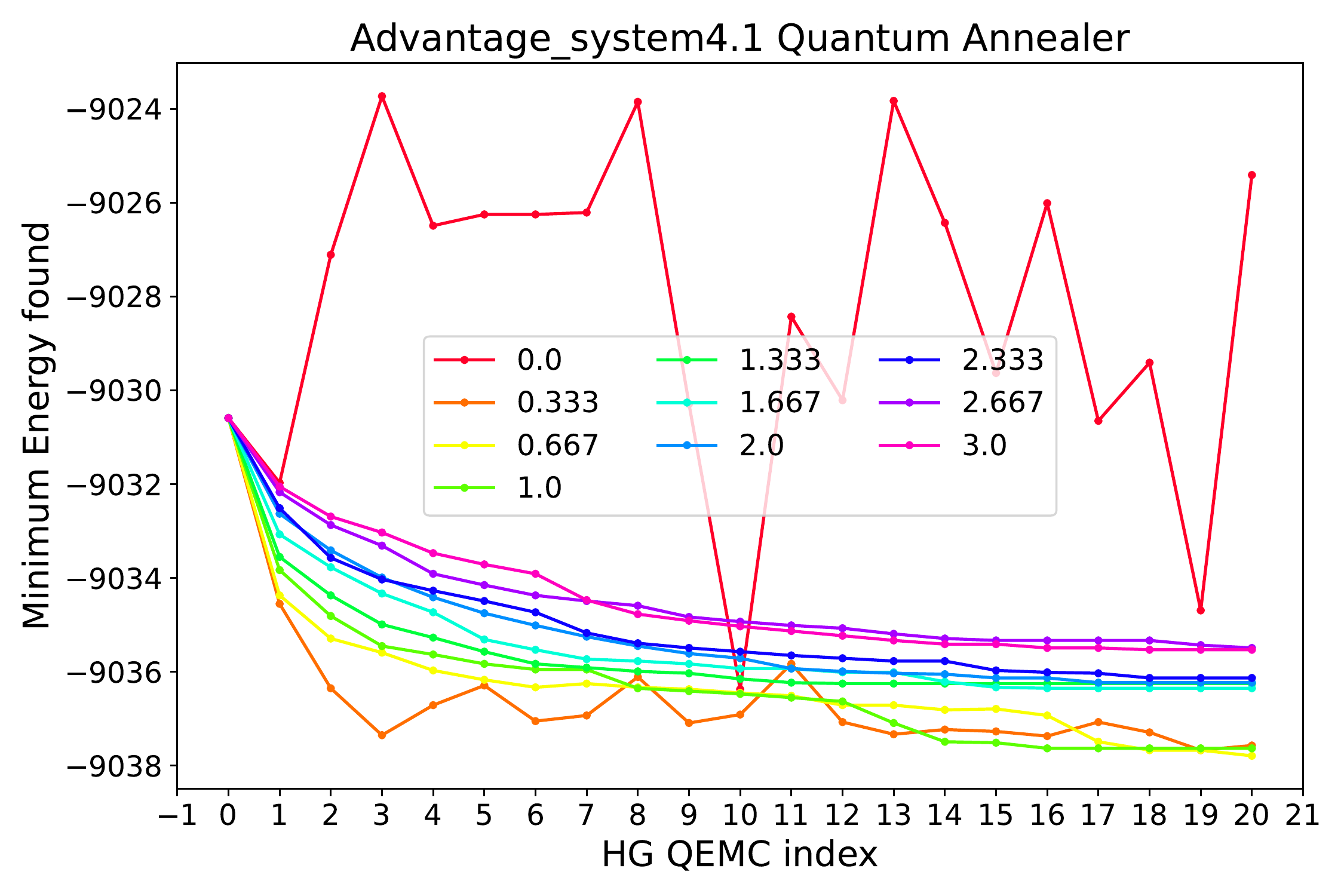}\\
    \includegraphics[width=0.32\textwidth]{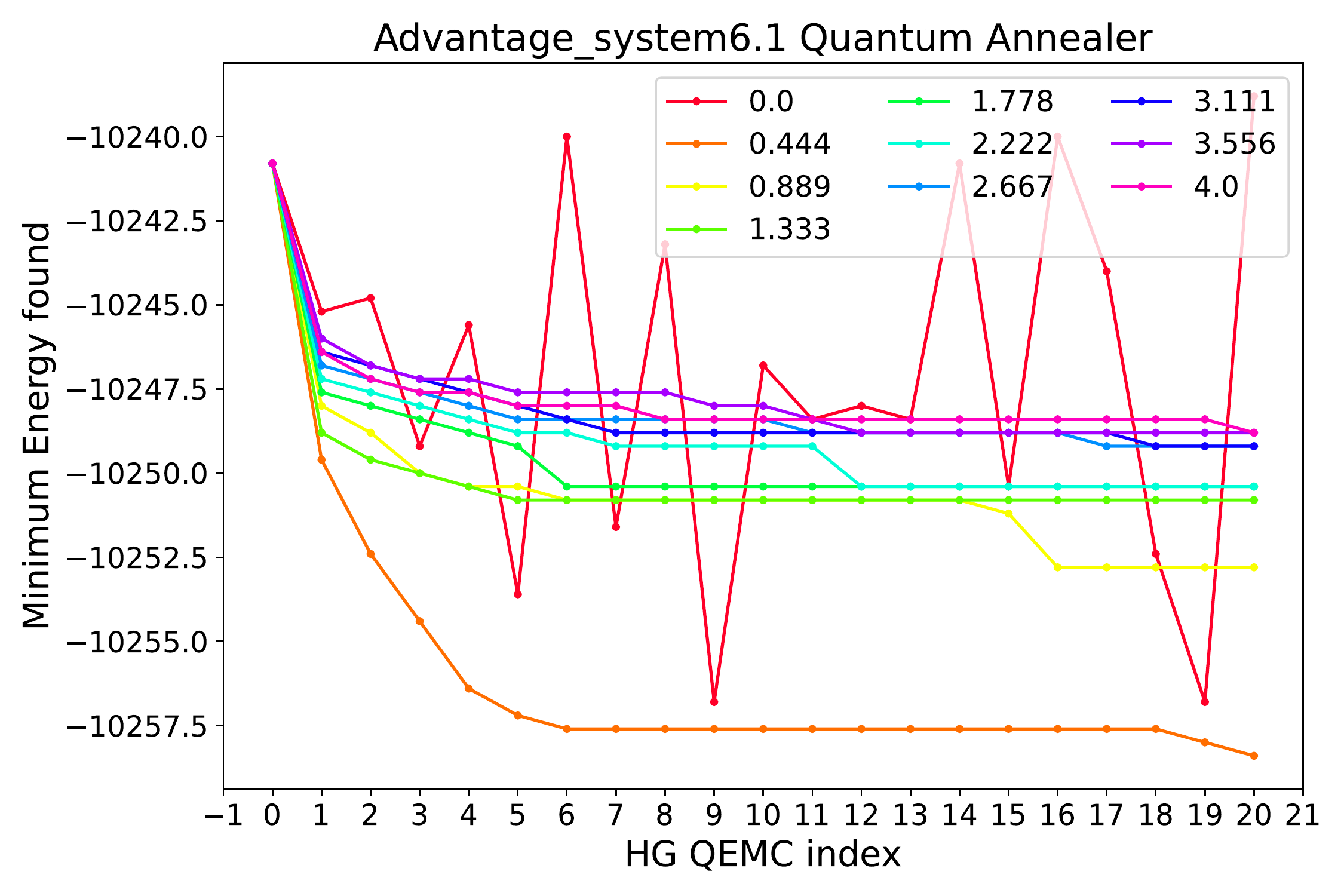}
    \includegraphics[width=0.32\textwidth]{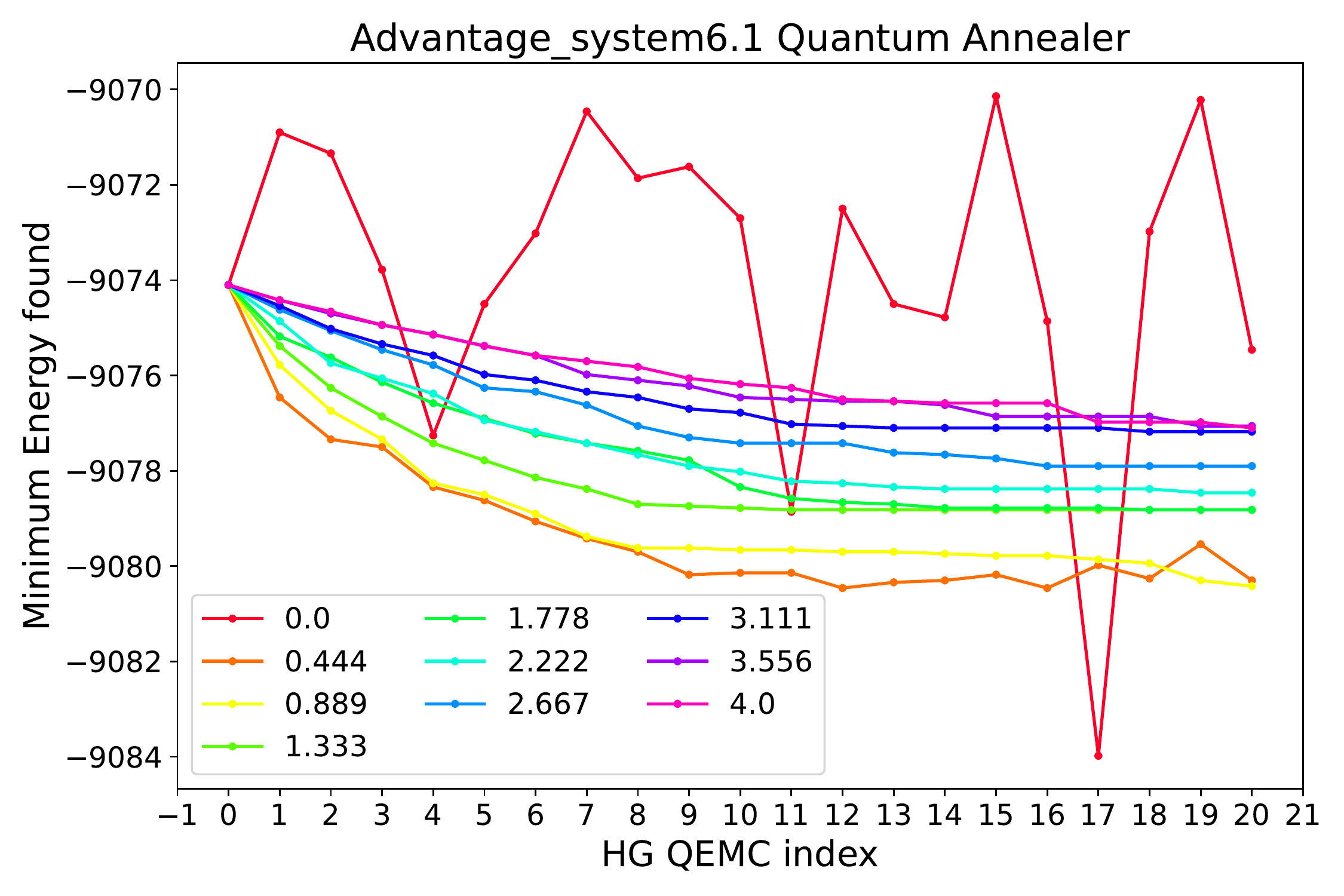}
    \includegraphics[width=0.32\textwidth]{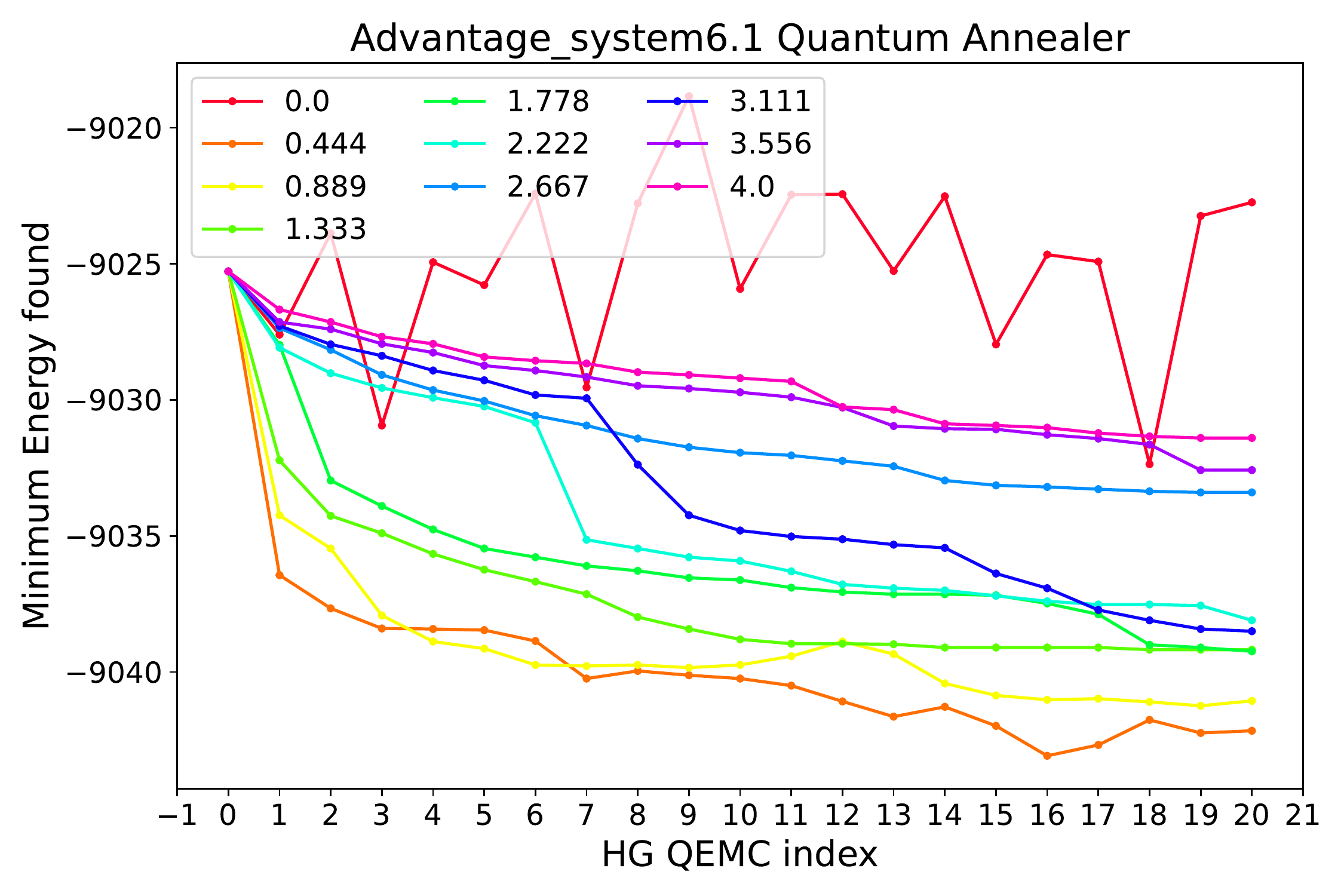}\\
    \includegraphics[width=0.32\textwidth]{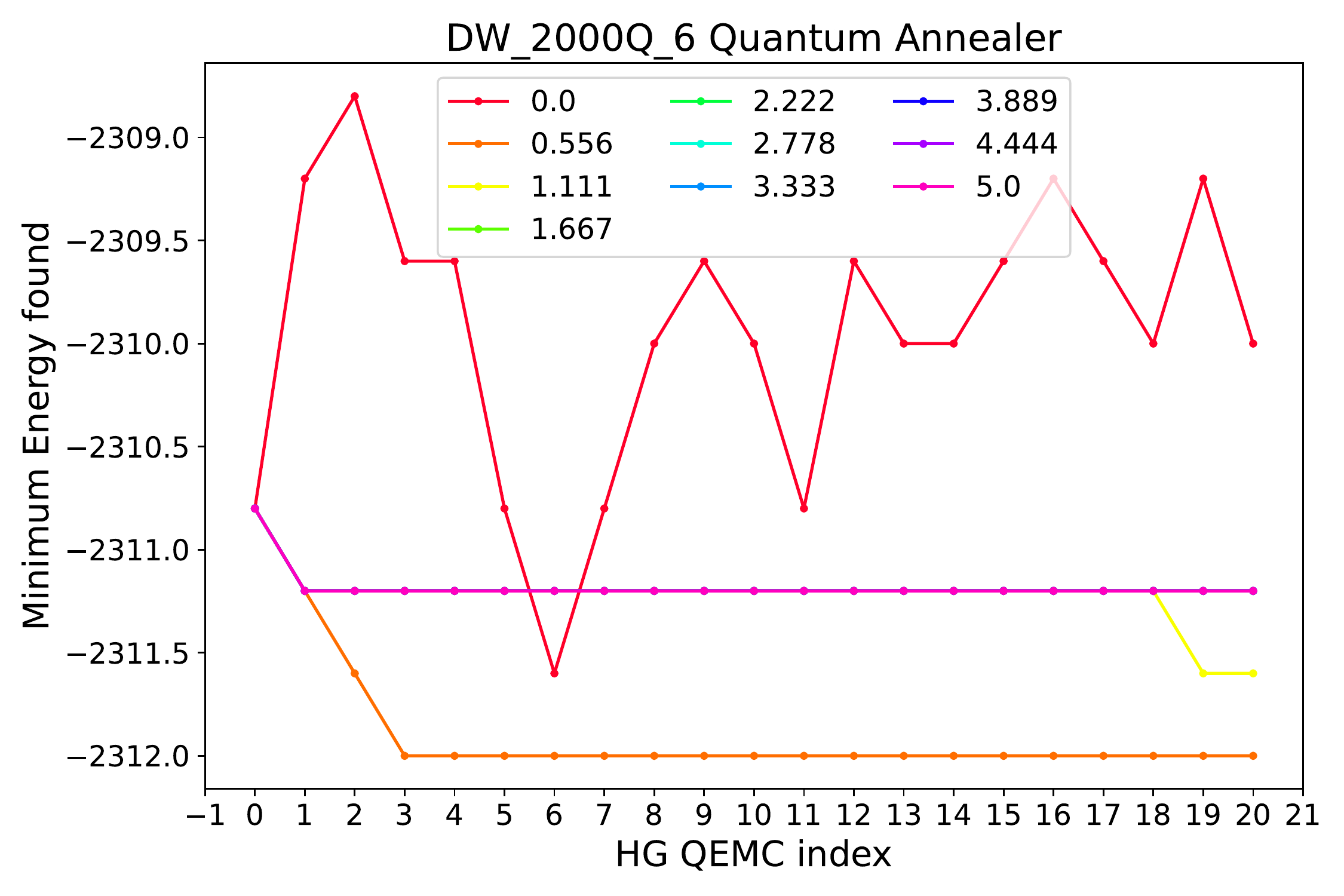}
    \includegraphics[width=0.32\textwidth]{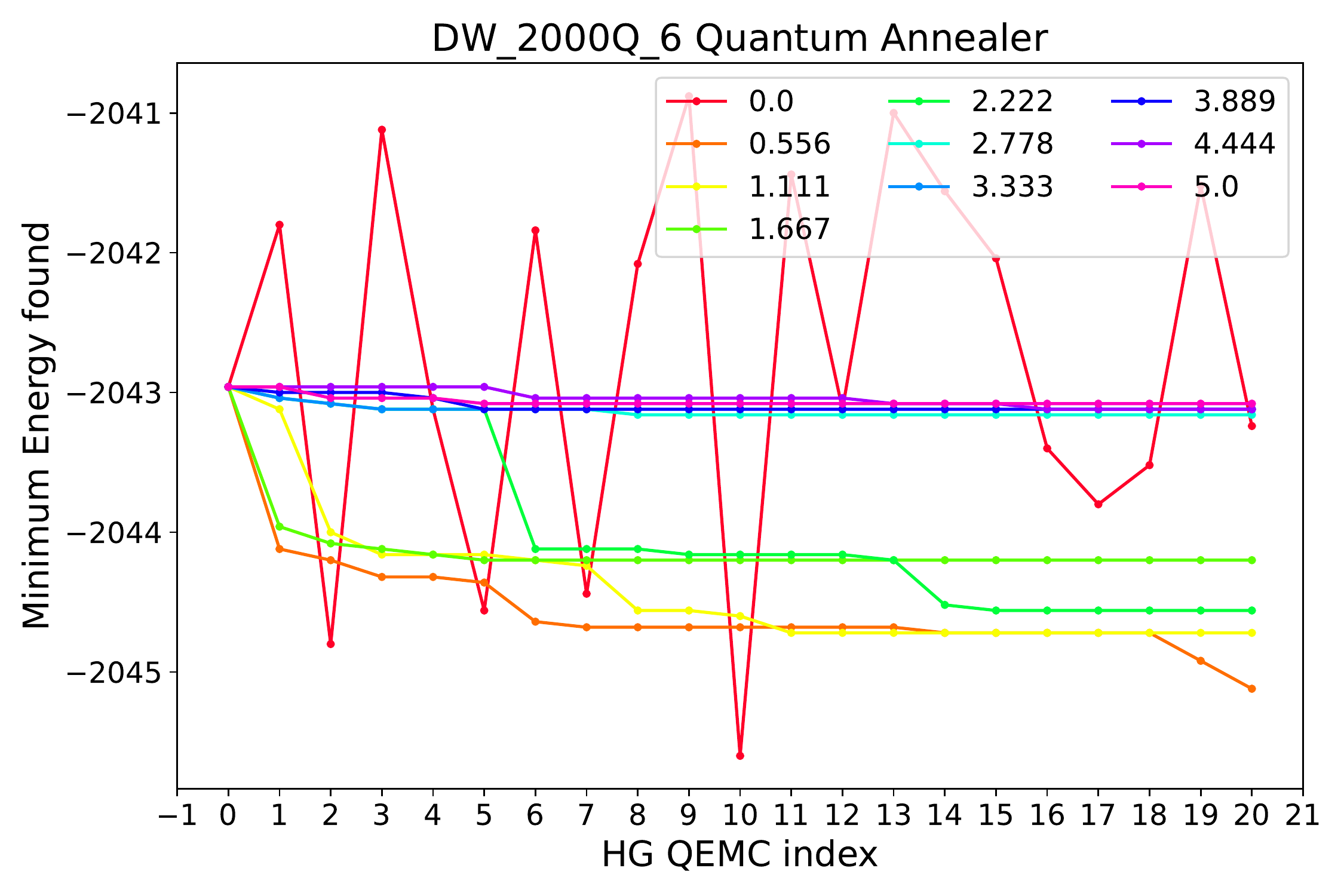}
    \includegraphics[width=0.32\textwidth]{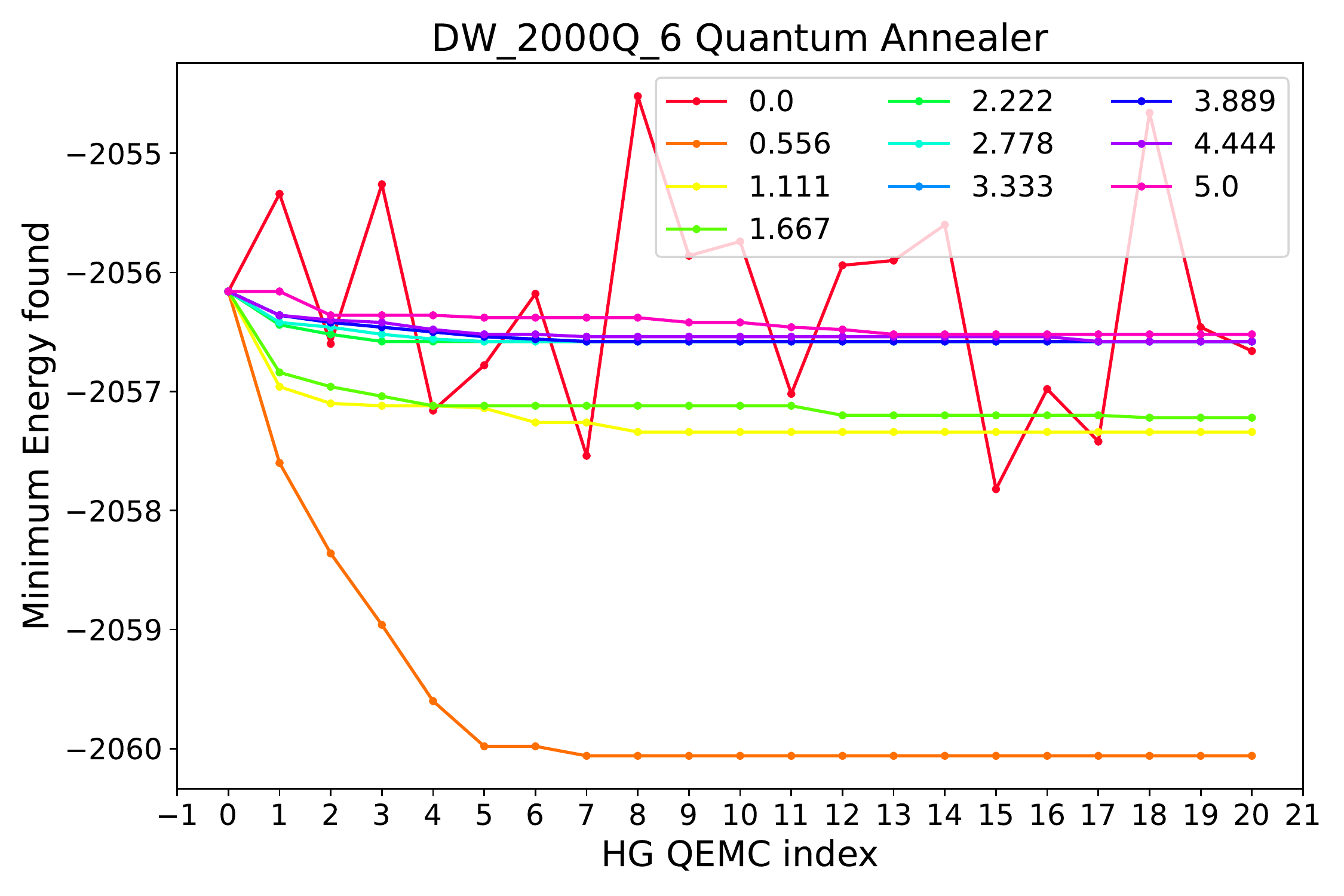}
    \caption{QEMC with h-gain initial state encoding. Random spin glasses on \texttt{Advantage\_system4.1} (top row), \texttt{Advantage\_system6.1} (middle row), and \texttt{DW\_2000Q\_6} (bottom row). Spin glasses generated with linearly spaced precision of $10$ (left column), $100$ (middle column), and $200$ (right column). The curves show different HG strengths $h$, given in the legend. The single point that is varied in the h-gain schedules used for these experiments is set to $10$ microseconds into the $100$ microsecond anneal. The maximum h-gain strength used was set to the maximum allowed on each D-Wave quantum annealer, which is $5$ for \texttt{DW\_2000Q\_6}, $4$ for \texttt{Advantage\_system6.1}, and $3$ for \texttt{Advantage\_system4.2}. }
    \label{fig:QEMC_h-gain_pos10}
\end{figure*}

\subsubsection{QEMC with reverse annealing}
\label{sec:QEMC_RA}
We first consider QEMC with RA only, meaning that we iteratively refine a solution by encoding the previously found best result using RA. In this experiment, the anneal fraction $s$ at which the pause occurs is varied; we utilize values $s \in \{ 0.2,0.25,0.3,\ldots,0.8 \}$. Figure~\ref{fig:RA_iterated_initial_state} shows our results for the three quantum annealers given in Table~\ref{table:hardware_summary} (rows) and spin glass generation mechanisms (columns). Each curve is a run with a different anneal fractions $s$ for RA, given in the legend. The figure displays the minimum energy found in the $1000$ anneals in each iteration and for each parameter. 

We observe that the behavior of QEMC with reverse annealing on \texttt{Advantage\_system4.1} is quite similar to the one on \texttt{Advantage\_system6.1}, while results on \texttt{DW\_2000Q\_6} differ. For the first two devices, lower anneal fractions are suboptimal, while anneal fractions around $s \in [0.3,0.4]$ perform best. For anneal fractions which are too high, we observe little improvement of the solution over the iterations of QEMC. For \texttt{DW\_2000Q\_6}, results display a higher degree of volatility, and the values of $s$ yielding the best converge behavior are slightly higher than for the newer annealer generations, which is around $s \in [0.5,0.6]$.

\subsubsection{QEMC with h-gain state encoding}
\label{sec:QEMC_h-gain}
Next, we repeat the same experiment in Figure~\ref{fig:QEMC_h-gain_pos10}, though now we use h-gain state encoding, comprising of the h-gain schedule and programming linear terms on the qubits corresponding to te encoded state, to plant the best previous solution sampled in each iteration. As before, the figure displays the minimum energy found in the $1000$ anneals in each iteration and for each parameter. The parameter $h$ is varied in the interval $[0,3]$ for \texttt{Advantage\_system4.1}, in $[0,4]$ for \texttt{Advantage\_system6.1}, and $[0,5]$ for \texttt{DW\_2000Q\_6}. The h-gain strength is set to $0$ at $10$ microseconds into the $100$ microsecond anneal, which means that the initial state is strongly encoded during the transverse field dominated portion of the anneal, and then is switched off for the rest of the anneal (see the top-right sub-figure of Figure \ref{fig:iterated_anneal_schedules}).

The results in Figure~\ref{fig:QEMC_h-gain_pos10} demonstrate that the the h-gain state encoding technique can be used to execute QEMC. We again observe that the newer generations of the D-Wave annealer, that is \texttt{Advantage\_system4.1} and \texttt{Advantage\_system6.1}, seem to be better suited for QEMC as they show a more stable convergence behavior. For \texttt{DW\_2000Q\_6}, convergence seems more unstable. 

\begin{figure*}[ht!]
    \centering
    \includegraphics[width=0.32\textwidth]{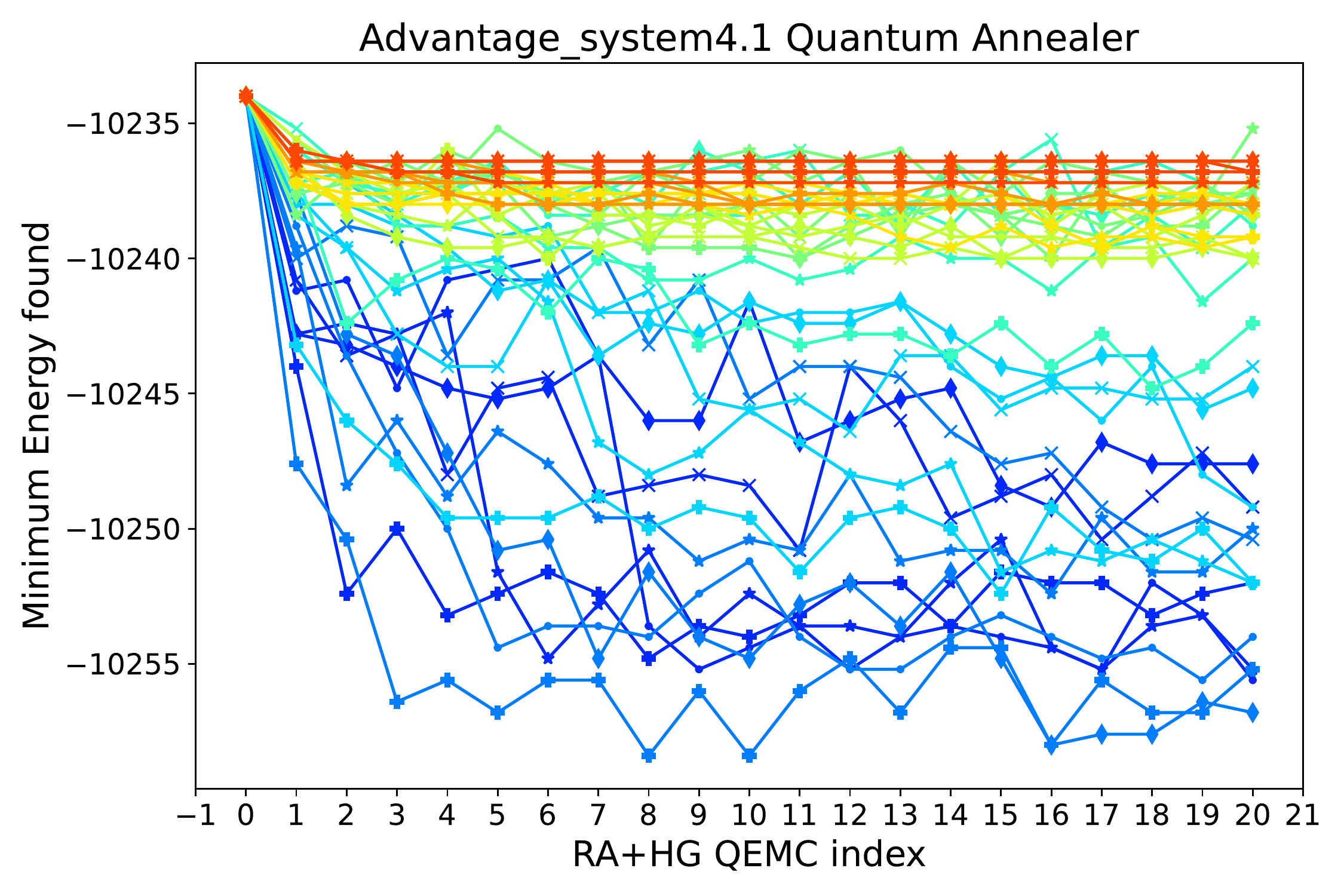}
    \includegraphics[width=0.32\textwidth]{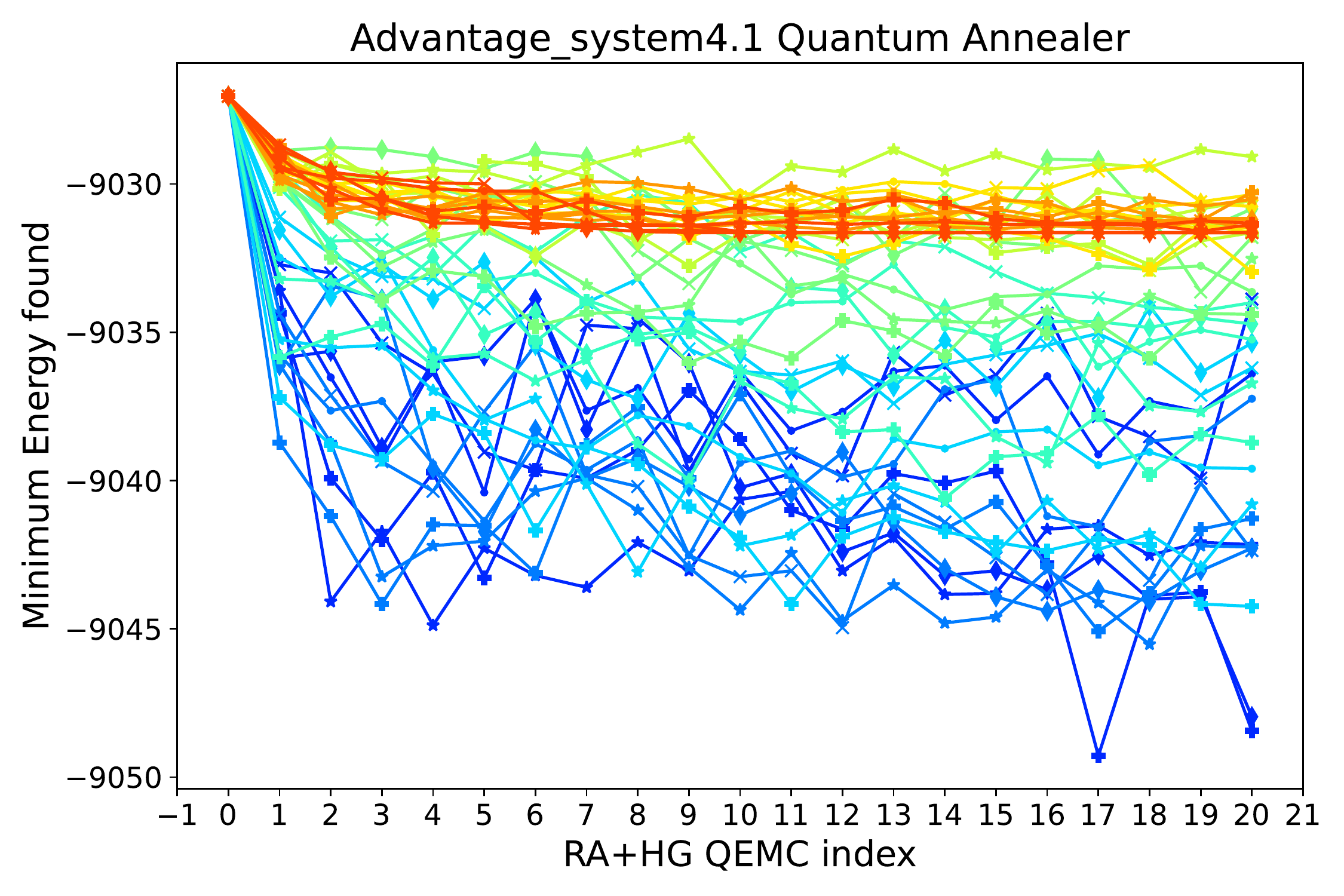}
    \includegraphics[width=0.32\textwidth]{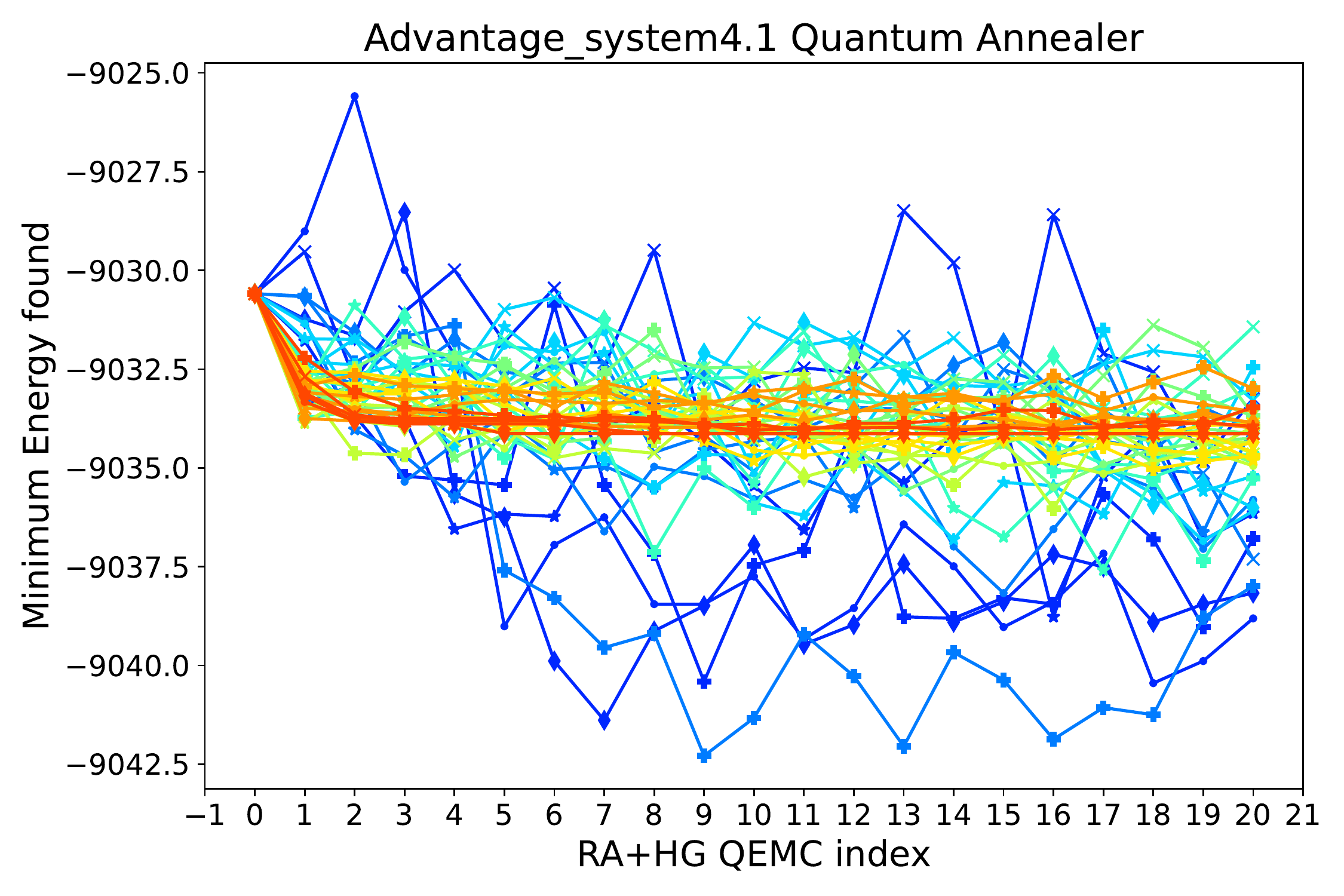}\\
    \includegraphics[width=0.32\textwidth]{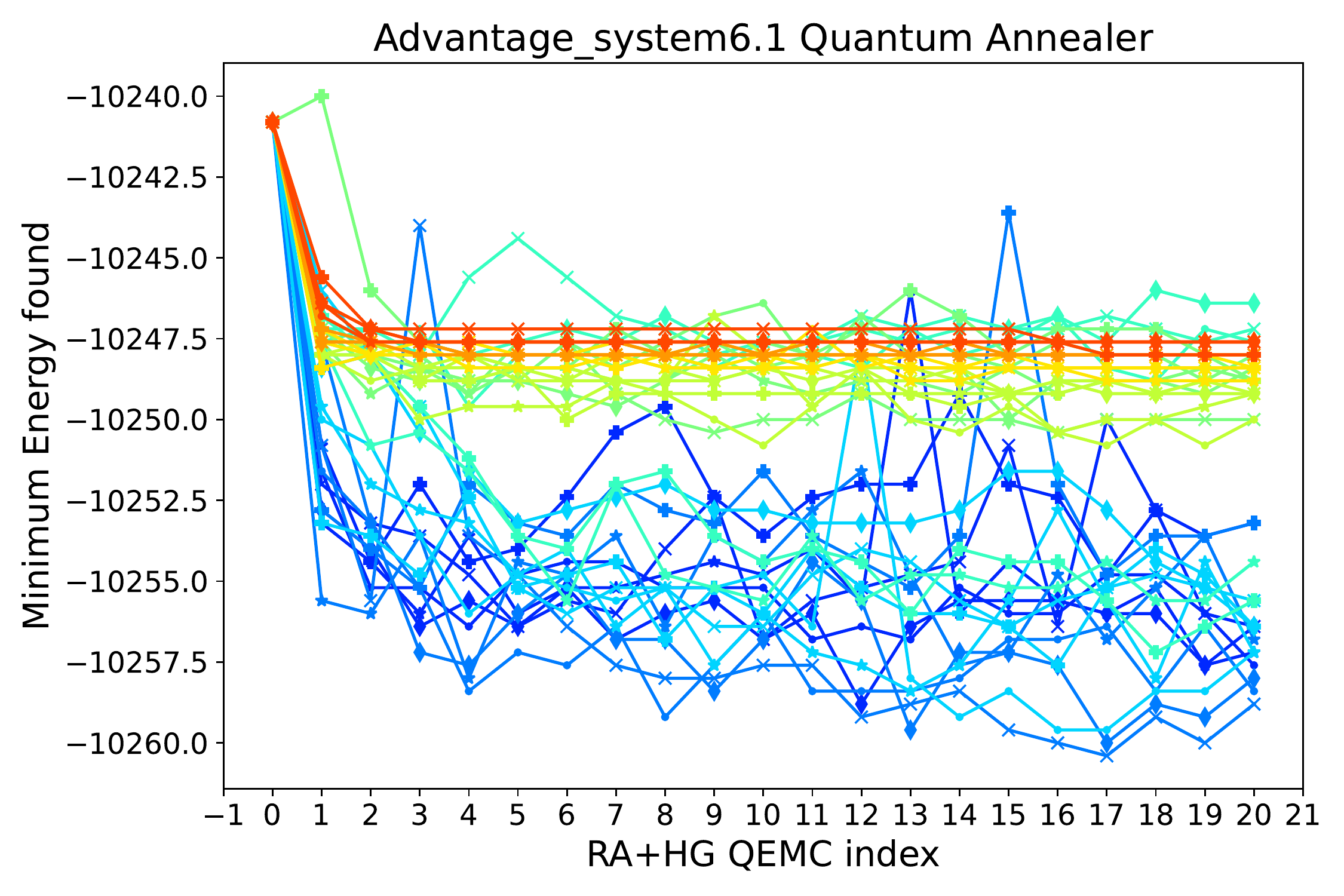}
    \includegraphics[width=0.32\textwidth]{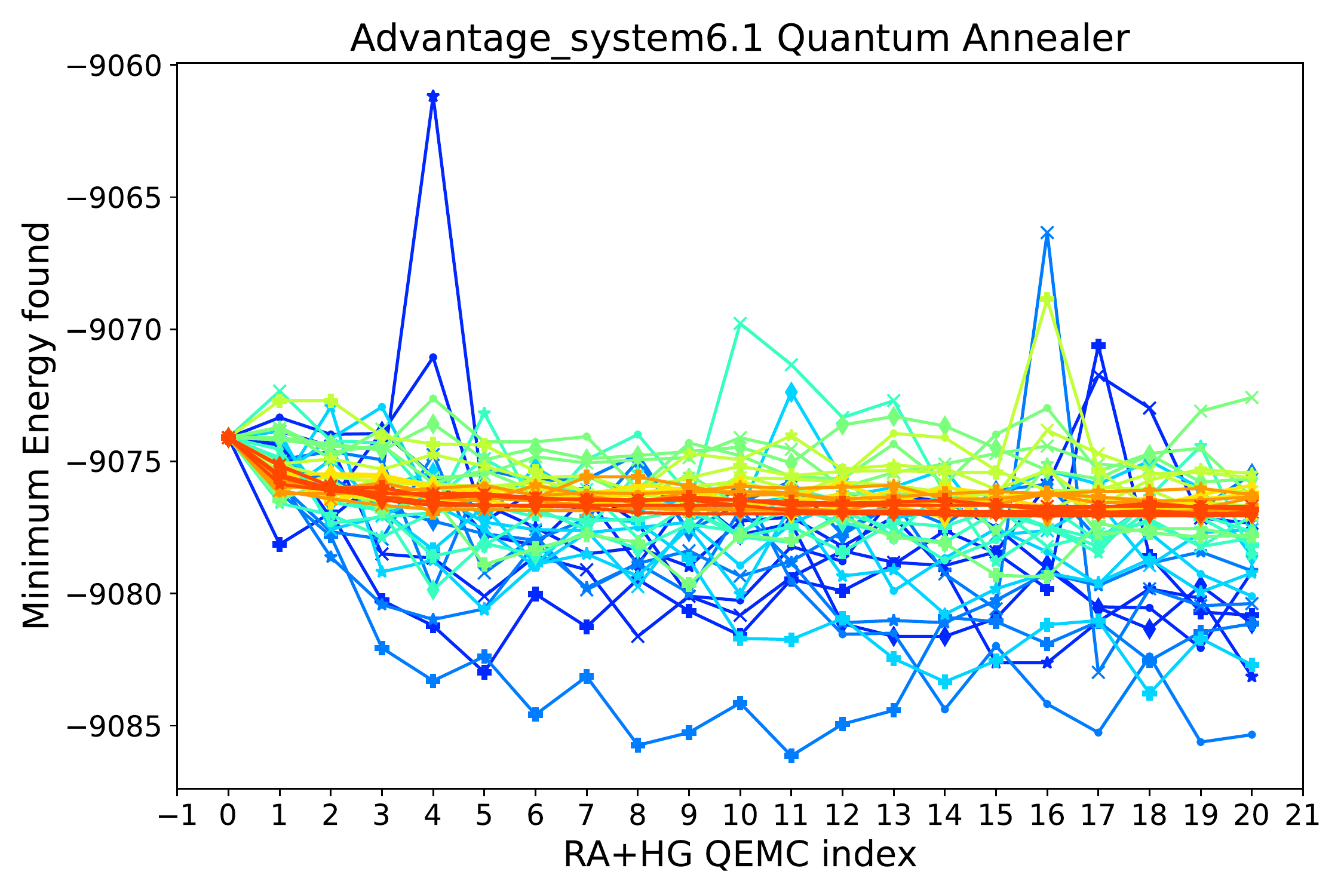}
    \includegraphics[width=0.32\textwidth]{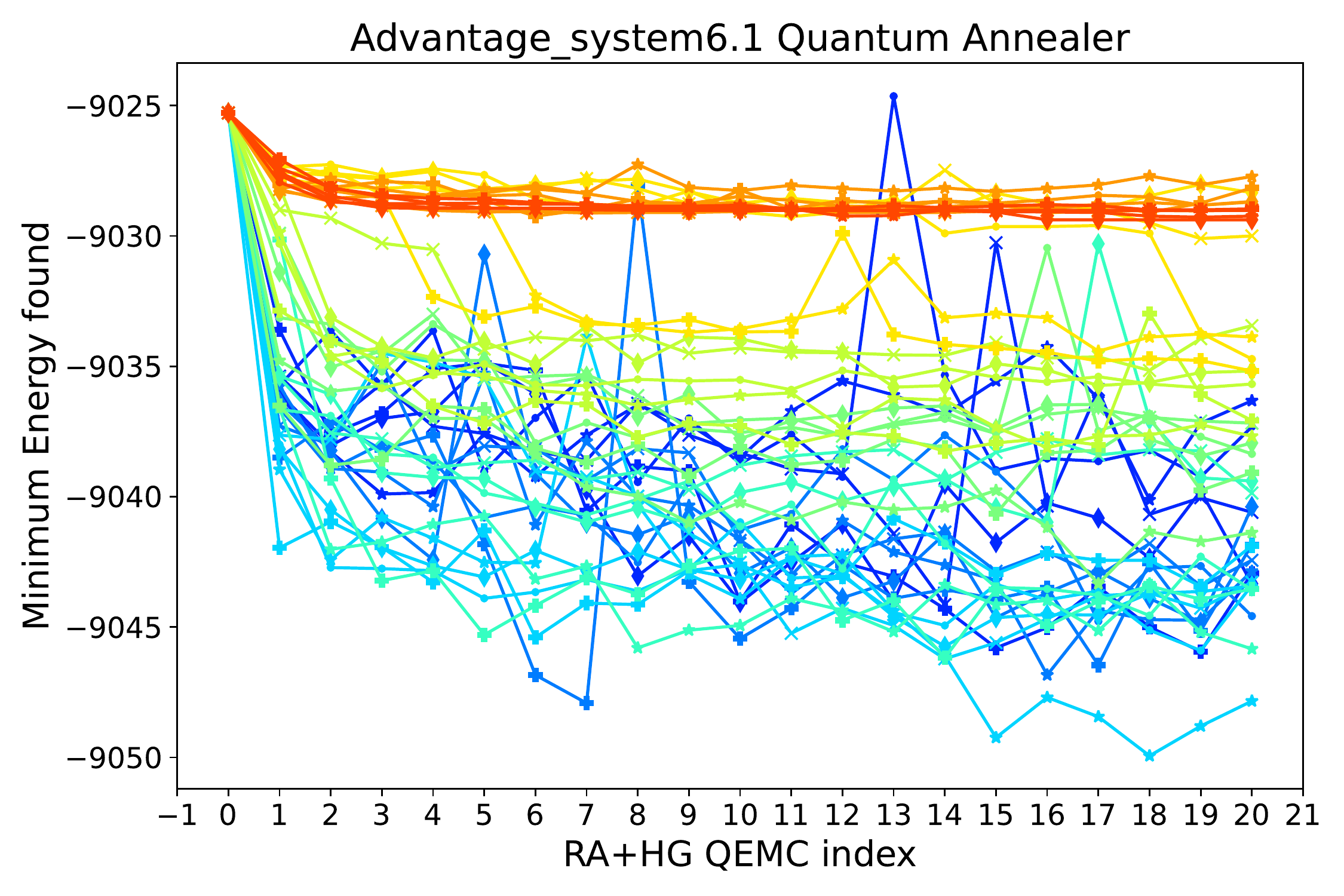}\\
    \includegraphics[width=0.32\textwidth]{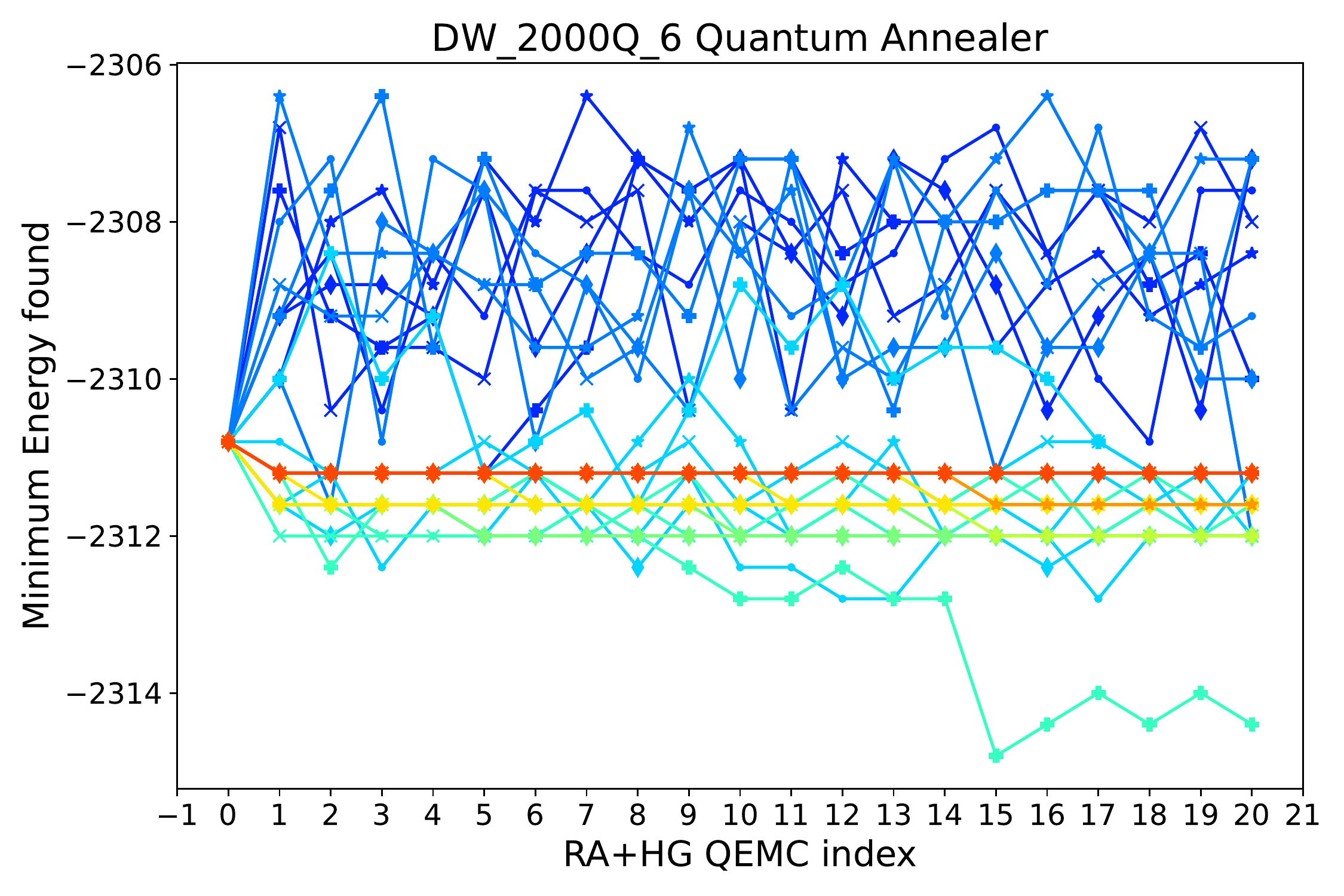}
    \includegraphics[width=0.32\textwidth]{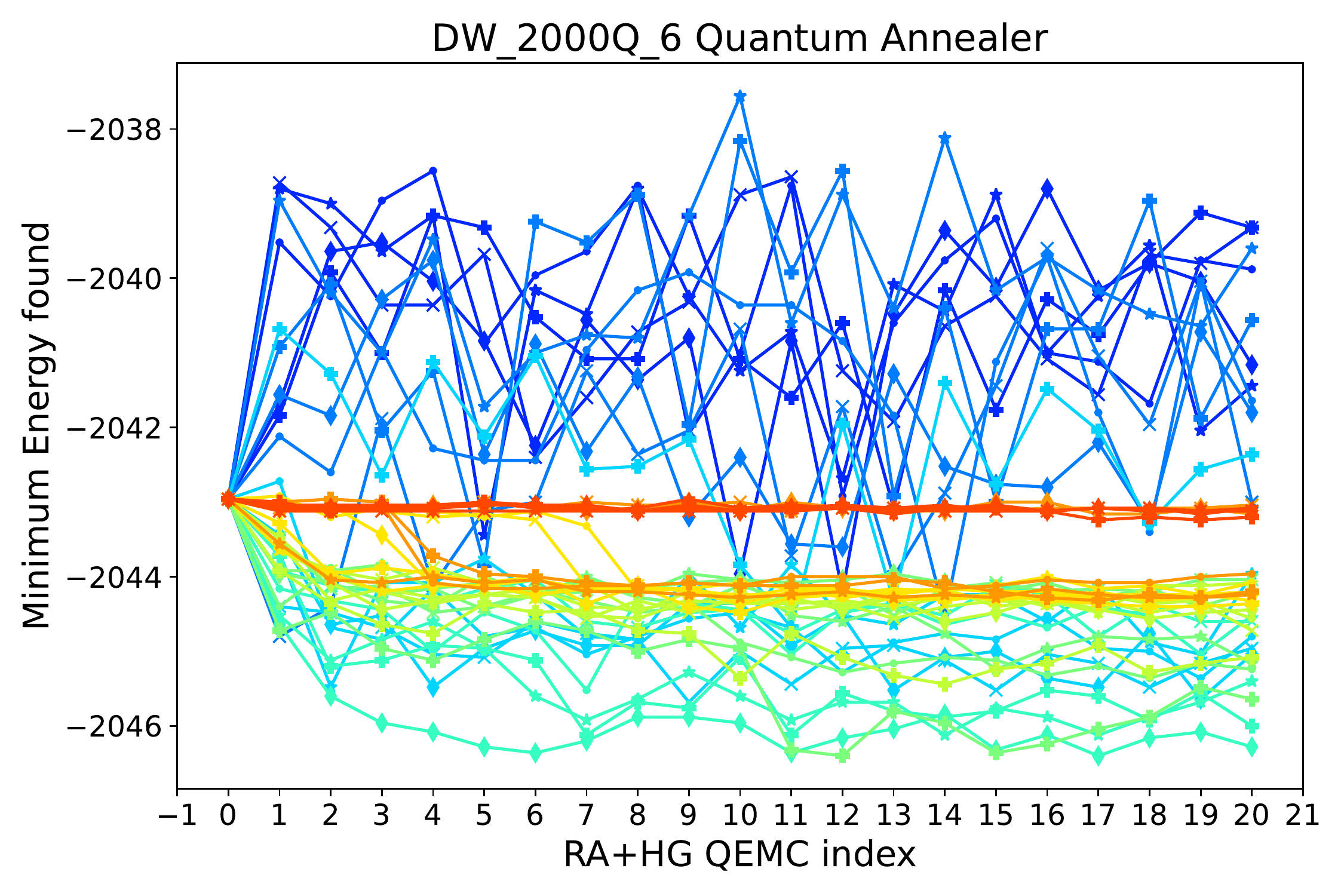}
    \includegraphics[width=0.32\textwidth]{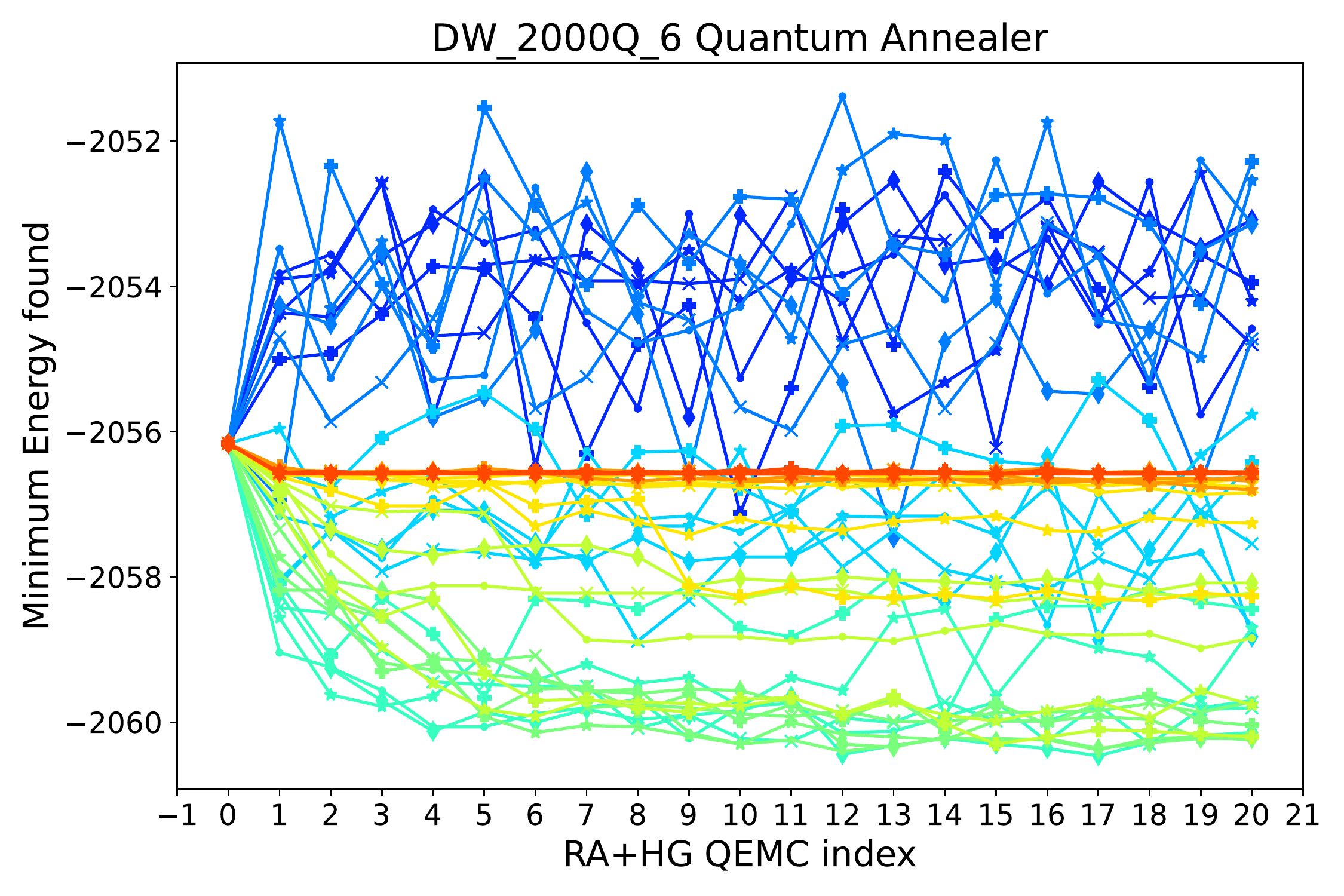}
    \includegraphics[width=0.59\textwidth]{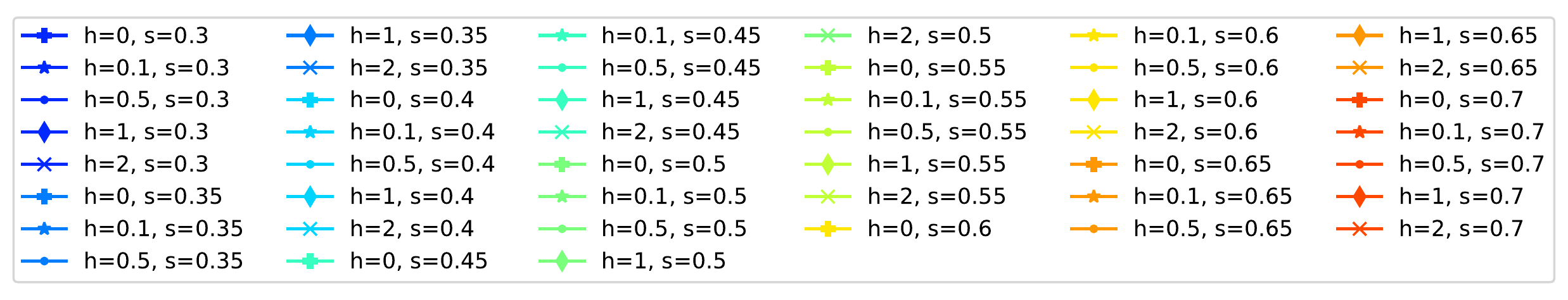}
    \caption{QEMC using reverse annealing and h-gain combined. Random spin glasses on \texttt{Advantage\_system4.1} (top row), \texttt{Advantage\_system6.1} (middle row), and \texttt{DW\_2000Q\_6} (bottom row). Spin glasses generated with linearly spaced precision of $10$ (left column), $100$ (middle column), and $200$ (right column). The curves show different combinations of both the symmetric pause anneal fractions $s$ for reverse annealing and the HG strength $h$, given in the legend. Note that the $h$ in the legend corresponds to the h-gain strength specified for at $10$ microseconds into the $100$ microsecond anneal, but the initial h-gain field is consistently set to be initialized an h-gain strength of $2$ (which in particular means these schedules are compatible with all three quantum annealers in Table \ref{table:hardware_summary}). }
    \label{fig:RA_and_h-gain_iterated_initial_state}
\end{figure*}

\subsubsection{QEMC with reverse annealing and h-gain}
\label{sec:QEMC_h-gain_RA}
As before, we can also combine RA and HG and use it to iteratively plant solutions before each anneal. This is investigated in Figure~\ref{fig:RA_and_h-gain_iterated_initial_state}, which displays the minimum energy found in the $1000$ anneals in each iteration and for each parameter combination of the anneal fraction $s$ for RA and HG strength $h$. Here, $s$ is varied in the interval $[0.3,0.7]$, and $h$ is varied in $[0,2]$ (which ensures that these schedules are comptaible with all $3$ of the tested quantum annealers).

The results in Figure~\ref{fig:RA_and_h-gain_iterated_initial_state} are consistent with the previous ones reported in Sections~\ref{sec:QEMC_RA} and \ref{sec:QEMC_h-gain}. The newer annealer generations seem more suited for an application of QEMC, and across both \texttt{Advantage\_system4.1} and \texttt{Advantage\_system6.1} we observe a consistent improvement of the initial solution when applying QEMC. Interestingly, the best parameter combinations seems to depend more on the choice of $s$ than the choice of $h$, as roughly $s \in [0.35,0.5]$ in connection with any $h$ yields best results. On \texttt{DW\_2000Q\_6}, the application of QEMC works less well, though an improved behavior can be seen for higher precision of the mapped Ising coefficients (right-hand column of plots).

\subsubsection{QEMC with forward annealing with a pause and an h-gain field}
\label{sec:QEMC_h-gain_forward_anneal}

\begin{figure*}[ht!]
    \centering
    \includegraphics[width=0.32\textwidth]{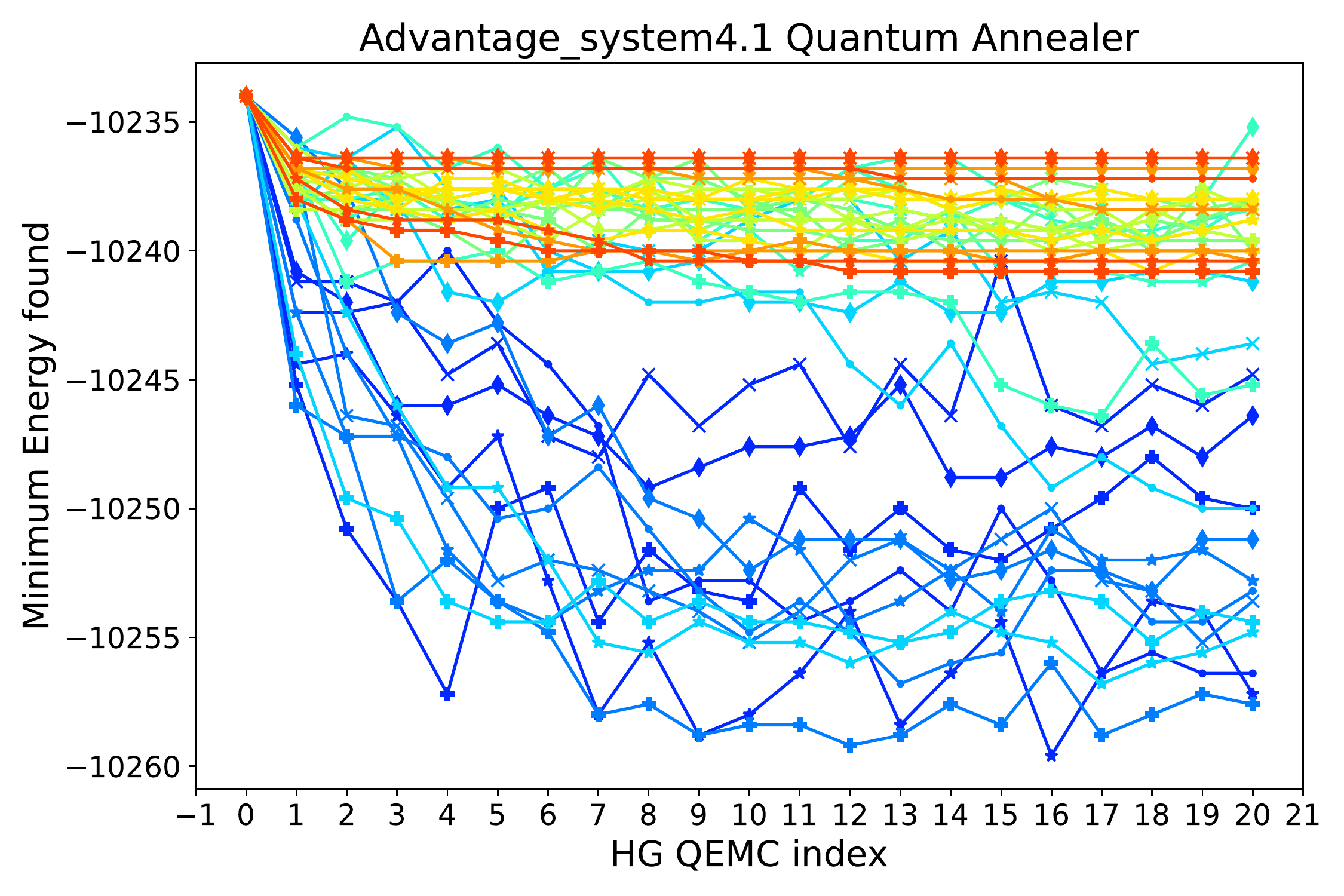}
    \includegraphics[width=0.32\textwidth]{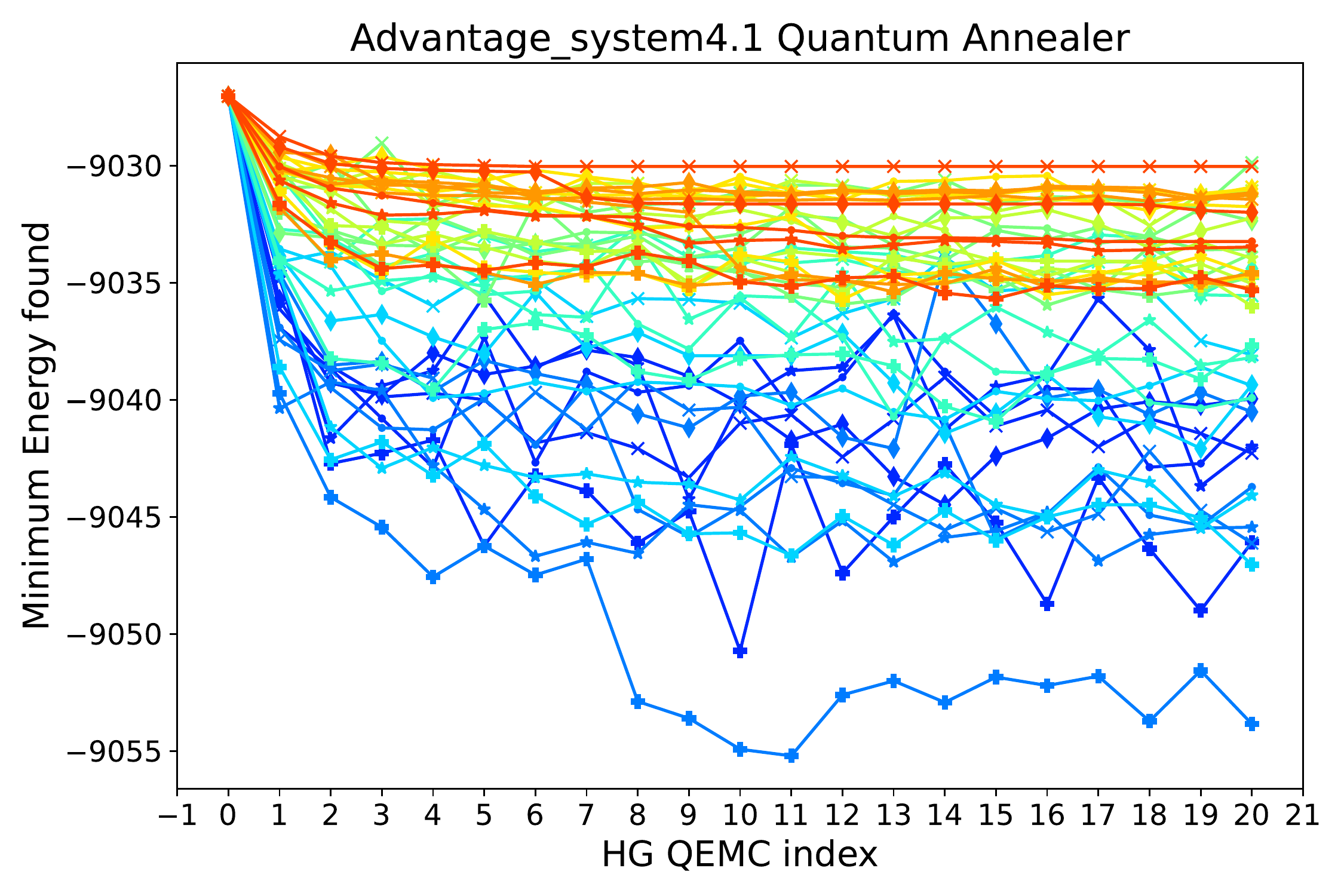}
    \includegraphics[width=0.32\textwidth]{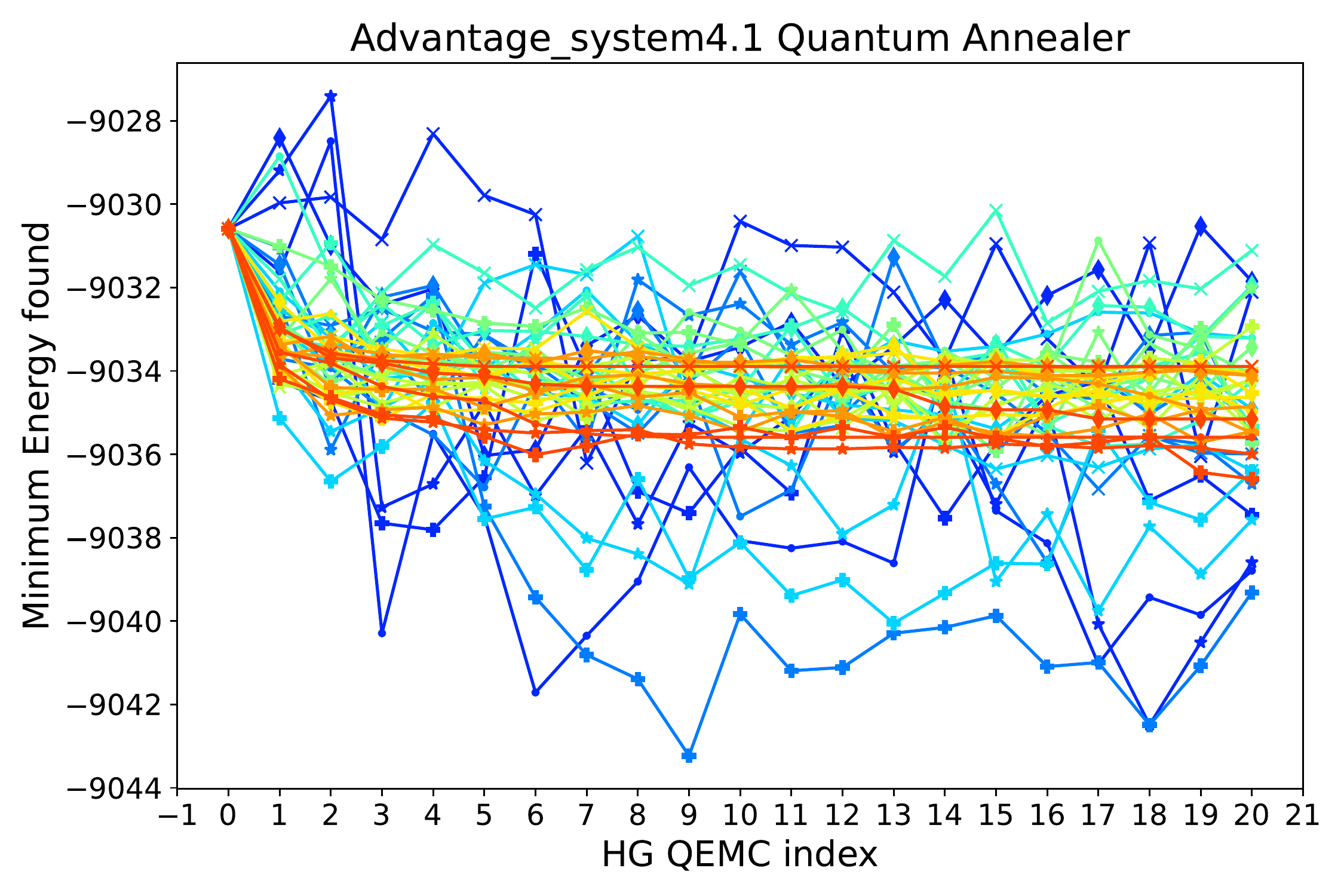}
    \includegraphics[width=0.32\textwidth]{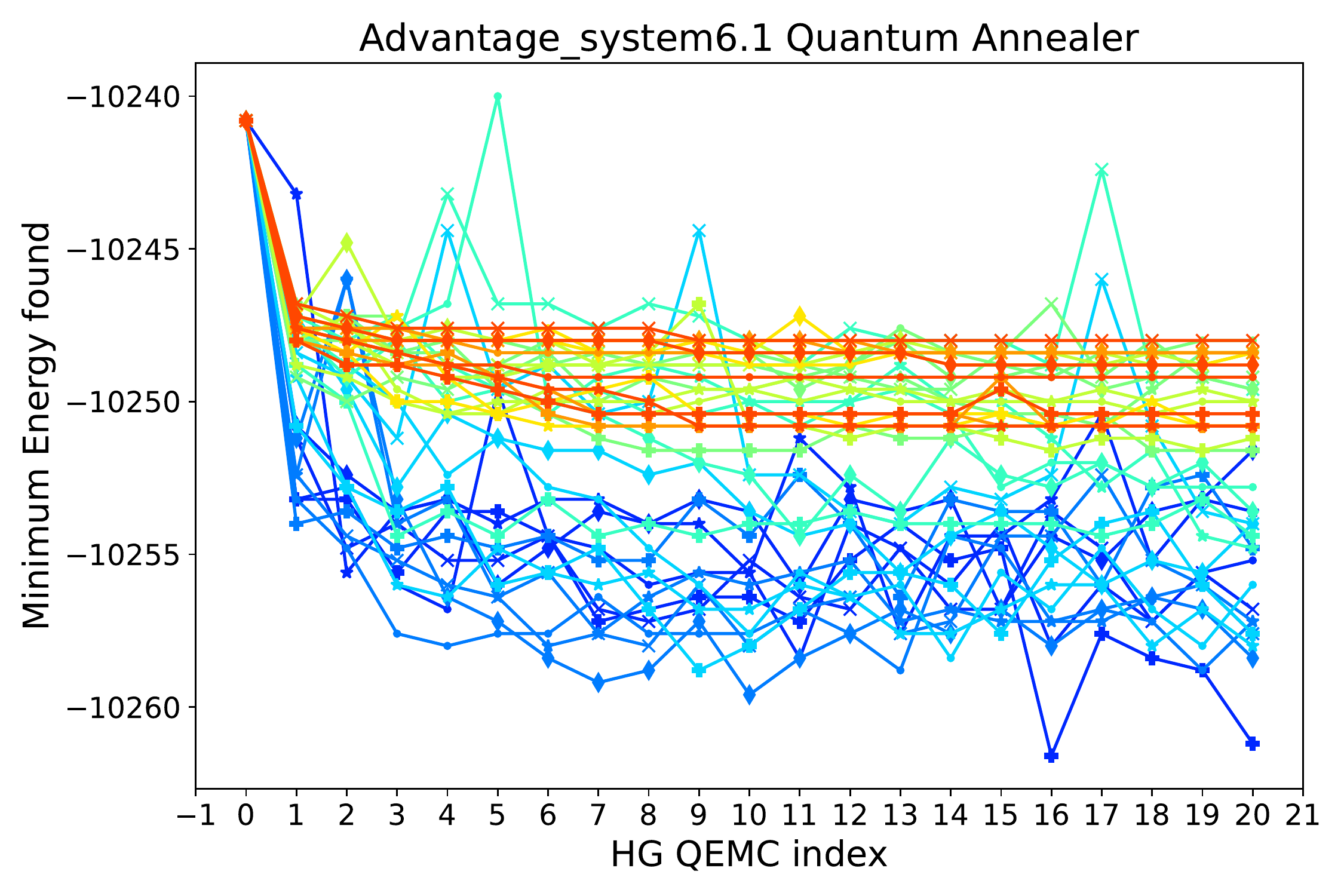}
    \includegraphics[width=0.32\textwidth]{fixed_init_QEMC_HG_forward_QA_pause_iters20_0_Advantage_system4.1_lin100.pdf}
    \includegraphics[width=0.32\textwidth]{fixed_init_QEMC_HG_forward_QA_pause_iters20_0_Advantage_system4.1_lin200.pdf}
    \includegraphics[width=0.32\textwidth]{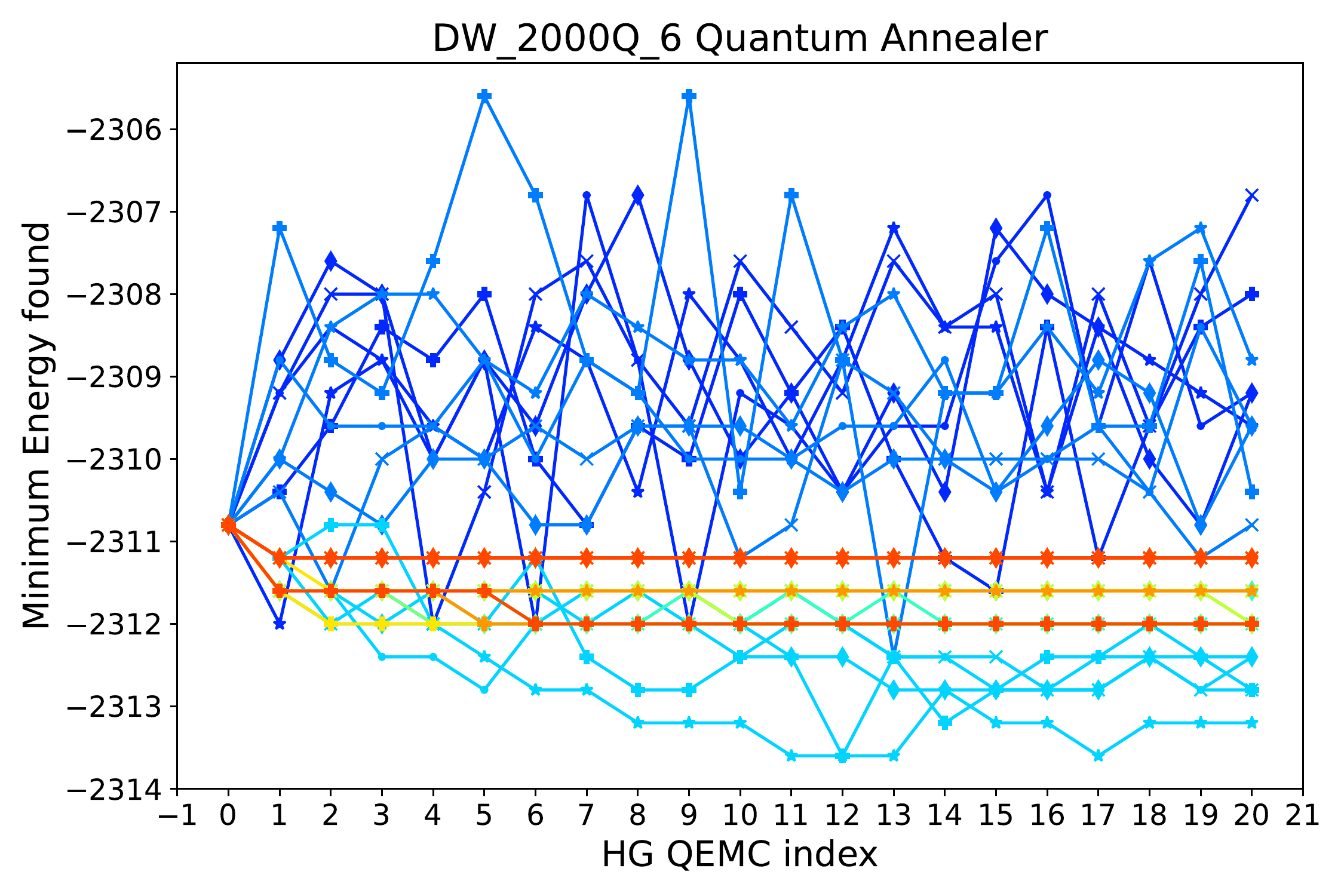}
    \includegraphics[width=0.32\textwidth]{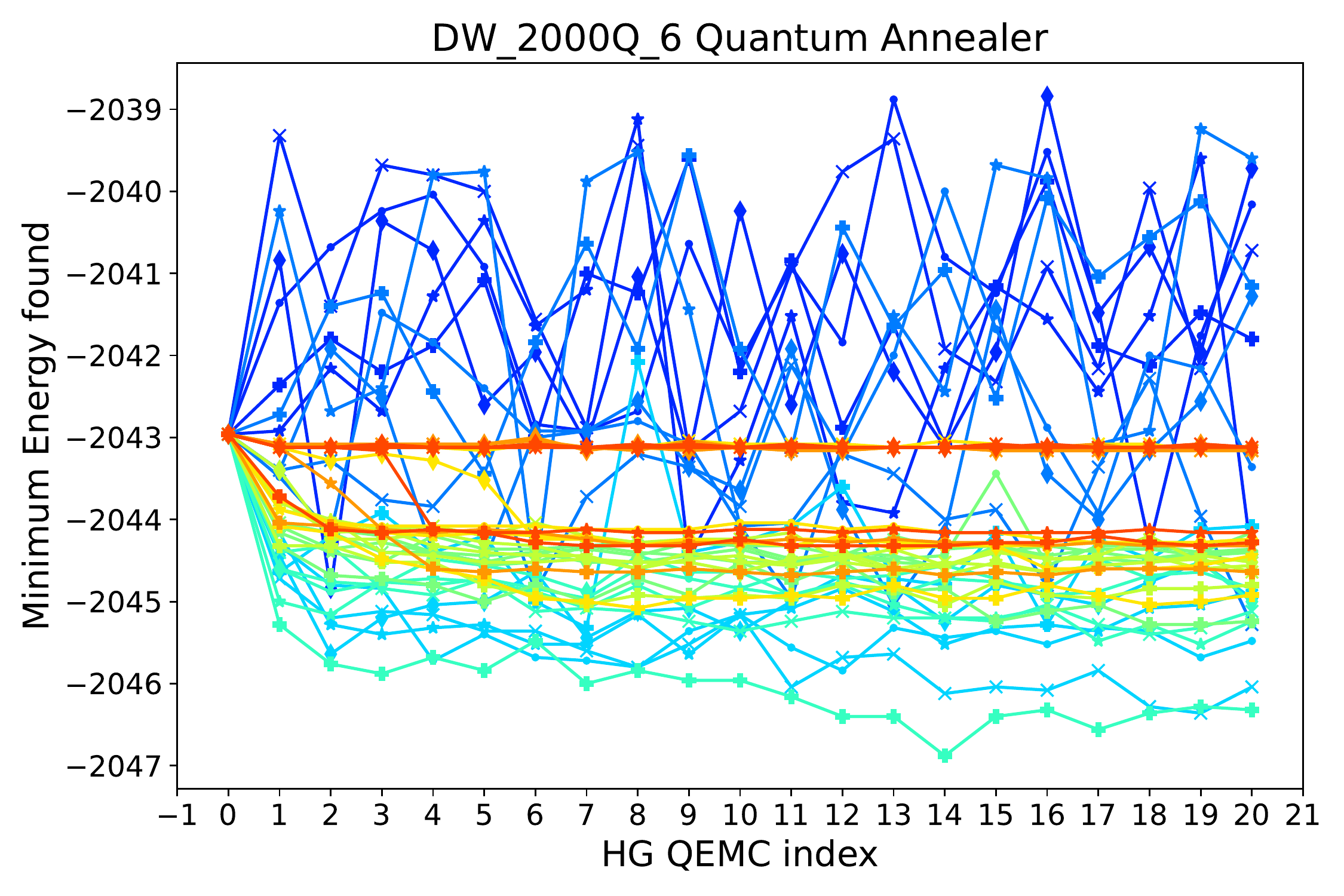}
    \includegraphics[width=0.32\textwidth]{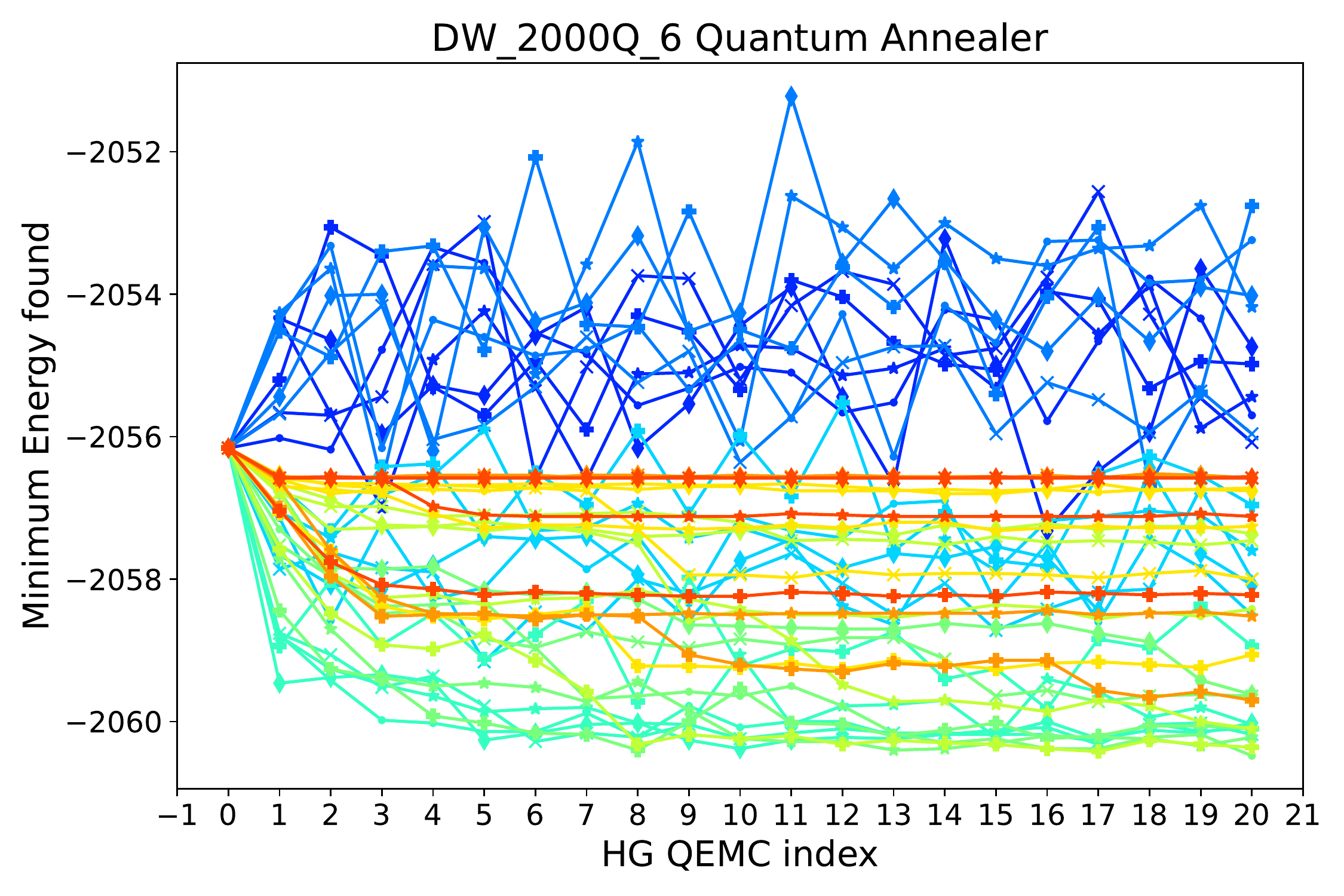}
    \includegraphics[width=0.59\textwidth]{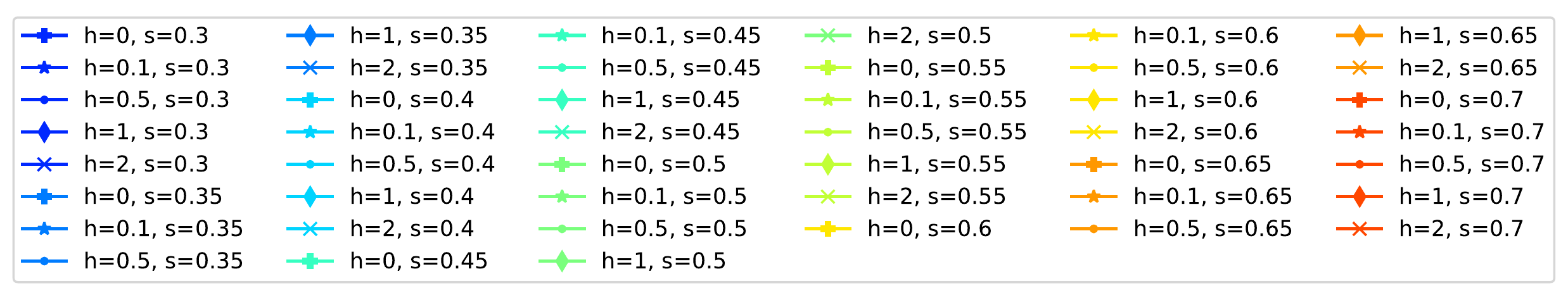}
    \caption{h-gain initial state encoding, with a forward anneal symmetric pause at varying anneal fractions. At each iteration, and parameter combination, the minimum energy found in the $1000$ anneals is plotted as a function of iteration. The h-gain strength at which the schedule begins to monotonically decrease (bottom-left subplot of Figure \ref{fig:iterated_anneal_schedules}) is varied, along with the anneal fraction at which the symmetric pause occurs (bottom-right subplot of Figure \ref{fig:iterated_anneal_schedules}) in the forward anneal schedule. Random spin glass on \texttt{Advantage\_system4.1} top row, \texttt{Advantage\_system6.1} middle row, \texttt{DW\_2000Q\_6} bottom row. Left column random spin glass has linearly spaced precision of $10$, middle column has linearly spaced precision of $100$, right column has linearly spaced precision of $200$. The forward anneal pause fraction is varied, along with the h-gain strength, the legend shows which of these parameters are varied for each of the different lines in the plots. Note that the $h$ in the legend corresponds to the h-gain strength specified for at $10$ microseconds into the $100$ microsecond anneal, but the initial h-gain field is consistently set to be initialized an h-gain strength of $2$ (which in particular means this schedule can applied to all three quantum annealers). }
    \label{fig:QEMC_h-gain_FA_pause}
\end{figure*}

The schedule of forward annealing can be modified to introduce a pause during the anneal. As shown in earlier works \cite{Marshall2019, Passarelli2019, Chen2020}, using such a pause in the anneal schedule has the potential to considerably increase the probability of successfully finding the ground state of the problem Hamiltonian.

In this case, we can use the h-gain initial state encoding method to specify an initial state, but then also use a forward annealing schedule with a pause, as opposed to the standard linearly interpolated schedule. The reasoning is that in other contexts, it has been shown that pausing can improve forward and reverse quantum annealing. Here we can utilize the h-gain field to reinforce that initial state while the pause is occurring in order to guide the iterative sampling towards monotonic solution improvement. The unknown thing a-priori is what h-gain field strength, and pause parameters, will work well for this - and we also do not know how this might compare against the combined reverse annealing and h-gain initial state encoding method. This method is the same as the method shown in Section \ref{sec:QEMC_h-gain}, but the anneal schedule is modified instead of using the standard linearly interpolated anneal schedule. 

The results for using this variant of iterative solution encoding is shown in Figure \ref{fig:QEMC_h-gain_FA_pause}. As before, $s$ is varied in the interval $[0.3,0.7]$, and $h$ is varied in $[0,2]$. Notably the optimal schedule combination of h-gain schedules and forward annealing schedules (with pauses) is not consistent across the different quantum annealers or Ising models. For \texttt{Advantage\_system4.1} sampling the precision $10$, $100$, and $200$ Ising models the best parameter choices include the pause anneal fraction being in the range of $0.3-0.4$, and the $h$ strength being set to $0$ at $10$ seconds. This shows that the initially very strong h-gain field that is applied in the first $10$ microseconds of the anneal while the transverse field terms are still dominating the system is enough to seed the anneal and guide towards lower energy states, and provides a reasonable iterative energy improvement when used with QEMC. A similar result is seen for \texttt{Advantage\_system6.1}. For \texttt{DW\_2000Q\_6} sampling the precision $10$ Ising models, the $s=0.4$ and $h=0.5$ to $h=2$ give reasonable improvements in energy as a function of iteration. For \texttt{DW\_2000Q\_6} sampling the precision $100$ and $200$ Ising models, the parameters of $s=0.45$ and $h=0$ to $h=2$ are best.

\section{Discussion}
\label{sec:discussion}
In this contribution we investigated two techniques suitable to encode an initial solution prior to the anneal on the D-Wave 2000Q quantum annealer. The two techniques are the reverse annealing feature of the D-Wave device, as well as our own method based on the h-gain feature.

Since the two techniques rely on a variety of tuning parameters, we conduct extensive testing to determine suitable annealing times, parameters, and schedules. Using optimized sets of parameters we compare both methods on both the Maximum Cut problem (whose Ising formulation does not have linear terms, thus making our h-gain technique directly applicable), as well as the Maximum Clique problem (for which we have to transform the Ising model first). Afterwards, we explore the suitability of both methods in a straightforward technique, iterated quantum annealing also referred to as QEMC. We summarize our findings as follows:

\begin{enumerate}
    \item The h-gain feature of D-Wave quantum annealers can be used to implement the QEMC algorithm. We demonstrate this using hardware-native spin glass instances. This type of iterative sampling on whole-chip quantum annealers has not been done before, but both the reverse annealing and h-gain version of this could be applied to quantum annealing chip native Ising benchmarking, such as what is demonstrated in ref. \cite{tasseff2022emerging}. 
    \item The best annealing durations for RA and HG seem to be very problem dependent. However, there is a consistent pattern in the anneal schedules for the weighted maximum clique and weighted maximum cut. Specifically, for graphs of lower density, RA schedules with an early and longer pause at a low anneal fraction are advantageous, whereas for higher densities a shorter pause at an anneal fraction of around $0.5$ seems better. For HG, the optimal schedules are close to the line connecting $(0,5)$ and $(1,0)$ independently of the density. There are important differences between the maximum cut and maximum clique problem formulations, specifically that the maximum clique QUBO formulation uses the complement of the problem graph edgeset, which means that the maximum clique QUBOs are sparser for denser graphs. This is not the case for maximum cut, and therefore this structural different could contribute to differences observed when sampling the two problems. 
    \item The scaling constants can be found successfully via Bayesian optimization.
    \item In the minor embedded weighted maximum clique experiments, the quantum annealer almost always returned samples with value for the slack variable $z=1$ for the optimized HG schedules. Having $z=1$ is necessary for our technique to work, but we did not expect it to happen so often. One possible explanation is that the HG bias helps guiding the anneals towards solutions with $z=1$. We also observe that $z=1$  occurs with much lower frequency for non-optimal HG schedules.
    \item We conclude that our technique to plant initial solutions with the help of the HG feature, as well as RA+HG, seem to be a viable alternative to reverse annealing.
\end{enumerate}

This article leaves considerable scope for future work:
\begin{enumerate}
    \item Determining where possible how close the quantum annealer comes to finding the global optimal solution for whole-lattice spin glass instances (or even just known-best solutions), for example what is studied in ref. \cite{tasseff2022emerging}, in particular with the addition of iterative sampling using reverse annealing or h-gain state encoding. 
    \item In this work we only considered RA schedules with two points defining a pause, and HG schedules with one or two points. However, more complicated schedules for both RA and HG are possible, including other annealing times, and RA+HG schedules with more points. 
    \item The h-gain initial state encoding technique we have proposed can be applied to many more interesting problems such as graph partitioning, the traveling salesman problem, minimum vertex cover, or graph coloring. Additionally, many of those problems themselves exist in different variants, including unweighted, vertex- or edge-weighted formulations.
    \item We used the Bayesian optimization framework of \cite{bayesianopt} in a rather ad-hoc way. Tuning the parameters of the Bayesian optimization, in particular with the aim to make the optimization more robust against the noise in the D-Wave samples, could further improve the optimized parameters and schedules we report.
    \item When applying the h-gain state encoding method to the minor embedded weighted maximum clique cases, in the cases where $z$ does not always equal 1, one could observe if the proportion of anneals where $z=1$ is higher for RA+HG in comparison to HG only, assuming all other variables are held constant. This is conjectured to be true because in RA+HG the value of $z$ is reinforced by the initial state of RA.
    \item Both the h-gain schedule and the anneal schedule can have many more points specified, allowing for very complex anneal schedules. Generally, complex anneal schedules have not been investigated in detail, for example by repeatedly turning off and on a specific state specified by the linear terms. 
    \item The Quantum alternating operator ansatz (QAOA) \cite{Hadfield_2019, https://doi.org/10.48550/arxiv.1411.4028} algorithm, which is a hybrid classical-quantum algorithm that operates on universal gate model quantum computers, can be warm-started \cite{Egger_2021}. Therefore, it can be used to iteratively improve solution quality in a similar fashion to QEMC.  
\end{enumerate}

\section*{Acknowledgments}
\label{section:acknowledgments}
This work was supported by the U.S. Department of Energy through the Los Alamos National Laboratory and the Laboratory Directed Research and Development program of Los Alamos National Laboratory under project number 20220656ER. Research presented in this article was supported by the NNSA's Advanced Simulation and Computing Beyond Moore's Law Program at Los Alamos National Laboratory. Los Alamos National Laboratory is operated by Triad National Security, LLC, for the National Nuclear Security Administration of U.S. Department of Energy (Contract No.~89233218CNA000001). This research used resources provided by the Los Alamos National Laboratory Institutional Computing Program. This work has been assigned the LANL technical report number LA-UR-23-22902.

The work of the H.D.\ was also supported by grant number KP-06-DB-11 of the Bulgarian National Science Fund. Funding for G.H.\ was provided through Cure Alzheimer's Fund, the National Institutes of Health [1R01 AI 154470-01; 2U01 HG 008685; R21 HD 095228 008976; U01 HL 089856; U01 HL 089897; P01 HL 120839; P01 HL 132825; 2U01 HG 008685; R21 HD 095228, P01HL132825], the National Science Foundation [NSF PHY 2033046; NSF GRFP 1745302], and a NIH Center grant [P30-ES002109].

\begin{figure*}[th!]
    \centering
    \includegraphics[width=0.4\textwidth]{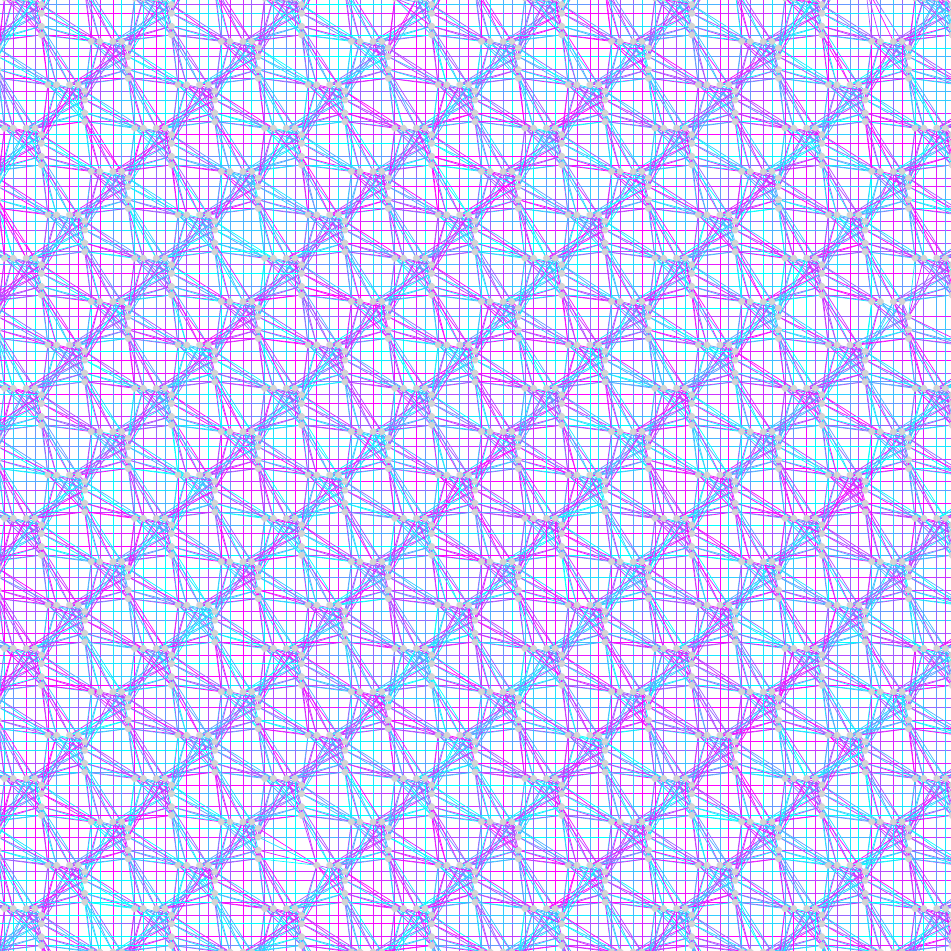}
    \includegraphics[width=0.4\textwidth]{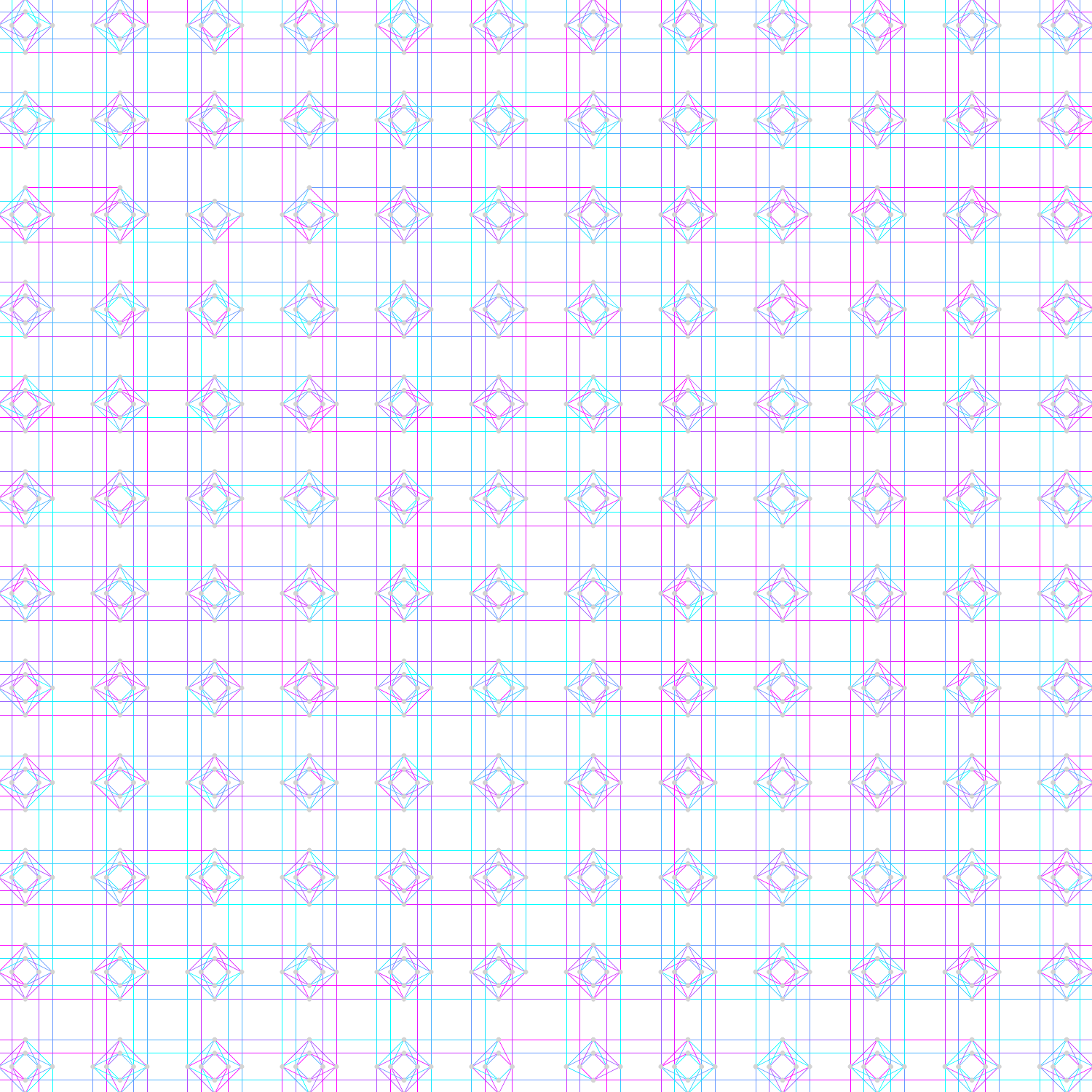}
    \caption{Cropped hardware graphs (edges of the chip are not shown) of the hardware connectivity for embedded spin glass instances on \texttt{Advantage\_system4.1} (left), both of which are Pegasus $P_{16}$ graphs with hardware defects, and \texttt{DW\_2000Q\_6} (right) which is a Chimera $C_{16}$ graph with hardware defects. The values of the quadratic coefficients are color coded from $-1$ (light blue) to $+1$ (purple). }
    \label{fig:hardware_graph_spin_glasses}
\end{figure*}

\appendix
\section{Scaling factors for the h-gain state encoding technique for minor embedded problems}
\label{sec:scaling_factors}
Table~\ref{tab:scaling_factors} shows the precise scaling factors $\alpha_1$ and $\alpha_2$ of eq.~\eqref{eq:H_final} we employ to apply the HG technique of Section~\ref{sec:hgain}, computed using the bayesian optimization procedure. Those scaling factors depend on the density of the graph that each problem is applied to. The Ising formulation of the Maximum Cut problem of Section~\ref{sec:experiments_maxcut} only requires one scaling factor, whereas the one of the Maximum Clique problem of Section~\ref{sec:experiments_maxclique} requires two scaling factors.

\begin{table}[h!]
    \centering
    \caption{Best scaling factor(s) for HG state encoding as a function of the random graph density, for the minor embedded weighted maximum clique and maximum cut problems. Left: Parameter $\alpha_1$ of the Maximum Cut problem. Right: Parameters $\alpha_1$ and $\alpha_2$ of the Maximum Clique problem. Data from \texttt{DW\_2000Q\_LANL}. 
    \label{tab:scaling_factors}}
    \begin{tabular}{c|c|cc}
    Density & Maximum Cut & \multicolumn{2}{|c}{Maximum Clique}\\
    & $\alpha_1$ & $\alpha_1$ & $\alpha_2$\\
    \hline
    0.1 & 0.16 & 0.3271 & 0.2997\\
    0.2 & 0.23 & 0.2709 & 0.8897\\
    0.3 & 0.51 & 0.1997 & 0.5401\\
    0.4 & 0.48 & 0.1542 & 0.9246\\
    0.5 & 0.41 & 0.3467 & 0.2473\\
    0.6 & 0.48 & 0.2602 & 0.7586\\
    0.7 & 0.63 & 0.0292 & 0.7455\\
    0.8 & 0.93 & 0.0334 & 0.0210\\
    0.9 & 0.33 & 0.0370 & 0.5121
    \end{tabular}
\end{table}

\section{Cropped hardware graphs of the natively embedded spin glasses}
\label{sec:hardware_graph_spin_glasses}
The random Ising models embedded onto the D-Wave hardware for the QEMC experiments are chosen to match the entire native hardware graphs of each of the three QPUs used (see Table~\ref{table:hardware_summary}). An example of such hardware graphs is given in Figure~\ref{fig:hardware_graph_spin_glasses}.

\section{Parameter tuning for the Weighted Maximum Cut Problem}
\label{sec:tuning_maxcut}
The following two subsections elaborate on the tuning of the scaling factors, the annealing time, and the computation of the anneal schedules for the experiments involving the weighted Maximum Cut problem of Section~\ref{sec:experiments_maxcut}.

\subsection{Setting scaling factors and annealing time}
\label{sec:experiments_maxcut_parameters}
We start by determining a suitable choice of the scaling factor $\alpha_1$ in eq.~\eqref{eq:H_final} for the Maximum Cut problem using the Bayesian optimization.

As a fitness function for the optimization, we use the improvement in the maximum cut over the baseline. Each time the optimizer issues a call to the fitness function, we supply the average of $10$ problems optimized with either RA, HG, or RA+HG (depending on which one is optimized) using the parameter set probed by the Bayesian framework. The fitness value is then the average maximum cut improvement over the baseline. We make the fitness function dependent on three parameters, the scaling factor $\alpha_1$ as well as the parameters $(h,t)$ determining the HG schedule (see Section~\ref{sec:parameters}). For this experiment the anneal time is set to $1$ microseconds.

\begin{figure}[th!]
    \centering
    \includegraphics[width=0.5\textwidth]{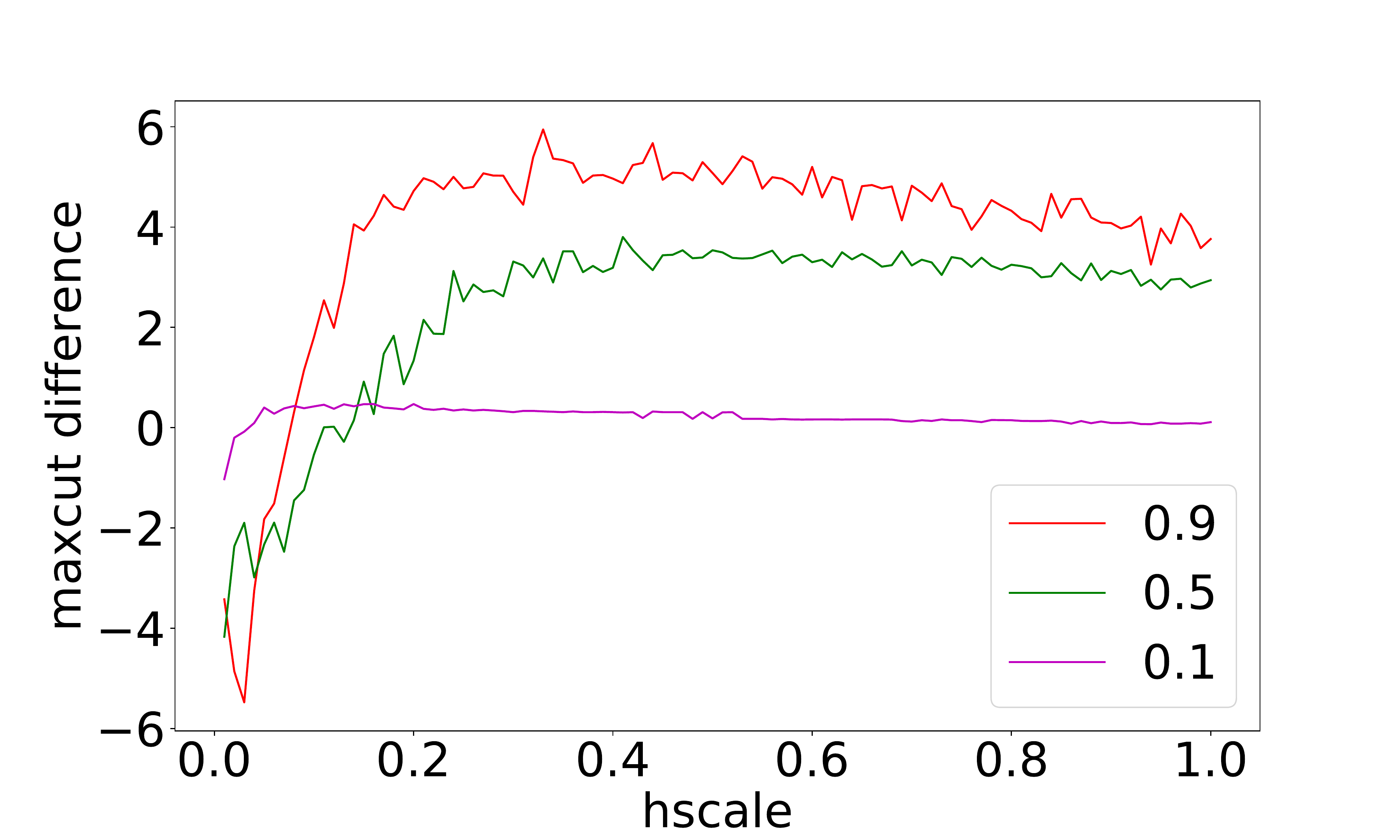}
    \caption{Maximum Cut problem. Difference in the maximum cut value to the baseline as a function of the HG scaling factor $\alpha_1$. The goal is for the value on the y-axis to be maximized, meaning that the tuned scaling factor improves the weight of the cut found. Data from \texttt{DW\_2000Q\_LANL}.  }
    \label{fig:maxcut_scaling}
\end{figure}

After obtaining the fittest values, we fix the schedule $(h,t)$ and the annealing time of $1$ microseconds, and cross check the scaling factor $\alpha_1$ on a linear grid on $[0.01,1]$ in increments of $0.01$. Results for three different densities are shown in Figure~\ref{fig:maxcut_scaling}, which displays the difference in the maximum cut value to the baseline as a function of $\alpha_1$. We observe that the best choice of $\alpha_1$ is very dependent on the graph density, with, e.g.,\ the best choice for density $0.9$ occurring at $\alpha_1 \approx 0.3$. Table~\ref{tab:scaling_factors} shows the precise optima we found for all nine densities $p \in \{0.1,\ldots,0.9\}$. We will be selecting the scaling factor depending on the underlying graph density in the remainder of this section.

Next, we determine a suitable anneal duration for RA, HG, as well as the combined RA+HG. For this, we fix the HG schedule to the three points $[0, 5], [0.5T, 2.5], [T, 0]$ and the RA schedule to $[0, 1]$, $[0.25T, 0.25]$, $[0.75T, 0.25]$, $[T, 1]$, where $T$ is the total annealing duration (given in Table~\ref{tab:maxcut}). We note that these schedules are not optimal. Instead, they merely divide up the variable range in an equidistant fashion. We choose the anneal fraction to be around $0.25$ as suggested in~\cite{McGeoch2018}.

\begin{table*}[ht!]
    \centering
    \caption{Evaluation of RA and HG, as well as RA+HG, for smallest and largest possible annealing times (in microseconds) possible under the bayesian optimization methodology (described in Section \ref{sec:parameters}. Maximal cut difference on Erd{\H o}s--R{\'e}nyi graphs of density ranging from $0.1$ to $0.9$. Here more positive values indicate that the solution planting method improved the mean cut-weight value found out of an average of the best solutions found from the $10$ fixed random graphs (per density). Data from \texttt{DW\_2000Q\_LANL}. \label{tab:maxcut}}
    \begin{tabular}{l|c||ccccccccc}
        & $T$ [ms] & 0.9 & 0.8 & 0.7 & 0.6 & 0.5 & 0.4 & 0.3 & 0.2 & 0.1\\
        \hline
        RA & 100 & 0.722 & -0.151 & 2.060 & 1.974 & 2.551 & 1.909 & 3.425 & 2.544 & 0.467\\
        RA & 2000 & 2.412 & -0.149 & 3.653 & 2.140 & 3.261 & 2.859 & 4.507 & 3.209 & 0.610\\
        HG & 1 & 5.173 & 2.084 & 3.72 & 2.632 & 3.127 & 2.799 & 2.876 & 2.082 & 0.338\\
        HG & 2000 & 3.963 & 1.266 & 2.216 & 1.421 & 2.047 & 1.617 & 1.705 & 1.525 & 0.185\\
        RA+HG & 100 & 3.853 & 2.832 & 4.867 & 3.022 & 3.478 & 3.170 & 4.32 & 2.726 & 0.526 \\ RA+HG & 2000 & 4.145 & 2.540 & 5.001 & 2.725 & 4.222 & 3.128 & 4.526 & 3.113 & 0.566
    \end{tabular}
\end{table*}

Table~\ref{tab:maxcut} shows maximum cut results for the smallest and largest possible anneal times as a function of the graph density. We observe that an annealing duration of $2000$ microseconds works best for RA, while a $1$ microseconds anneal is best for HG. The combined technique of RA+HG does not seem to be as affected by the annealing duration, but since an annealing time of $2000$ microseconds yields slightly better results, we decide to employ RA+HG in connection with a $2000$ microseconds anneal in this section.

\subsection{Schedule computation via Bayesian optimization}
\label{sec:experiments_maxcut_bayes}
After having fixed the annealing duration for all three methods, we proceed by determining the parameters of the anneal schedule (see Section~\ref{sec:parameters}) via Bayesian optimization. For each density, we carry out a single run of the Bayesian optimizer.

\begin{figure*}[ht]
    \centering
    \includegraphics[width=0.32\textwidth]{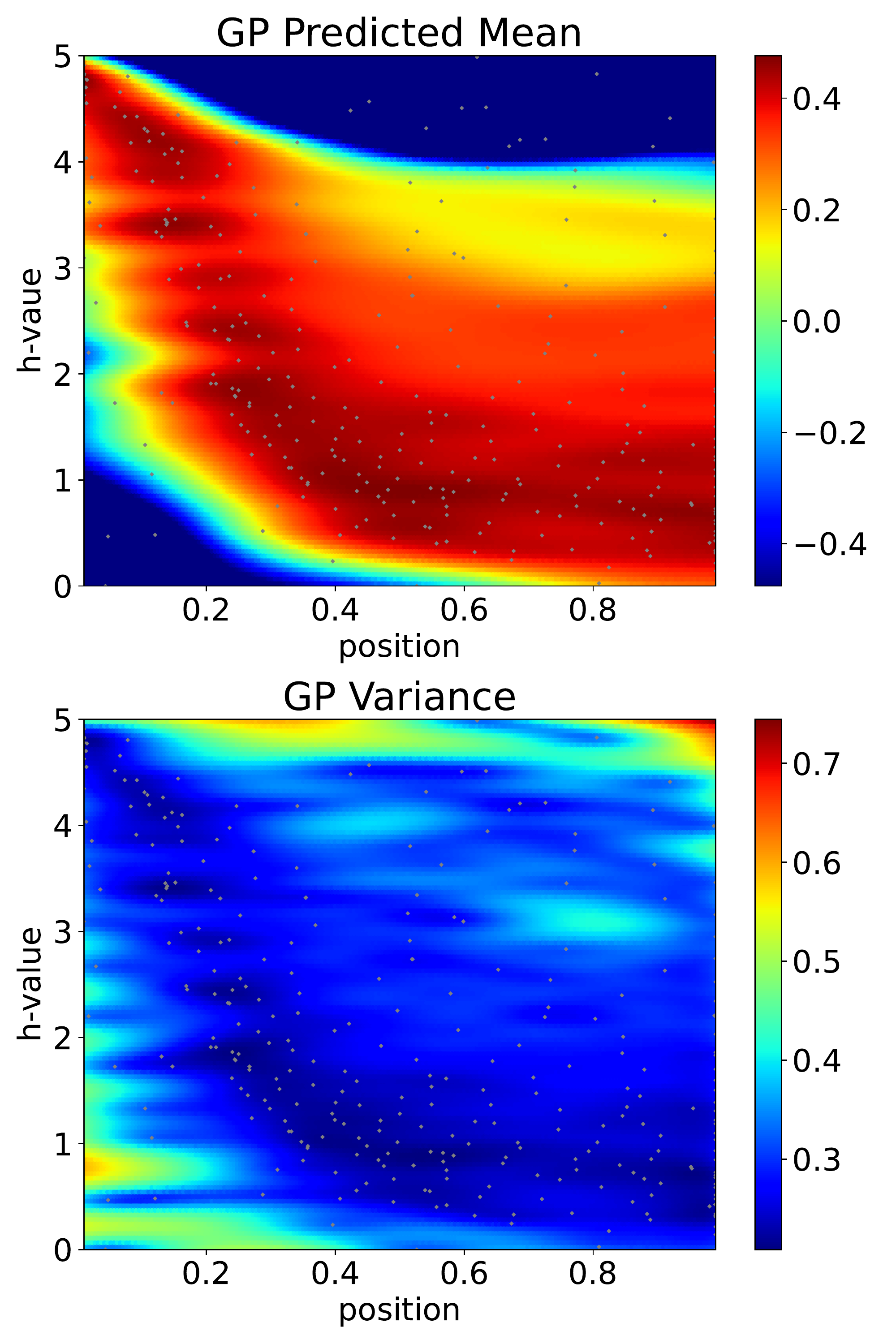}
    \includegraphics[width=0.32\textwidth]{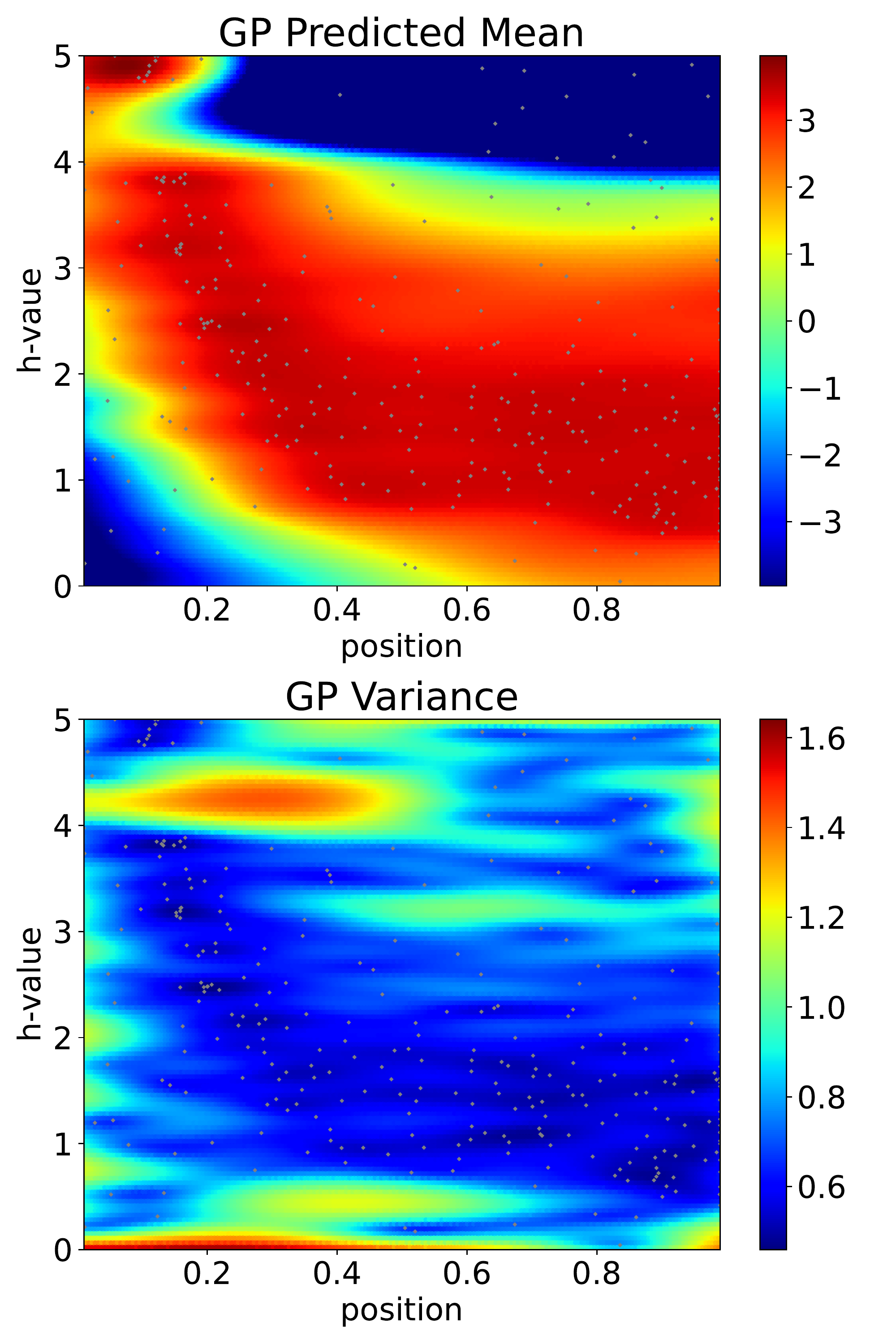}
    \includegraphics[width=0.32\textwidth]{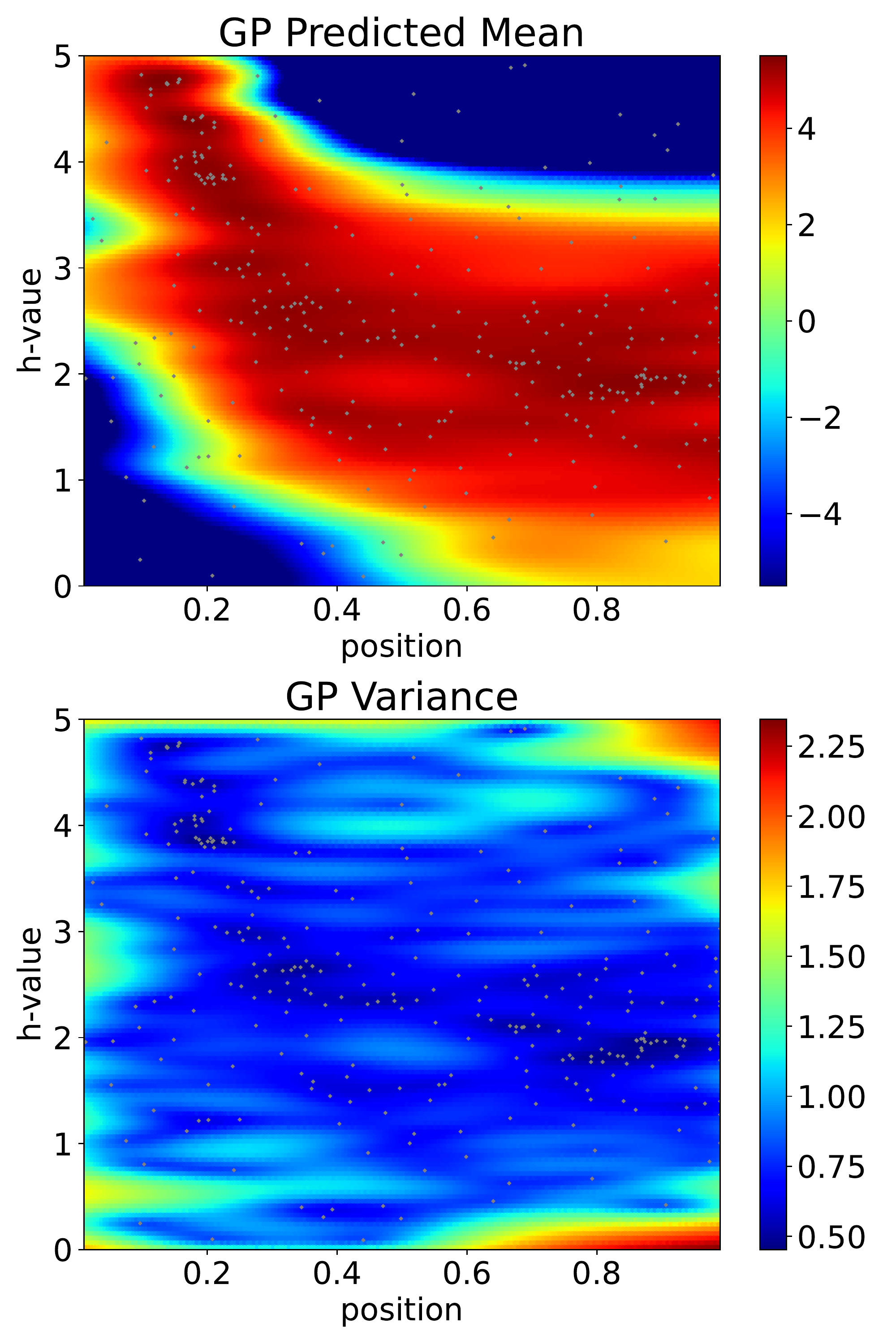}
    \caption{Bayesian optimization of the h-gain schedule for the weighted maximum cut problem on random graphs of density $0.1$ (left column), $0.5$ (middle column) and $0.9$ (right column). Mean (top sub-plots) and variance (bottom sub-plots) of the Gaussian process (GP) used by the Bayesian optimizer, where the goal is to maximize the objective function (e.g. red regions mean that the constructed h-gain schedule resulted in better cut values being sampled). The parameters of the HG schedule for Maximum Cut are visualized as a heatmap for and annealing time $1$ microsecond. Top shows the maximum cut size improvement as a function of $g(t) \in [0,5]$ (see eq.~\eqref{eq:hgain}) on the y-axis, where $t$ is the position in the schedule (x-axis). The small grey dots indicate the points where the objective function was evaluated. Each heatmap is computed from a total of $300$ function calls ($100$ random points and $200$ optimization steps). Data from \texttt{DW\_2000Q\_LANL}. }
    \label{fig:hgain_heat_map_maxcut}
\end{figure*}

Since the schedule of HG has two parameters determining the midpoint in the anneal schedule (see Section~\ref{sec:parameters}), we can visualize its optimization as a heatmap in Figure~\ref{fig:hgain_heat_map_maxcut}. In particular, Figure~\ref{fig:hgain_heat_map_maxcut} shows the color coded improvement in cut size over the baseline for each possible midpoint in the HG schedule. As described in Section~\ref{sec:parameters}, this point consists of a position in the anneal and a value of the HG function $g(t) \in [0,5]$, see eq.~\eqref{eq:hgain}. Notice that the region of good h-gain schedules for the tested maximum cut problem instances can be directly read off of Figure \ref{fig:hgain_heat_map_maxcut} (as opposed to the singular best schedules shown in Figure \ref{fig:maxcut_schedules}), shown by the dark red region of the top sub-plot, where the y-axis h-value corresponds to the h-gain strength of the middle point of the h-gain schedule and the position on the x-axis corresponds to a proportion of the total annealing time. 

The figure shows that the best choice of the HG value, defined as the one yielding the best improvement in maximum cut difference (red values), roughly decreases with the position in the anneal. We determine the maximum in this way for each density $p \in \{0.1,\ldots,0.9\}$. The schedules for RA (3 parameters) and RA+HG (five parameters) are fitted in a similar way, one schedule per density $p \in \{0.1,\ldots,0.9\}$.

\section{Parameter tuning for the Maximum Clique problem}
\label{sec:tuning_maxclique}
The following two subsections elaborate on the tuning of the scaling factors, the annealing time, and the computation of the anneal schedules for the experiments involving the Maximum Clique problem of Section~\ref{sec:experiments_maxclique}.

\subsection{Setting scaling factors and annealing time}
\label{sec:experiments_maxclique_parameters}
We again focus first on the HG feature and repeat the tuning of Section~\ref{sec:experiments_maxcut_parameters}. In particular, to determine the two scaling factors $\alpha_1$ for $h(\bm x)$ and $\alpha_2$ for $z$, we run Bayesian optimization to fit both the schedule and the scaling factors simultaneously (see Section~\ref{sec:parameters}). For this, we fix the anneal time at $1$ microsecond.

Each time the Bayesian optimizer requests a new point, we return the average maximum clique improvement over the baseline (using $1000$ anneals) for $10$ graphs for each density. If no solutions are found, i.e.\ $z=-1$ for all $1000$ anneals, we return a large negative constant (we use $-1000$) to the optimizer.

After having obtained the result from the Bayesian optimization run, we fix the best schedule found. After initializing the Bayesian optimization algorithm with the parameters of the previous best solution (the previously found scaling constants $\alpha_1$ and $\alpha_2$ for the fixed schedule), we re-fit $\alpha_1$ and $\alpha_2$ with the help of the Bayesian optimization.

\begin{figure}[ht!]
    \centering
    \includegraphics[width=0.5\textwidth]{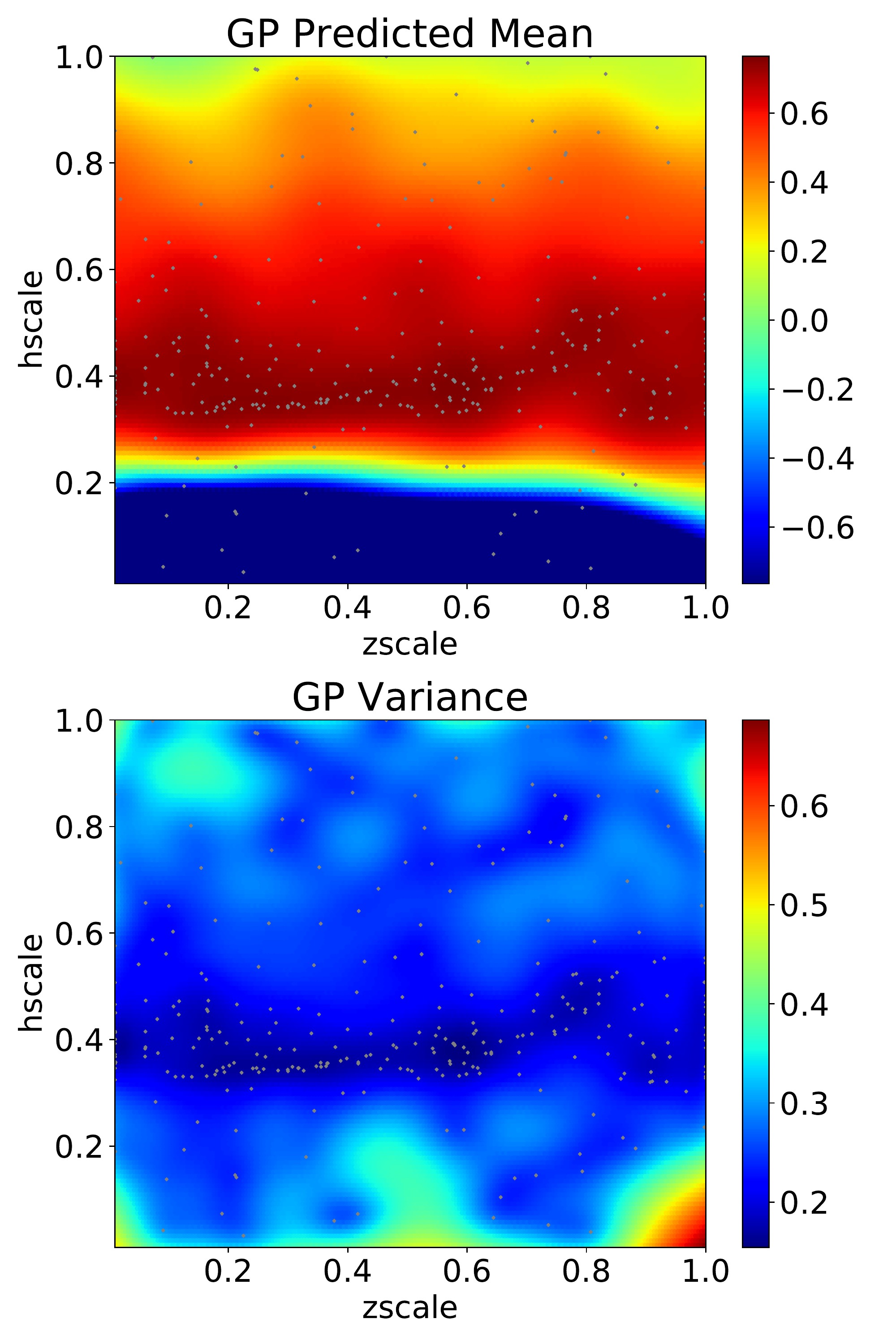}
    \caption{Mean (top) and variance (bottom) of the Gaussian process (GP) used by the Bayesian optimizer. Landscape encoded in $\alpha_1$ (h-scale on the y-axis) and $\alpha_2$ (z-scale on the x-axis) for a graph density of $p=0.5$. The small grey dots indicate the points where the objective function was evaluated. Data from \texttt{DW\_2000Q\_LANL}. }
    \label{fig:scaling_factor_heat_maps}
\end{figure}

Figure~\ref{fig:scaling_factor_heat_maps} shows the result of the Bayesian optimization run with a fixed annealing duration of $1$ microseconds and a graph density of $p=0.5$, as well as our fixed optimized schedule. We see that the best values for the scaling constants are essentially in a band around $\alpha_1=0.4$ (h-scale), with various maxima for $\alpha_2$ (z-scale). The precise optimal scaling factors for $p=0.5$ returned by the Bayesian optimization are $\alpha_1=0.35$ (for h-scale) and $\alpha_2 = 0.25$ (for z-scale), which we fix for the remainder of this section.

We repeat this procedure for the other values of $p \in \{0.1,0.2,\ldots,0.9\}$ as well, and use individual scaling constants for each density in the remainder of this section as done for the Maximum Cut problem.

\begin{table*}[ht!]
    \centering
    \caption{Evaluation of RA and HG, as well as combined RA+HG, for smallest and largest possible anneal times (in microseconds) allowed using the bayesian optimization method to compute good h-gain and anneal schedules (described in Section \ref{sec:parameters}). Maximum clique difference on Erd{\H o}s--R{\'e}nyi graphs for densities ranging from $0.1$ to $0.9$. Here more positive values indicate that the solution planting method improved the mean clique-weight value found out of an average of the best solutions found from the $10$ fixed random graphs (per density). \label{tab:maxclique}}
    \begin{tabular}{l|c||ccccccccc}
        & anneal [ms] & 0.9 & 0.8 & 0.7 & 0.6 & 0.5 & 0.4 & 0.3 & 0.2 & 0.1\\
        \hline
        RA & 100 & 0.366 & 0.142 & 0.031 & 0.068 & -0.309 & -0.124 & -0.391 & -0.364 & -0.322\\
        RA & 2000 & 0.481 & 0.289 & -0.041 &  -0.060 & -0.322 & -0.171 & -0.474 & -0.408 & -0.309\\
        HG & 1 & 1.195 & 1.307 & 1.263 & 0 & 0 & 0 & 0 & 0 & 0\\
        HG & 2000 & 0.908 & 1.004 & 1.018 & 0 & 0 & 0 & 0 & 0 & 0\\
        RA+HG & 100 & 0.780 & -0.797 & -3.874 & 0.104 & 0 & 0.659 & 0.442 & 0 & 0 \\
        RA+HG & 2000 & 1.127 & 0.050 & -2.167 & 0.011 & 0 & 0.610 & 0.309 & 0.039 & 0
    \end{tabular}
\end{table*}

After having tuned the HG feature, we now focus again on the three techniques (RA, HG and RA+HG). Similarly to Section~\ref{sec:experiments_maxcut_parameters}, we determine a suitable annealing duration for all three techniques by testing them for the shortest and longest annealing durations possible under the constraints of the bayesian optimization we employed on Erd{\H o}s--R{\'e}nyi graphs of varying values of $p \in \{0,1,\ldots,0.9\}$.

Results are displayed in Table~\ref{tab:maxclique}, showing that RA works better for longer annealing times (especially for denser graphs), and HG works consistently better for shorter annealing durations. We will thus employ RA with an annealing time of $2000$ microseconds in the remainder the minor embedded bayesian optimization experiments, and HG with an annealing time of $1$ microseconds. For RA+HG we fix the annealing duration at $2000$ microseconds.

\subsection{Schedule computation via Bayesian optimization}
\label{sec:experiments_maxclique_bayes}
The experiments of the previous sections allowed us to fix the anneal times, as well as the (density dependent) schedules and scaling constants for HG, RA, as well as RA+HG. Using the calibrations of the three methods, we now evaluate HG, RA, and RA+HG on graphs of varying density with respect to the improvement of the maximum clique weight over the baseline.

\begin{figure*}
    \centering
    \includegraphics[width=0.5\textwidth]{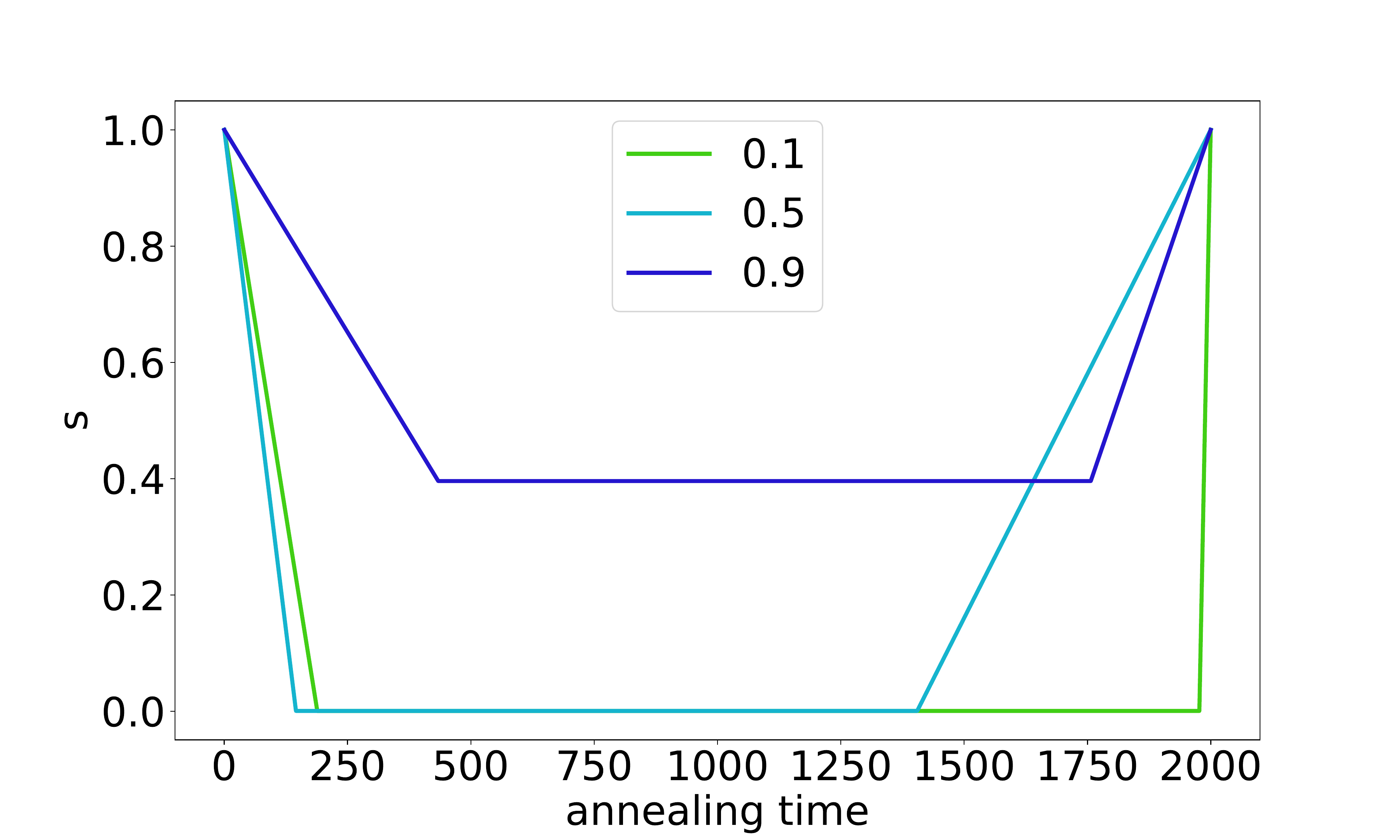}\hfill
    \includegraphics[width=0.5\textwidth]{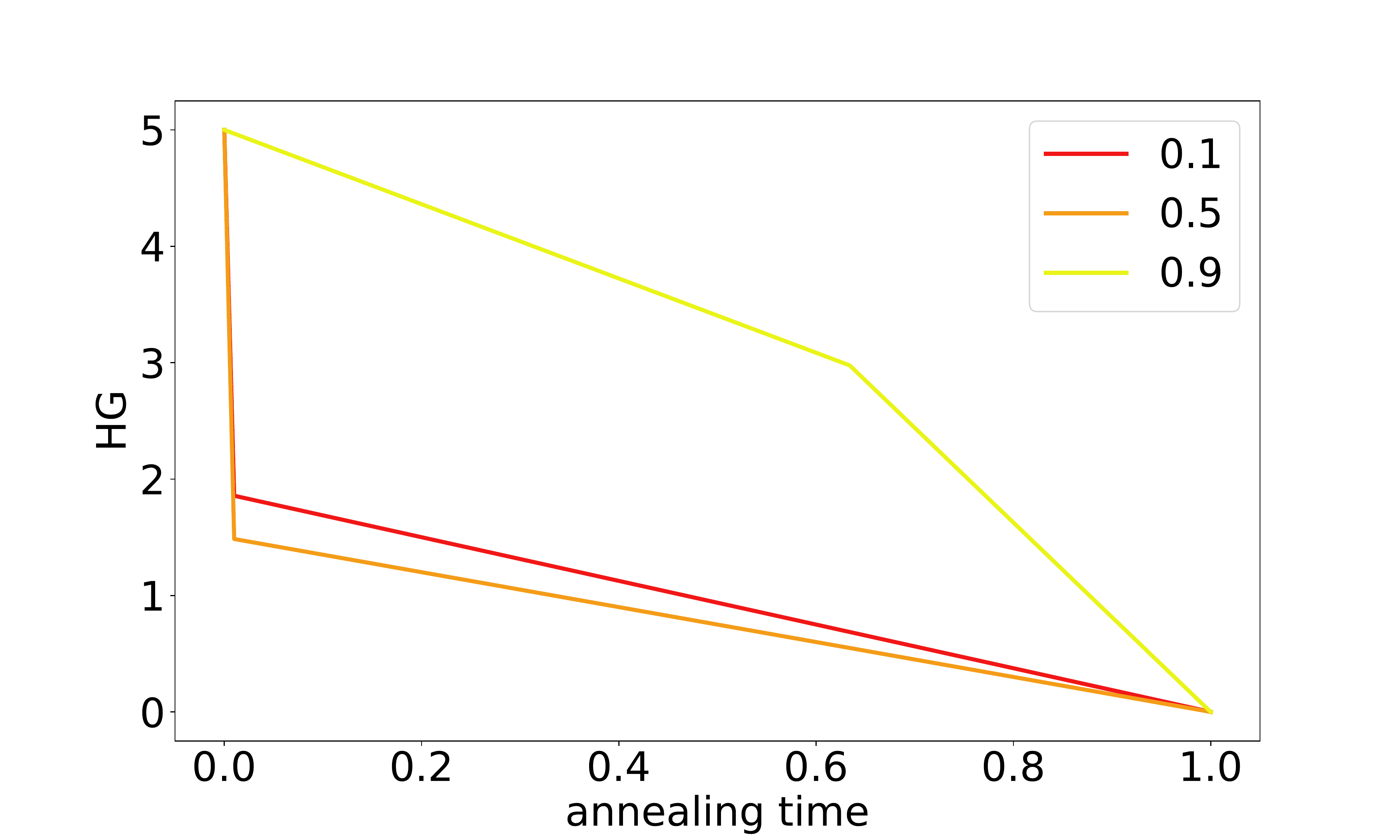}\hfill
    \caption{Weighted Maximum Clique problem. Best schedules for RA (left), and HG (right) optimized for maximum clique weight. Each line is the best schedule for one density. These schedules were computed using the bayesian optimization approach, on \texttt{DW\_2000Q\_LANL}. }
    \label{fig:maxclique_schedules}
\end{figure*}

Similarly to Figure~\ref{fig:maxcut_schedules} (left), we also report the best schedules found by the Bayesian optimization in the case of the Maximum Clique problem. All schedules are optimized to maximize the maximum clique weight over the baseline in a standard forward anneal.

Figure~\ref{fig:maxclique_schedules} shows the resulting schedules. For RA we see that, in contrast to the schedules for the Maximum Cut problem, for low densities the optimal RA schedules decrease the anneal fraction at the start of the anneal to very small values, and perform a pause until almost the full annealing time. As the graph under consideration becomes denser, the anneal fraction is only decreased down to roughly $0.4$, and the pause occurs in the center having roughly a duration of half the annealing time. The h-gain schedules (Figure~\ref{fig:maxclique_schedules}, right) resemble the ones observed for the Maximum Cut problem.

\begin{figure*}
    \centering
    \includegraphics[width=0.32\textwidth]{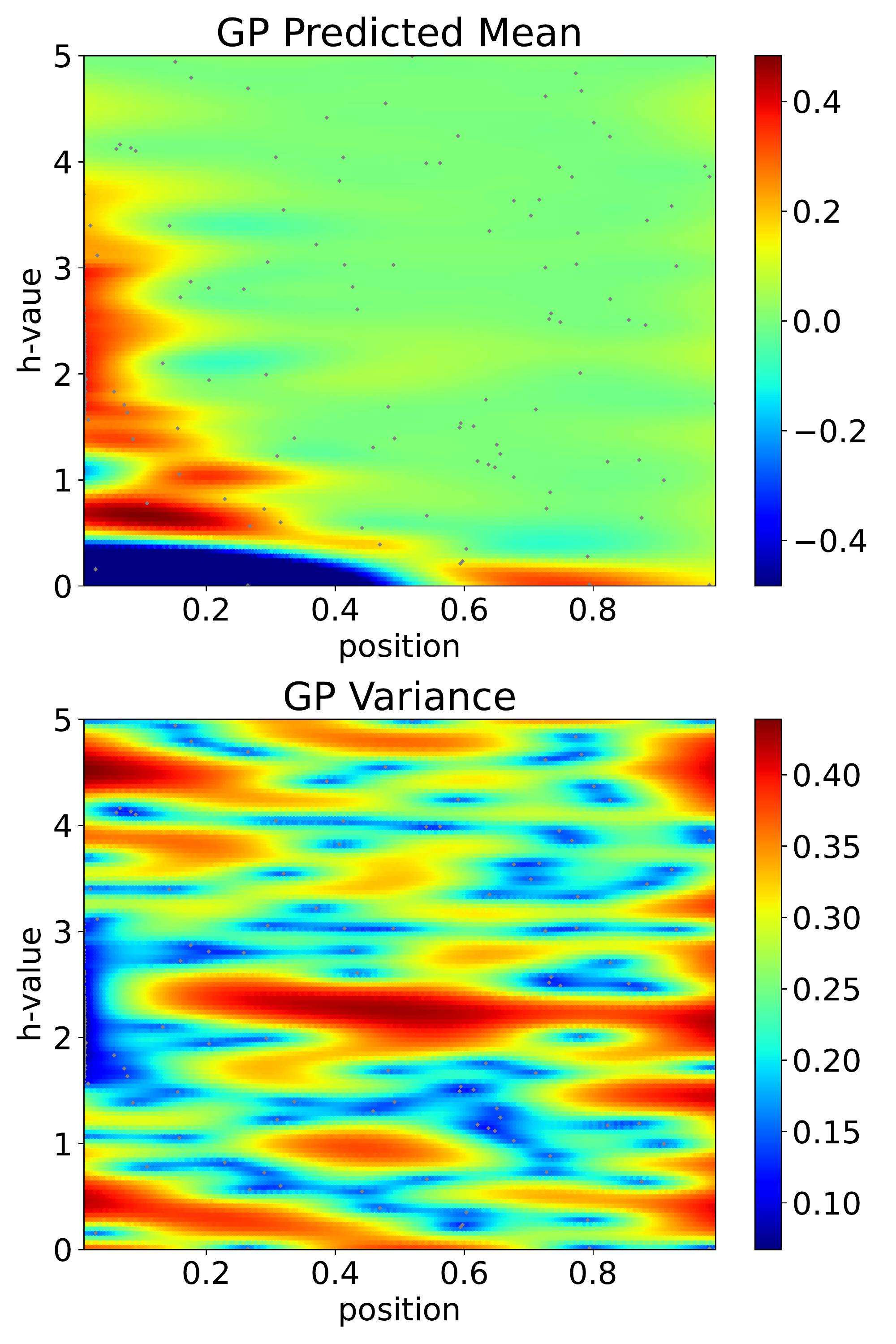}
    \includegraphics[width=0.32\textwidth]{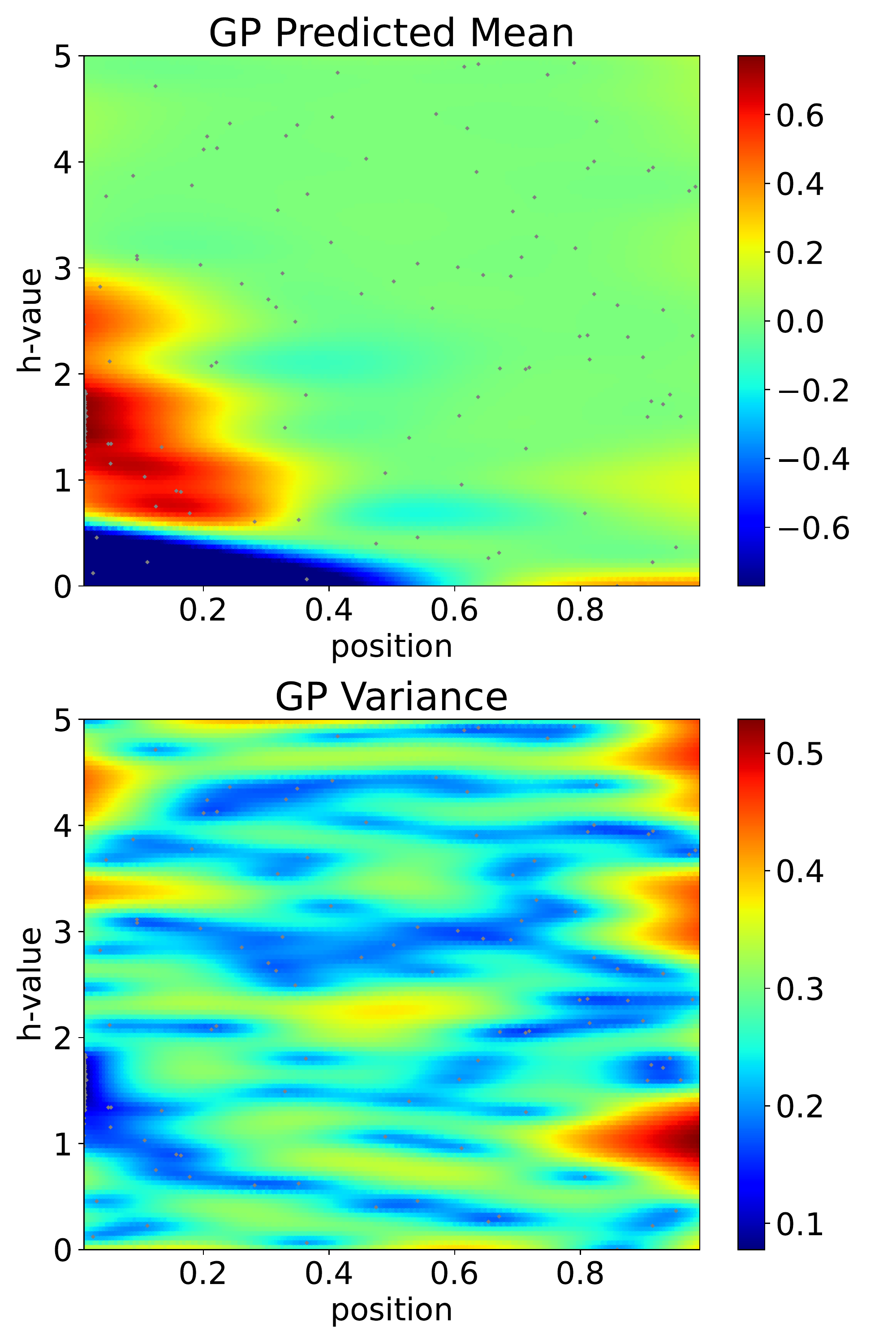}
    \includegraphics[width=0.32\textwidth]{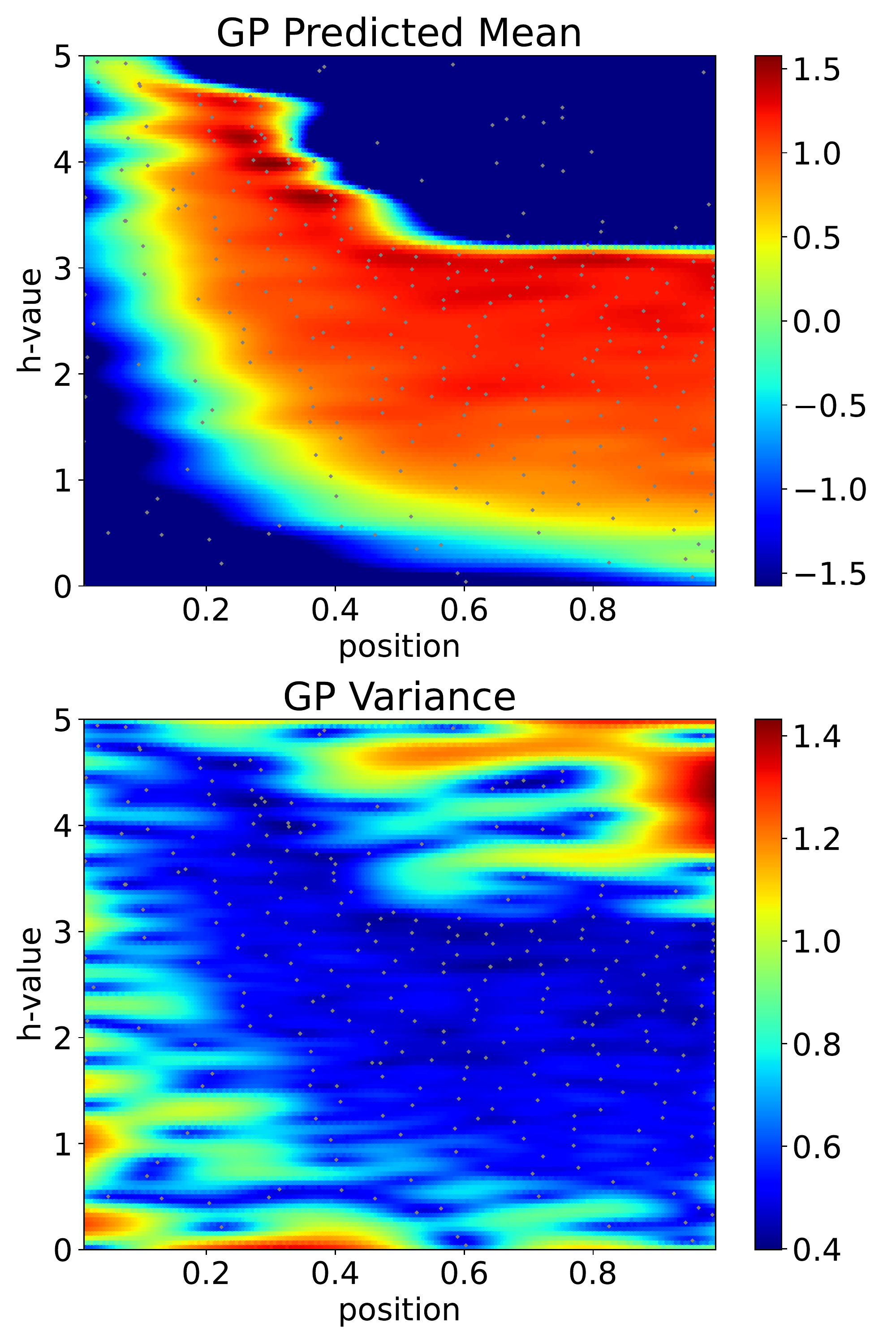}
    \caption{Bayesian optimization of the h-gain schedule for the weighted maximum clique problem on random graphs of density $0.1$ (left column), $0.5$ (middle column) and $0.9$ (right column). Mean (top sub-plots) and variance (bottom sub-plots) of the Gaussian process (GP) used by the Bayesian optimizer, where the goal is to maximize the objective function (e.g. red regions mean that the constructed h-gain schedule resulted in better cut values being sampled). The parameters of the HG schedule for Maximum Cut are visualized as a heatmap for and annealing time $1$ microsecond. Top plots show the maximum clique weight improvement as a function of $g(t) \in [0,5]$ (see eq.~\eqref{eq:hgain}) on the y-axis, where $t$ is the position in the schedule (x-axis). The small grey points indicate the points where the objective function was evaluated. Each heatmap is computed from a total of $300$ function calls ($100$ random points and $200$ optimization steps). Data from \texttt{DW\_2000Q\_LANL}. }
    \label{fig:hgain_heat_maps_max_clique}
\end{figure*}

Similarly to Figure~\ref{fig:hgain_heat_map_maxcut}, we again visualize the optimization of the HG schedule as a heatmap in Figure~\ref{fig:hgain_heat_maps_max_clique}. We see that the optimal point occurs at roughly position $0.2$ (anneal fraction) and has a HG value of around $1$. Notice that the region of good h-gain schedules for the tested maximum clique problem instances can be directly read off of Figure \ref{fig:hgain_heat_maps_max_clique} (as opposed to the singular best schedules shown in Figure \ref{fig:maxclique_schedules}), shown by the red region of the top sub-plot, where the y-axis h-value corresponds to the h-gain strength of the middle point of the h-gain schedule and the position on the x-axis corresponds to a proportion of the total annealing time.

\clearpage

\setlength\bibitemsep{0pt}
\printbibliography

@article{Johnson2011,
    author = {Johnson, Mark and Amin, Mohammad and Gildert, S and Lanting, Trevor and Hamze, F and Dickson, N and Harris, R and Berkley, Andrew and Johansson, J and Bunyk, Paul and Chapple, E and Enderud, C and Hilton, Jeremy and Karimi, Kamran and Ladizinsky, E and Ladizinsky, Nicolas and Oh, T and Perminov, I and Rich, C and Rose, Geordie},
    year = {2011},
    month = {05},
    pages = {194-8},
    title = {Quantum annealing with manufactured spins},
    volume = {473},
    journal = {Nature},
    doi = {10.1038/nature10012}
}

@inproceedings{Hahn2017ReducingBQ,
    title = {{Reducing Binary Quadratic Forms for More Scalable Quantum Annealing}},
    author = {Georg Hahn and Hristo N. Djidjev},
    booktitle = {IEEE International Conference on Rebooting Computing (ICRC)},
    year = {2017},
    pages = {1-8}
}

@article{Lucas2014,
    author = {Lucas, A.},
    title = {Ising formulations of many {NP} problems},
    year =  {2014},
    journal = {Front Phys},
    volume = {2},
    number = {5},
    pages = {1--27}
}

@article{Chapuis2019,
    author = {Guillaume Chapuis and Hristo Djidjev and Georg Hahn and Guillaume Rizk},
    title = {{Finding Maximum Cliques on the D-Wave Quantum Annealer}},
    year = {2019},
    journal = {Journal of Signal Processing Systems},
    volume = {91},
    number = {3-4},
    pages = {363--377}
}

@misc{TechnicalDescriptionDwave,
    title = {{\href{https://docs.dwavesys.com/docs/latest/doc_qpu.html}{Technical Description of the D-Wave Quantum Processing Unit}}},
    author = {D-Wave},
    year = {2020}
}

@misc{minorminer,
    title = {{\href{https://docs.ocean.dwavesys.com/projects/minorminer/en/latest/}{D-Wave Ocean Software Documentation: Minorminer}}},
    author = {D-Wave},
    year = {2020}
}

@article{ErdosRenyi1960,
    author = {Erd{\H o}s, P. and R{\'e}nyi, A.},
    title = {{On the Evolution of Random Graphs}},
    year = {1960},
    journal = {Publication of the Math Inst of the Hungarian Academy of Sciences},
    volume = {5},
    pages = {17--61}
}

@misc{bayesianopt,
    author = {Fernando Nogueira},
    title = {\href{https://github.com/fmfn/BayesianOptimization}{Bayesian Optimization: Open source constrained global optimization tool for Python}},
    year = {2014}
}

@article{Harris2018,
    title = {Phase transitions in a programmable quantum spin glass simulator},
    author = {Harris, R. and Sato, Y. and Berkley, A.J. and Reis, M. and Altomare, F. and Amin, M.H. and Boothby, K. and Bunyk, P. and Deng, C. and Enderud, C. and Huang, S. and Hoskinson, E. and Johnson, M.W. and Ladizinsky, E. and Ladizinsky, N. and Lanting, T. and Li, R. and Medina, T. and Molavi, R. and Neufeld, R. and Oh, T. and Pavlov, I. and Perminov, I. and Poulin-Lamarre, G. and Rich, C. and Smirnov, A. and Swenson, L. and Tsai, N. and Volkmann, M. and Whittaker, J. and Yao, J.},
    journal = {Science},
    year = {2018},
    volume = {361},
    number = {6398},
    pages = {162--165}
}

@article{Yamashiro2019,
    title = {Dynamics of reverse annealing for the fully connected p-spin model},
    author = {Yu Yamashiro and Masaki Ohkuwa and Hidetoshi Nishimori and Daniel A. Lidar},
    journal = {Phys. Rev. A},
    volume = {100},
    number = {5},
    pages = {052321},
    year = {2019}
}

@article{Ohkuwa2018,
    title = {Reverse annealing for the fully connected p-spin model},
    author = {Masaki Ohkuwa and Hidetoshi Nishimori and Daniel A. Lidar},
    journal = {Phys. Rev. A},
    volume = {98},
    number = {2},
    pages = {022314},
    year = {2018}
}

@article{Passarelli2020,
    title = {Reverse quantum annealing of the p-spin model with relaxation},
    author = {Gianluca Passarelli and Ka-Wa Yip and Daniel A. Lidar and Hidetoshi Nishimori and Procolo Lucignano},
    journal = {Phys. Rev. A},
    volume = {101},
    number = {2},
    pages = {022331},
    year = {2020}
}

@article{Venturelli2019,
    title = {Reverse quantum annealing approach to portfolio optimization problems},
    author = {Davide Venturelli and Alexei Kondratyev},
    journal = {Quantum Mach. Intell.},
    volume = {1},
    pages = {17--30},
    year = {2019}
}

@misc{McGeoch2018,
    title = {\href{https://www.dwavesys.com/sites/default/files/2_Wed_Am_PerfTips.pdf}{Performance Tuning for D-Wave Quantum Processors}},
    author = {Catherine McGeoch},
    year = {2018}
}

@article{King2018,
    author = {A. D. King and J. Carrasquilla and I. Ozfidan and J. Raymond and E. Andriyash and A. Berkley and M. Reis and T. M. Lanting and R. Harris and G. Poulin-Lamarre and A. Y. Smirnov and C. Rich and F. Altomare and P. Bunyk and J. Whittaker and L. Swenson and E. Hoskinson and Y. Sato and M. Volkmann and E. Ladizinsky and M. Johnson and J. Hilton and M. H. Amin},
    title = {Observation of topological phenomena in a programmable lattice of 1,800 qubits},
    journal = {Nature},
    volume = {560},
    pages = {456--460},
    year = {2018}
}

@article{King_2021_scaling,
    author = {Andrew D. King and Jack Raymond and Trevor Lanting and Sergei V. Isakov and Masoud Mohseni and Gabriel Poulin-Lamarre and Sara Ejtemaee and William Bernoudy and Isil Ozfidan and Anatoly Yu. Smirnov and Mauricio Reis and Fabio Altomare and Michael Babcock and Catia Baron and Andrew J. Berkley and Kelly Boothby and Paul I. Bunyk and Holly Christiani and Colin Enderud and Bram Evert and Richard Harris and Emile Hoskinson and Shuiyuan Huang and Kais Jooya and Ali Khodabandelou and Nicolas Ladizinsky and Ryan Li and P. Aaron Lott and Allison J. R. MacDonald and Danica Marsden and Gaelen Marsden and Teresa Medina and Reza Molavi and Richard Neufeld and Mana Norouzpour and Travis Oh and Igor Pavlov and Ilya Perminov and Thomas Prescott and Chris Rich and Yuki Sato and Benjamin Sheldan and George Sterling and Loren J. Swenson and Nicholas Tsai and Mark H. Volkmann and Jed D. Whittaker and Warren Wilkinson and Jason Yao and Hartmut Neven and Jeremy P. Hilton and Eric Ladizinsky and Mark W. Johnson and Mohammad H. Amin},
	title = {{Scaling advantage over path-integral Monte Carlo in quantum simulation of geometrically frustrated magnets}},
	journal = {Nature Communications},
    year = {2021},
	volume = {12},
	number = {1},
    doi = {10.1038/s41467-021-20901-5}
}

@article{King_2021,
	author = {Andrew D. King and Cristian D. Batista and Jack Raymond and Trevor Lanting and Isil Ozfidan and Gabriel Poulin-Lamarre and Hao Zhang and Mohammad H. Amin},
	title = {{Quantum Annealing Simulation of Out-of-Equilibrium Magnetization in a Spin-Chain Compound}},
	journal = {PRX Quantum},
	year = 2021,
	volume = {2},
	number = {3},
    doi = {10.1103/prxquantum.2.030317}
}

@article{Mockus1974,
    author = {Jonas Mockus},
    title = {{On Bayesian Methods for Seeking the Extremum}},
    journal = {Optimization Techniques},
    year = {1974},
    pages = {400--404}
}

@inproceedings{Mockus1977,
    author = {Jonas Mockus},
    title = {{On Bayesian Methods for Seeking the Extremum and their Application}},
    booktitle = {IFIP Congress},
    year = {1977},
    pages = {195--200}
}

@book{Mockus1989,
    author = {Jonas Mockus},
    title = {{Bayesian Approach to Global Optimization: Theory and Applications}},
    publisher = {Springer Netherlands},
    year = {1989}
}

@article{Perdomo2011,
    author = {A. Perdomo-Ortiz and S. E. Venegas-Andraca and A. Aspuru-Guzik},
    title = {A study of heuristic guesses for adiabatic quantum computation},
    journal = {Quantum Information Processing},
    volume = {10},
    year = {2011},
    pages = {33--52}
}

@article{Chancellor_2017,
    author = {Nicholas Chancellor},
	title = {Modernizing quantum annealing using local searches},
    journal = {New Journal of Physics},
	volume = {19},  
	number = {2},
	pages = {023024},
    year = {2017},
    doi = {10.1088/1367-2630/aa59c4}
}

@inproceedings{Pelofske_2020,
	author = {Elijah Pelofske and Georg Hahn and Hristo N. Djidjev}, 
	title = {Advanced anneal paths for improved quantum annealing},
	booktitle = {2020 {IEEE} International Conference on Quantum Computing and Engineering ({QCE})},
	year = {2020},
	publisher = {IEEE},
    doi = {10.1109/qce49297.2020.00040}
}

@article{PhysRevApplied.17.054033,
    title = {{Breakdown of the Weak-Coupling Limit in Quantum Annealing}},
    author = {Bando, Yuki and Yip, Ka-Wa and Chen, Huo and Lidar, Daniel A. and Nishimori, Hidetoshi},
    journal = {Phys. Rev. Applied},
    volume = {17},
    number = {5},
    pages = {054033},
    year = {2022},
    doi = {10.1103/PhysRevApplied.17.054033}
}

@article{PhysRevResearch.3.033006,
    title = {Mean field analysis of reverse annealing for code-division multiple-access multiuser detection},
    author = {Arai, Shunta and Ohzeki, Masayuki and Tanaka, Kazuyuki},
    journal = {Phys. Rev. Research},
    volume = {3},
    number = {3},
    pages = {033006},
    year = {2021},
    doi = {10.1103/PhysRevResearch.3.033006}
}

@article{PRXQuantum.1.020320,
    title = {{Simulating the Shastry-Sutherland Ising Model Using Quantum Annealing}},
    author = {Kairys, Paul and King, Andrew D. and Ozfidan, Isil and Boothby, Kelly and Raymond, Jack and Banerjee, Arnab and Humble, Travis S.},
    journal = {PRX Quantum},
    volume = {1},
    number = {2},
    pages = {020320},
    year = {2020},
    doi = {10.1103/PRXQuantum.1.020320}
}

@article{Hadfield_2019,
    author = {Stuart Hadfield and Zhihui Wang and Bryan O{\textquotesingle}Gorman and Eleanor Rieffel and Davide Venturelli and Rupak Biswas},
	title = {{From the Quantum Approximate Optimization Algorithm to a Quantum Alternating Operator Ansatz}},
	journal = {Algorithms},
	year = {2019},
	volume = {12},
	number = {2},
	doi = {10.3390/a12020034}
}

@article{Sack_2021,
	author = {Stefan H. Sack and Maksym Serbyn},
	title = {Quantum annealing initialization of the quantum approximate optimization algorithm},
	journal = {Quantum},
    year = {2021},
	volume = {5},
	pages = {491},
	doi = {10.22331/q-2021-07-01-491}
}

@misc{https://doi.org/10.48550/arxiv.1411.4028,
    author = {Farhi, Edward and Goldstone, Jeffrey and Gutmann, Sam},  
    title = {A Quantum Approximate Optimization Algorithm},
    publisher = {arXiv},
    year = {2014},
    doi = {10.48550/ARXIV.1411.4028}
}

@article{Egger_2021,
    author = {Daniel J. Egger and Jakub Mare{\v{c}}ek and Stefan Woerner},
	title = {Warm-starting quantum optimization},
	journal = {Quantum},
	year = {2021},	  
	publisher = {Verein zur Forderung des Open Access Publizierens in den Quantenwissenschaften},
	volume = {5},
	pages = {479},
    doi = {10.22331/q-2021-06-17-479}
}

@misc{https://doi.org/10.48550/arxiv.2207.05089,
    author = {Cain, Madelyn and Farhi, Edward and Gutmann, Sam and Ranard, Daniel and Tang, Eugene},    
    title = {The QAOA gets stuck starting from a good classical string},
    publisher = {arXiv},
    year = {2022},
    doi = {10.48550/ARXIV.2207.05089}
}

@article{ash2020warm,
    title = {On warm-starting neural network training},
    author = {Ash, Jordan and Adams, Ryan P},
    journal = {Advances in Neural Information Processing Systems},
    volume = {33},
    pages = {3884--3894},
    year = {2020}
}

@misc{https://doi.org/10.48550/arxiv.2301.01853,
    author = {Lopez-Bezanilla, Alejandro and Raymond, Jack and Boothby, Kelly and Carrasquilla, Juan and Nisoli, Cristiano and King, Andrew D.},
    title = {Kagome qubit ice},
    publisher = {arXiv},
    year = {2023},
    doi = {10.48550/ARXIV.2301.01853}
}

@article{Lanting2014,
    title = {{Entanglement in a Quantum Annealing Processor}},
    author = {Lanting, T. and Przybysz, A. J. and Smirnov, A. Yu. and Spedalieri, F. M. and Amin, M. H. and Berkley, A. J. and Harris, R. and Altomare, F. and Boixo, S. and Bunyk, P. and Dickson, N. and Enderud, C. and Hilton, J. P. and Hoskinson, E. and Johnson, M. W. and Ladizinsky, E. and Ladizinsky, N. and Neufeld, R. and Oh, T. and Perminov, I. and Rich, C. and Thom, M. C. and Tolkacheva, E. and Uchaikin, S. and Wilson, A. B. and Rose, G.},
    journal = {Phys Rev X},
    volume = {4},
    issue = {2},
    pages = {021041},
    year = {2014},
    doi = {10.1103/PhysRevX.4.021041}
}

@article{Das2008,
    title = {{Colloquium: Quantum annealing and analog quantum computation}},
    author = {Das, Arnab and Chakrabarti, Bikas K},
    journal = {Reviews of Modern Physics},
    volume = {80},
    number = {3},
    pages = {1061},
    year = {2008},
    doi = {10.1103/revmodphys.80.1061}
}

@article{Morita2008,
    title = {Mathematical foundation of quantum annealing},
    author = {Morita, Satoshi and Nishimori, Hidetoshi},
    journal = {Journal of Mathematical Physics},
    volume = {49},
    number = {12},
    pages = {125210},
    year = {2008},
    doi = {10.1063/1.2995837}
}

@article{Kadowaki1998,
    author = {Tadashi Kadowaki and Hidetoshi Nishimori},
    title = {{Quantum annealing in the transverse Ising model}},
    journal = {Physical Review E},
    year = {1998},
	volume = {58},
	number = {5},
	pages = {5355-5363},
    doi = {10.1103/physreve.58.5355}
}

@article{Boixo2013,
    title = {Experimental signature of programmable quantum annealing},
    author = {Boixo, Sergio and Albash, Tameem and Spedalieri, Federico M and Chancellor, Nicholas and Lidar, Daniel A},
    journal = {Nature communications},
    year = {2013},
    volume = {4},
    number = {1},
    pages = {1-8},
    doi={10.1038/ncomms3067}
}

@article{Boixo2014,
    author = {Sergio Boixo and Troels F. R{\o}nnow and Sergei V. Isakov and Zhihui Wang and David Wecker and Daniel A. Lidar and John M. Martinis and Matthias Troyer},
    title = {Evidence for quantum annealing with more than one hundred qubits},
    journal = {Nature Physics},
    year = {2014},  
    volume = {10},
    number = {3},
    pages = {218-224},
    doi = {10.1038/nphys2900}
}

@article{Mandra2016,
    author = {Salvatore Mandr{\`a} and Zheng Zhu and Wenlong Wang and Alejandro Perdomo-Ortiz and Helmut G. Katzgraber},
    title = {{Strengths and weaknesses of weak-strong cluster problems: A detailed overview of state-of-the-art classical heuristics versus quantum approaches}},
    journal = {Physical Review A},
    year = {2016},
    volume = {94},  
    number = {2},
    doi = {10.1103/physreva.94.022337}
}

@article{Hauke2020,
    title = {{Perspectives of quantum annealing: Methods and implementations}},
    author = {Hauke, Philipp and Katzgraber, Helmut G and Lechner, Wolfgang and Nishimori, Hidetoshi and Oliver, William D},
    journal = {Reports on Progress in Physics},
    volume = {83},
    number = {5},
    pages = {054401},
    year = {2020},
    doi = {10.1088/1361-6633/ab85b8}
}

@article{Chakrabarti2023,
    title = {Quantum annealing and computation: challenges and perspectives},
    author = {Chakrabarti, Bikas K and Leschke, Hajo and Ray, Purusattam and Shirai, Tatsuhiko and Tanaka, Shu},
    journal = {Philosophical Transactions of the Royal Society A},
    volume = {381},
    number = {2241},
    pages = {20210419},
    year = {2023},
    doi = {10.1098/rsta.2021.0419}
}

@misc{tasseff2022emerging,
    title = {{On the Emerging Potential of Quantum Annealing Hardware for Combinatorial Optimization}}, 
    author = {Byron Tasseff and Tameem Albash and Zachary Morrell and Marc Vuffray and Andrey Y. Lokhov and Sidhant Misra and Carleton Coffrin},
    year = {2022},
    note = {arXiv:2210.04291},
    doi = {https://doi.org/10.48550/arXiv.2210.04291}
}

@misc{Cai2014,
    title = {A practical heuristic for finding graph minors}, 
    author = {Jun Cai and William G. Macready and Aidan Roy},
    year = {2014},
    note = {arXiv:1406.2741},
    doi = {https://doi.org/10.48550/arXiv.1406.2741}
}

@article{King2022,
	author = {Andrew D. King and Sei Suzuki and Jack Raymond and Alex Zucca and Trevor Lanting and Fabio Altomare and Andrew J. Berkley and Sara Ejtemaee and Emile Hoskinson and Shuiyuan Huang and Eric Ladizinsky and Allison J. R. MacDonald and Gaelen Marsden and Travis Oh and Gabriel Poulin-Lamarre and Mauricio Reis and Chris Rich and Yuki Sato and Jed D. Whittaker and Jason Yao and Richard Harris and Daniel A. Lidar and Hidetoshi Nishimori and Mohammad H. Amin},  
	title = {Coherent quantum annealing in a programmable 2,000 qubit Ising chain},
	journal = {Nature Physics},
    year = {2022},  
	volume = {18},
	number = {11},
	pages = {1324-1328},
  	doi = {10.1038/s41567-022-01741-6}
}

@misc{Farhi2000,
    author = {Farhi, Edward and Goldstone, Jeffrey and Gutmann, Sam and Sipser, Michael},    
    title = {{Quantum Computation by Adiabatic Evolution}},    
    year = {2000},
    note = {arXiv:quant-ph/0001106},
    doi = {https://doi.org/10.48550/arXiv.quant-ph/0001106}
}

@article{Passarelli2022,
    title = {Standard quantum annealing outperforms adiabatic reverse annealing with decoherence},
    author = {Gianluca Passarelli and Ka-Wa Yip and Daniel A. Lidar and Procolo Lucignano},
    journal = {Phys. Rev. A},
    volume = {105},
    number = {032431},
    year = {2022}
}

@article{Golden2021,
    title = {{Reverse annealing for nonnegative/binary matrix factorization}},
    author = {John Golden and Daniel O'Malley},
    year = {2021},
    journal = {PLoS ONE},
    volume = {16},
    number = {1},
    pages = {e0244026}
}

@article{Rocutto2021,
    title = {{Quantum Semantic Learning by Reverse Annealing of an Adiabatic Quantum Computer}},
    author = {Lorenzo Rocutto and Claudio Destri and Enrico Prati},
    journal = {Advanced Quantum Technologies},
    volume = {4},
    number = {2},
    year = {2021},
    pages = {2000133}
}

@article{Grant2021,
    title = {Benchmarking Quantum Annealing Controls with Portfolio Optimization},
    author = {Grant, Erica and Humble, Travis S. and Stump, Benjamin},
    journal = {Phys. Rev. Appl.},
    volume = {15},
    issue = {1},
    pages = {014012},
    numpages = {14},
    year = {2021},
    publisher = {American Physical Society},
    doi = {10.1103/PhysRevApplied.15.014012},
    url = {https://link.aps.org/doi/10.1103/PhysRevApplied.15.014012}
}

@article{Marshall2019,
    title = {Power of Pausing: Advancing Understanding of Thermalization in Experimental Quantum Annealers},
    author = {Marshall, Jeffrey and Venturelli, Davide and Hen, Itay and Rieffel, Eleanor G.},
    journal = {Phys. Rev. Appl.},
    volume = {11},
    issue = {4},
    pages = {044083},
    numpages = {23},
    year = {2019},
    publisher = {American Physical Society},
    doi = {10.1103/PhysRevApplied.11.044083},
    url = {https://link.aps.org/doi/10.1103/PhysRevApplied.11.044083}
}

@article{Passarelli2019,
    title = {Improving quantum annealing of the ferromagnetic $p$-spin model through pausing},
    author = {Passarelli, G. and Cataudella, V. and Lucignano, P.},
    journal = {Phys. Rev. B},
    volume = {100},
    issue = {2},
    pages = {024302},
    numpages = {11},
    year = {2019},
    publisher = {American Physical Society},
    doi = {10.1103/PhysRevB.100.024302},
    url = {https://link.aps.org/doi/10.1103/PhysRevB.100.024302}
}

@article{Chen2020,
    title = {Why and When Pausing is Beneficial in Quantum Annealing},
    author = {Chen, Huo and Lidar, Daniel A.},
    journal = {Phys. Rev. Appl.},
    volume = {14},
    issue = {1},
    pages = {014100},
    numpages = {21},
    year = {2020},
    publisher = {American Physical Society},
    doi = {10.1103/PhysRevApplied.14.014100},
    url = {https://link.aps.org/doi/10.1103/PhysRevApplied.14.014100}
}

@article{Katzgraber2014,
    title = {Glassy Chimeras Could Be Blind to Quantum Speedup: Designing Better Benchmarks for Quantum Annealing Machines},
    author = {Katzgraber, Helmut G. and Hamze, Firas and Andrist, Ruben S.},
    journal = {Phys. Rev. X},
    volume = {4},
    issue = {2},
    pages = {021008},
    numpages = {8},
    year = {2014},
    month = {4},
    publisher = {American Physical Society},
    doi = {10.1103/PhysRevX.4.021008},
    url = {https://link.aps.org/doi/10.1103/PhysRevX.4.021008}
}

@article{Mandra2017,
    title = {The pitfalls of planar spin-glass benchmarks: raising the bar for quantum annealers (again)},
    author = {Salvatore Mandr{\`a} and Helmut G. Katzgraber and Creighton Thomas},
    year = {2017},
    journal = {Quantum Sci. Technol.},
    volume = {2},
    number = {3},
    pages = {038501},
    year = {2017}
}

@article{QianHeng2013,
    title = {{An Alternative Approach to Construct the Initial Hamiltonian of the Adiabatic Quantum Computation}},
    author = {Duan Qian-Heng and Zhang Shuo and Wu Wei and Chen Ping-Xing},
    year = {2013},
    journal = {Chinese Physics Letters},
    volume = {30},
    number = {1},
    pages = {010302}
}

@article{Grass2019,
    title = {{Quantum Annealing with Longitudinal Bias Fields}},
    author = {Tobias Gra{\ss}},
    journal = {Phys. Rev. Lett.},
    volume = {123},
    pages = {120501},
    year = {2019}
}

@article{PhysRevX.5.019901,
    title = {Erratum: Glassy Chimeras could be blind to quantum speedup: Designing better benchmarks for quantum annealing machines [Phys. Rev. X 4, 021008 (2014)]},
    author = {Weigel, Martin and Katzgraber, Helmut G. and Machta, Jonathan and Hamze, Firas and Andrist, Ruben S.},
    collaboration = {Octomore Collaboration},
    journal = {Phys. Rev. X},
    volume = {5},
    issue = {1},
    pages = {019901},
    numpages = {2},
    year = {2015},
    month = {1},
    publisher = {American Physical Society},
    doi = {10.1103/PhysRevX.5.019901},
    url = {https://link.aps.org/doi/10.1103/PhysRevX.5.019901}
}

@misc{jauma2023exploring,
    title={Exploring Quantum Annealing Architectures: A Spin Glass Perspective}, 
    author={Gabriel Jaumà and Juan José García-Ripoll and Manuel Pino},
    year={2023},
    eprint={2307.13065},
    archivePrefix={arXiv},
    primaryClass={quant-ph}
}

@article{Callison2022,
    title = {Hybrid quantum-classical algorithms in the noisy intermediate-scale quantum era and beyond},
    author = {Callison, Adam and Chancellor, Nicholas},
    journal = {Phys. Rev. A},
    volume = {106},
    issue = {1},
    pages = {010101},
    numpages = {17},
    year = {2022},
    publisher = {American Physical Society},
    doi = {10.1103/PhysRevA.106.010101},
    url = {https://link.aps.org/doi/10.1103/PhysRevA.106.010101}
}

@misc{dattani2019pegasus,
    title={Pegasus: The second connectivity graph for large-scale quantum annealing hardware}, 
    author={Nike Dattani and Szilard Szalay and Nick Chancellor},
    year={2019},
    eprint={1901.07636},
    archivePrefix={arXiv},
    primaryClass={quant-ph}
}

@misc{boothby2020nextgeneration,
    title={Next-Generation Topology of D-Wave Quantum Processors}, 
    author={Kelly Boothby and Paul Bunyk and Jack Raymond and Aidan Roy},
    year={2020},
    eprint={2003.00133},
    archivePrefix={arXiv},
    primaryClass={quant-ph}
}

@inproceedings{zbinden2020embedding,
    title={Embedding algorithms for quantum annealers with chimera and pegasus connection topologies},
    author={Zbinden, Stefanie and B{\"a}rtschi, Andreas and Djidjev, Hristo and Eidenbenz, Stephan},
    booktitle={International Conference on High Performance Computing},
    pages={187--206},
    year={2020},
    organization={Springer}
}

@misc{lobe2021minor,
    title={Minor Embedding in Broken Chimera and Pegasus Graphs is NP-complete}, 
    author={Elisabeth Lobe and Annette Lutz},
    year={2021},
    eprint={2110.08325},
    archivePrefix={arXiv},
    primaryClass={quant-ph}
}

@article{Lobe_2021,
    author = {Elisabeth Lobe and Lukas Schürmann and Tobias Stollenwerk},
    title = {Embedding of complete graphs in broken Chimera graphs},
    journal = {Quantum Information Processing},
    year = {2021},
	volume = {20},  
	number = {7},
    doi = {10.1007/s11128-021-03168-z}
}

@article{boothby2016fast,
    title={Fast clique minor generation in Chimera qubit connectivity graphs},
    author={Boothby, Tomas and King, Andrew D and Roy, Aidan},
    journal={Quantum Information Processing},
    volume={15},
    pages={495--508},
    year={2016},
    publisher={Springer}
}

\end{document}